\documentclass[a4paper,twocolumn,11pt,accepted=2022-03-23]{quantumarticle}
\pdfoutput=1

\usepackage[utf8]{inputenc}
\usepackage[T1]{fontenc}
\usepackage{graphicx}
\usepackage[centertags]{amsmath}
\usepackage{amsfonts}
\usepackage{amssymb}
\usepackage[english]{babel}
\usepackage{csquotes}
\usepackage{textcomp}
\usepackage{bm}
\usepackage{euscript}
\usepackage{multirow,tabularx}
\usepackage[dvipsnames]{xcolor}
\definecolor{darkblue}{rgb}{0,0,.6}

\usepackage{tikz}
\usepackage{lipsum}

\usepackage[kerning=true,protrusion=true,expansion=true]{microtype}
\graphicspath{ {./images/} }

\usepackage[backend=biber,style=phys,articletitle=true,biblabel=brackets,pageranges=true,arxiv=none,mincitenames=1,maxcitenames=1,maxnames=5,uniquename=false,uniquelist=false]{biblatex}
\usepackage{hyperref}
\hypersetup{colorlinks=true, linkcolor=darkblue, citecolor=darkblue}

\newcommand{\coefficient}{\mathfrak}
\newcommand*{\sfG}{\mathsf{G}}

\newcommand*{\sfM}{\mathsf{M}}
\newcommand*{\s}[1]{\ensuremath{_\text{#1}}}
\newcommand*{\up}[1]{\ensuremath{^\text{#1}}}
\renewcommand*{\bold}[1]{\mathbf{#1}}
\newcommand*{\ket}[1]{\left|#1\right\rangle}

\newcommand*{\dyad}[2]{\left|#1\middle\rangle\middle\langle#2\right|}

\newcommand*{\coeffL}{\mathsf{L}}
\newcommand{\comm}[2]{\left[ #1 , #2 \right]}
\newcommand{\qdr}[1]{\hat{\mathbf{#1}}}
\newcommand{\ii}{\ensuremath{\mathrm{i}}}
\newcommand{\mean}[1]{\ensuremath{\left\langle #1 \right\rangle}}
\renewcommand{\vec}[1]{\mathbf{#1}}

\colorlet{acolor}{red!75!black}
\colorlet{mcolor}{blue!75!black}
\def\BY{\begin{eqnarray}}
\def\EY{\end{eqnarray}}
\def\BE{\begin{equation}}
\def\EE{\end{equation}}
\def\BEA{\begin{eqnarray}}
\def\EEA{\end{eqnarray}}

\def\L{\label}
\def\nn{\nonumber}
\def\({\left (}
\def\){\right )}
\def\[{\left [}
\def\]{\right]}
\def\<{\langle}
\def\>{\rangle}

\newcolumntype{Y}{>{\centering\arraybackslash}X}

\newcommand*{\coeffK}{\mathsf{K}}
\newcommand*{\coeffM}{\mathsf{M}}

\colorlet{acolor}{red!60!black}

\begin{document}

%-----------------------------------------------------------
\title{Atom-Mechanical Hong-Ou-Mandel Interference}% \vspace{1cm}
\newcommand{\UPOL}{Department of Optics, Palack{\'y} University, 17. Listopadu 12, 771 46 Olomouc, Czech Republic}
\date{}
\author{A. D. Manukhova}
\email{alisamanukhova@gmail.com
%, alisadmitrievna.manukhova@upol.cz
}
\orcid{0000-0003-3406-9826}
\affiliation{\UPOL}
\author{A.~A.~Rakhubovsky}
\email{rakhubovsky@optics.upol.cz}
\orcid{0000-0001-8643-670X}
\affiliation{\UPOL}
\author{R.~Filip}
\email{filip@optics.upol.cz}
\orcid{0000-0003-4114-6068}
\affiliation{\UPOL}
\maketitle

\begin{abstract}
%%%%%%%%%%%%%%%%%%%%%%%%%%%%%%%%%%%%%%%%%%%%%%%%%%%%%%%%%%%%

Quantum coupling between mechanical oscillators and atomic gases  generating entanglement has been recently experimentally demonstrated using their subsequent interaction with light.
The next step is to build a hybrid atom-mechanical quantum gate showing bosonic interference effects of single quanta in the atoms and oscillators.
We propose an experimental test of Hong-Ou-Mandel interference between single phononic excitation and single collective excitation of atoms using the optical connection between them.
A single optical pulse is sufficient to build a hybrid quantum-nondemolition gate to observe the bunching of such different quanta.
The output atomic-mechanical state exhibits a probability of a hybrid bunching effect that proves its nonclassical aspects.
This proposal opens a feasible road to broadly test such advanced quantum bunching phenomena in hybrid systems with different specific couplings.

%%%%%%%%%%%%%%%%%%%%%%%%%%%%%%%%%%%%%%%%%%%%%%%%%%%%%%%%%%%%
\end{abstract}

%%%%%%%%%%%%%%%%%%%%%%%%%%%%%%%%%%%%%%%%%%%%%%%%%%%%%%%%%%%%
\section{Introduction\L{I}}
%%%%%%%%%%%%%%%%%%%%%%%%%%%%%%%%%%%%%%%%%%%%%%%%%%%%%%%%%%%%

Hybridization of matter quantum platforms using light as an intermediary is currently growing in directions in quantum technology.
The aim of this development is to understand the compatibility of different experimental platforms and combine the advantages and capabilities of different parts into one hybrid system.
A pioneering road connects atomic ensembles with mechanical oscillators of optomechanical cavities~\cite{aspelmeyer_cavity_2014}.
The outstanding degree of quantum control over atoms makes them an excellent platform for quantum information~\cite{julsgaard_experimental_2004,sangouard_quantum_2011,PhysRevA.103.062426,Yu2020,Masalaeva_2020,SOKOLOV2020126762,0d5b4c3194cc4260b8712d40a35abd11,19840a96f37e4257a691e279bc03f5ef}, quantum memory~\cite{Cho:16},
and quantum simulations~\cite{gross_quantum_2017}.
Mechanical oscillators, having huge quality factors~\cite{hoj_ultracoherent_2021,norte_mechanical_2016,tsaturyan_ultracoherent_2017,Gärtner2018,Ma2020}, appear suitable for quantum sensing~\cite{monteiro2020,Ahn2020,Ranjit2016,Magrini2021}
and fundamental physics tests~\cite{Kaltenbaek2016,Pikovski2012,PhysRevLett.121.220404}.
Importantly, mechanical systems offer an access to quantum nonlinearities in continuous-variable regime~\cite{ochs_amplification_2021,PhysRevA.104.013501,Ma2020} not easily accessible in atomic systems.
Recently, coupling between mechanical oscillators and atoms reached a new phase of experimental development.

The most recent experiments show
that a spin mode of the warm atomic ensemble can interact with a mechanical mode of a distant optomechanical cavity using light as a mediator.
In~\cite{Karg2020}, the authors reported Einstein-Podolsky-Rosen-type~(EPR) correlations in a hybrid system consisting of a mechanical oscillator and a spin oscillator.
A vibrational mode of a highly stressed dielectric membrane, which was embedded in a free-space optical cavity, constituted the mechanical oscillator.
The spin oscillator had been prepared in a warm ensemble of optically pumped atoms confined in a spin-preserving microcell.
The two oscillators were coupled to an itinerant light in a cascaded fashion, that is the light interacted with the mechanics between its two interactions with atoms.
The authors have shown $5.5$~dB of two-mode squeezing of thermal fluctuations in both oscillators which is an important step towards quantum EPR-type entanglement.

In~\cite{Thomas2020}, the authors realized a similar long-distance interaction using a laser beam in a loop geometry.
Free-space laser beam coupled a collective atomic spin and a micromechanical membrane, both in room-temperature environment.
Through the loop the systems could exchange light photons, realizing a bidirectional interaction.
The loop led to an interference of quantum noise introduced by the light field ---
for any system that couples to the light twice and with opposite phase, quantum noise interferes destructively and associated decoherence is suppressed.
The versatility of light-mediated interactions is demonstrated this way.
The authors engineered a beam-splitter and a parametric-gain coupling between atoms and mechanics and could switch from these couplings to a dissipative one by applying a phase shift to the light field between the systems.
In both works, the distance between atomic ensemble and mechanics was of the order of one meter.
Thus, the basic possibility to couple atomic ensemble with a mechanical mode is conclusively proven.

The next step is to turn the hybrid entangling coupling to the pulsed hybrid gate and test its performance.
Pulsed operation brings a number of advantages, including working with modern tools of quantum optics~\cite{PhysRevA.103.043701}
and, compared to a continuous-wave driving, a possibility to get rid of thermal decoherence by operating on shorter timescales.
For the applications, it is advantageous to build a hybrid quantum nondemolition gate that allows to use geometric phase effects~\cite{Vostrosablin2017,Vostrosablin2018}.
Quantum nondemolition gate is basic continuous-variable gate capable to build not only all up-to-quadratic nonlinearities~\cite{PhysRevLett.88.097904}  but
also higher-order nonlinearities~\cite{PhysRevA.97.022329}.
Such hybrid gates need to be tested at the level of single quanta before they will be used.
They can materialize new hybrid bunching between phononic and atomic excitations.
A phonon of mechanical oscillations can change its nature and add to an atomic excitation, and simultaneously, an atomic excitation can be transferred and increase the number of phonons.
These two effects can superpose as it happens in the optical Hong-Ou-Mandel experiment for a pair of photons interfering at a balanced beam-splitter~\cite{HOM1987, Bouchard}.
Hong-Ou-Mandel interference effect rises despite the phase insensitive nature of single photons.
The ideal bunching superposes the photon pairs at one or either output with a probability that will never appear for classical phase-insensitive states~\cite{Furusawa2017}.
This proves that bunching for nonclassical states goes beyond the interference effects known for classical continuous waves and emphasizes truly quantum nature of excitations.

At the moment, numerous proposals are put forward to test the Hong-Ou-Mandel effect with different platforms besides optical photons.
Some of these are already implemented experimentally.
It is worth mentioning such bosonic platforms as surface plasmon polaritons, i.e. the quanta of the surface plasma waves~\cite{Heeres2013,Dieleman2017};
phonons, the quantized excitation of mechanical motion~\cite{Toyoda2015,Tamura2020};
collective atomic excitations, where the HOM effect is obtained using the Rydberg blockade~\cite{Li2016}.
Massive particles such as atoms also are able to provide two-particle interferences~\cite{Kaufman2014,Preiss2015,Lopes2015,Kaufman2018}.
Besides, the HOM effect has been proposed using quantum memory cell instead of a beam-splitter~\cite{ManukhovaSPb2020}.
Finally,  not only bosons but also fermions, namely, electrons can interfere in an HOM-like arrangement~\cite{Freulon2015}, and
the anti-HOM effect by interfering bosonic and fermionic wavefunctions of entangled photons has been recently experimentally demonstrated~\cite{vetlugin2021anti,Li2021}.
The Hong-Ou-Mandel effect itself can be used for applications like quantum metrology and sensing~\cite{BasiriEsfahani2015}.
Two-photon output states emerging from the HOM interferometer are an example of NOON states known for their capability of achieving metrological sensitivity superior to classical states~\cite{thekkadath_quantumenhanced_2020}.

This paper proposes a feasible atom-mechanical Hong-Ou-Mandel experiment capable of proving such a nonclassical interference regime for the hybrid quantum nondemolition gate.
This hybrid gate uses a single pulse of squeezed light interacting sequentially with atoms and mechanical oscillator.
The pulse is subsequently measured by homodyne detection whose output controls the atomic state.
First, we analyze nonclassical atom-optical and optomechanical Hong-Ou-Mandel effects separately.
We propose experiments to demonstrate them on current experimental platforms.
Finally, we present their combination in the atom-mechanical gate and derive conditions for a successful demonstration of nonclassical atom-phonon bunching.
An experimental test of our proposal will prove the new level of quantum control for hybrid systems and stimulate proposals and verifications of hybrid bunching between other bosonic platforms.

%%%%%%%

%%%%%%%%%%%%%%%%%%%%%%%%%%%%%%%%%%%%%%%%%%%%%%%%%%%%%%%%%%%%
\section{Results}
%%%%%%%%%%%%%%%%%%%%%%%%%%%%%%%%%%%%%%%%%%%%%%%%%%%%%%%%%%%%

\begin{figure}[ht]
\begin{center}
\includegraphics[width=1.\linewidth]{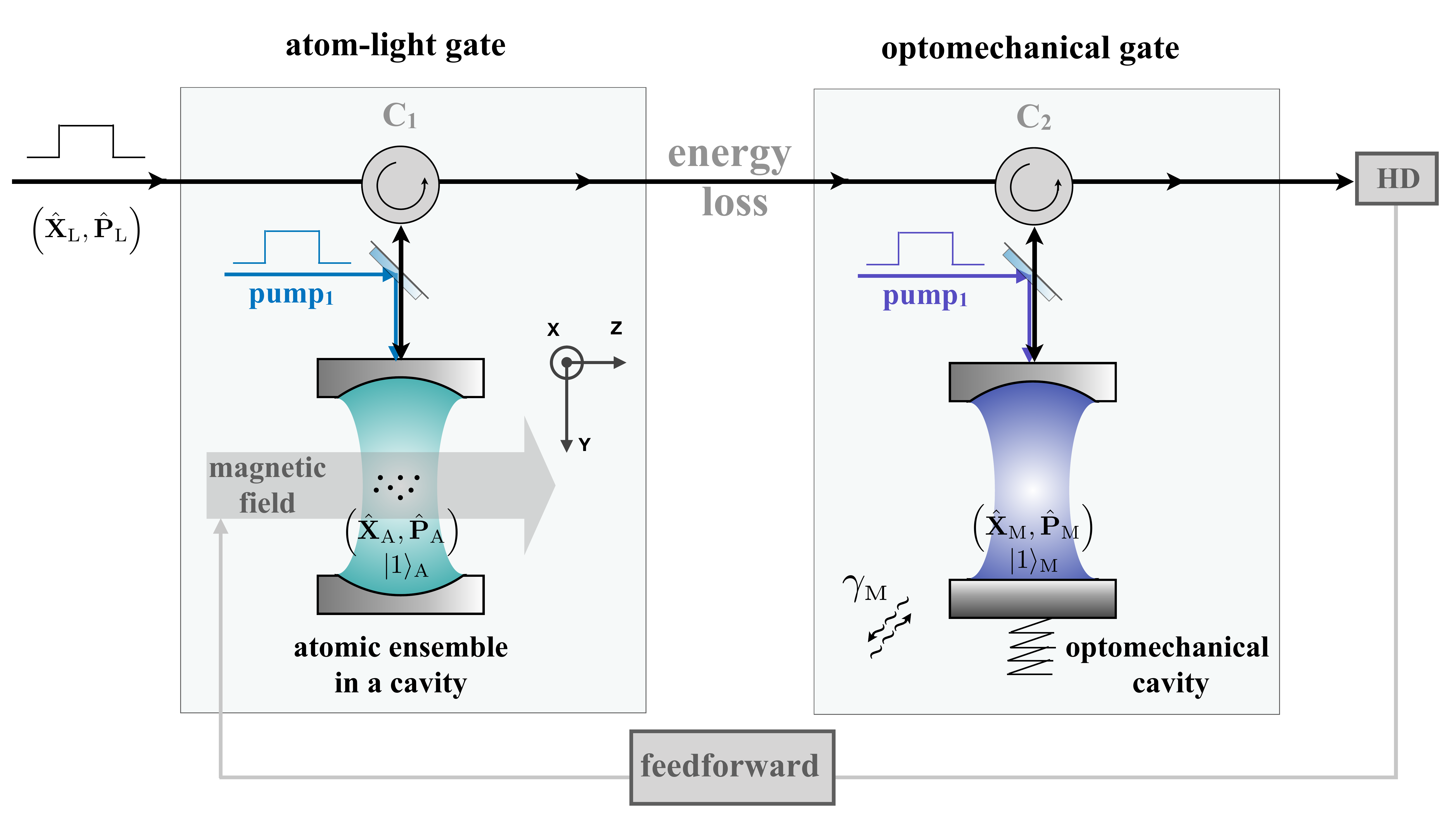}
\caption{HOM effect via QND gate between an atomic ensemble and a mechanical oscillator excited by single quanta $|1\rangle\s{\tiny{A}}$ and $|1\rangle\s{\tiny{M}}$.
A quantum light pulse with a rectangular temporal profile sequentially passes the atomic ensemble in a cavity
and then the optomechanical cavity and then goes to the homodyne detector (HD).
Routing of the pulse is enabled by the circulators C$_{1,2}$.
Within the cavities the optical pulse is coupled to atoms and mechanics respectively via QND interactions enabled by strong classical optical pumps.
The homodyne detection data are used to control the optical feedforward procedure after the detection to shift the atomic quadratures.
The homodyne measurement and magnetic feedforward control via magnetic field phase shifter are optimized to perform the atom-mechanical QND interaction and the squeezed light is used to achieve large entangling power.
}
\label{Fig1}
\end{center}
\end{figure}

In this article, we demonstrate the possibility to observe an analogue of the Hong-Ou-Mandel~(HOM) interference
using a quantum nondemolition~(QND) gate between a cloud of atoms and a mechanical mode of an optomechanical cavity.
The schematic diagram of the setup that allows realization of the gate is shown in the Fig.~\ref{Fig1} and is considered in detail in~\cite{Manukhova2020}.
Note that such an atomic-mechanical system consists of two parts,
which in turn perform QND gates coupling the mediating optical mode to the atomic or to the mechanical mode.
We will show that each of the three gates (atom-optical, optomechanical, and atom-mechanical) is capable of providing the HOM effect given feasible experimental parameters.

For its operation and observation of the HOM interference, the gate requires the physical parameters (including the coupling rates and cavities linewidths) that are within reach from the ones used in~\cite{Thomas2020,Karg2020}.
In both these works, the atoms are in free space, however, the presence of an atomic cavity is not essential for our treatment.
Very same results can be obtained with atoms in free space, and we assume a presence of the cavity for atoms for the convenience of single-mode description.
Two specifics are critical for the observation of HOM effect in the hybrid system~\cite{Manukhova2020}.
First, both interactions are of QND type  (in~\cite{Thomas2020}, another type of interaction was implemented for the optomechanical part).
Another important requirement is cooling and isolation of the mechanical mode from its thermal environment
since the thermal noises can totally destroy the HOM interference.
In~\cite{Karg2020}, mechanics is at the room temperature, however cooling of a membrane oscillator~\cite{jockel_sympathetic_2015,peterson_laser_2016,christoph_combined_2018} and operation at low temperatures, in particular, in cryogenic environment~\cite{jayich_cryogenic_2012,purdy_cavity_2012} has been reported previously.

We consider the HOM effect as a bunching of two excitations, initially in two different interacting subsystems, in one of them.
Our aim is to demonstrate the nonclassical HOM effect, i.e.  the buildup of the bunched state via the second-order interference not achievable by phase-randomized classical waves after the same interaction.
As in optical HOM effect at the beam-splitter, we assume incoherent mixture of ground (vacuum) and single-boson states at the input of the QND gate,
investigate the dependences of the HOM matrix element of the bunched state on the parameters of the gate
and compare it to the HOM element corresponding to the classical phase-random coherent cases.
For classical phase-stabilized states the bunching can be obtained in part via the first-order phase-sensitive interference between the amplitudes.
It has been shown that the classical states are capable of producing nearly perfect visibility of two-photon interference~\cite{kim_conditions_2013,sadana_near100_2019}.
To eliminate this phase effect (which is not available to single-boson states as not posessing a well-defined phase) we use phase randomized coherent states to determine the classical thresholds.

%----------------------------------------------------------------
\subsection{Atom-Light Hong-Ou-Mandel Interference}\label{secA}
%----------------------------------------------------------------

We first examine a system comprising a pulse of traveling light and an atomic ensemble for the capability to demonstrate a HOM effect via a QND coupling.
This coupling can be naturally observed in such systems as has been reported in~\cite{Vasilakis,Novikova}.
We briefly reiterate the strategy to achieve a QND coupling in the system and then derive the input-output relations using which we evaluate the output statistics of light and atoms.

The basic principle to realize the QND gate between an atomic ensemble and light is the following~\cite{BOOK}.
A pulse of quantum signal field, accompanied by the classical driving, passes through the atomic ensemble, located in the cavity with optical decay rate~$\kappa_\text{\tiny{A}}$.
Both fields are the pulses with rectangular time profiles, of duration~$\tau$.
The time profiles of the light fields can be considered as an additional degree of freedom that allows to control the gain more subtly and enhance the overall coupling efficiency~\cite{Masalaeva_2020}.
In this manuscript, we use rectangular pulse shapes for simplicity.
To describe the atomic subsystem we consider the state of an ensemble of atoms at room temperature, each having two stable ground states.
We assume a strong magnetic driving along the $Z$-axis for the atomic ensemble that allows us to apply the Holstein-Primakoff transformation
and consider normalized collective spins~$(\hat{\bold{X}}_\text{\tiny{A}},\hat{\bold{P}}_\text{\tiny{A}})$ as very long-lived canonical atomic variables ($\comm{\qdr X_\text{\tiny{A}}}{ \qdr P_\text{\tiny{A}}} = 2 \ii$).
The phase of the driving is chosen in a way that the effective Hamiltonian for the atom-light interaction is~$\hat{H}\s{\text{\tiny{LA}}} = \hbar g_\text{\tiny{A}} \hat{\bold{X}}_\text{\tiny{A}}\hat{p}_\text{c}$,
where $g_\text{\tiny{A}}$ is the coupling strength
and $\hat{p}_\text{c}$ is the canonical phase quadrature of the intracavity light.
The light-matter coupling thus interfaces a \emph{single} quantum mode of atomic ensemble to a \emph{single} quantum mode of light.
This marks a departure from the traditional HOM effect where the mode structure of input photons can be complicated and directly influence the visibility of interference.
The single-mode character of the light-matter coupling is typical for single-rail encoding in linear-optical quantum information processing~\cite{kok_linear_2007}.
After the interaction with atoms, the signal leaves the atomic cavity and at the output can be derived using the input-output relations.
At this stage we also take into account the loss that occurs during the coupling process.

After the interaction, the initial quadratures $(\hat{\bold{X}}_\text{\tiny{A}}^0,\hat{\bold{P}}_\text{\tiny{A}}^0,\hat{\bold{X}}_\text{\tiny{L}}^{0},\hat{\bold{P}}_\text{\tiny{L}}^{0})^T$
transform to the final quadratures $(\hat{\bold{X}}\up{out}_\text{\tiny{A}},\hat{\bold{P}}\up{out}_\text{\tiny{A}},\hat{\bold{X}}{}\up{out}_\text{\tiny{L}},\hat{\bold{P}}{}\up{out}_\text{\tiny{L}})^T$ as follows:
\begin{align}
& \hat{\bold{X}}_\text{\tiny{A}}\up{out} = \hat{\bold{X}}_\text{\tiny{A}}^0+\hat{\mathbf{N}}_{\text{X}_\text{\tiny{A}}},\\
& \hat{\bold{P}}_\text{\tiny{A}}\up{out}=\hat{\bold{P}}_\text{\tiny{A}}^0-\mathsf{G}_\text{\tiny{A}}\hat{\bold{P}}^{0}_\text{\tiny{L}}+\hat{\mathbf{N}}_{\text{P}_\text{\tiny{A}}}, \\
& \hat{\bold{X}}{}\up{out}=\mathsf{T}_\text{\tiny{L}}\hat{\bold{X}}^{0}_\text{\tiny{L}}+\mathsf{G}_\text{\tiny{L}} \hat{\bold{X}}_\text{\tiny{A}}^0+\hat{\mathbf{N}}_{\text{X}_\text{\tiny{L}}}, \label{atom1}\\
& \hat{\bold{P}}{}\up{out}=\mathsf{T}_\text{\tiny{L}}\hat{\bold{P}}^{0}_\text{\tiny{L}}+\hat{\mathbf{N}}_{\text{P}_\text{\tiny{L}}} \label{atom4},
\end{align}
where $(\hat{\bold{X}}_\text{\tiny{L}}^{0},\hat{\bold{P}}_\text{\tiny{L}}^{0})$ are the canonical quadratures of the signal light pulse.
Transfer factor $\mathsf{T}_\text{\tiny{L}}$ and the excess noises $\hat{\mathbf{N}}_\bullet{}$
are complex functions of the physical parameters of the system~--- the gate efficiency $\eta$, the coupling constant $g_\text{\tiny{A}}$, the pulse duration $\tau$, and the decay rate of the atomic cavity $\kappa_\text{\tiny{A}}$.
The interaction gains $\mathsf{G}_\text{\tiny{A}}$ and $\mathsf{G}_\text{\tiny{L}}$ characterize the coupling strength between the atomic oscillator and the light:
\begin{align}
& \mathsf{G}_\text{\tiny{A}}=g_\text{\tiny{A}}\sqrt{\frac{2\tau}{\kappa_\text{\tiny{A}}}},\\
 & \mathsf{G}_\text{\tiny{L}} =
  g_\text{\tiny{A}} \sqrt{ \frac{ 2 \tau }{ \kappa_\text{\tiny{A}} }} \times \sqrt{ \eta } \left[ 1 - \frac{ 1 - e^{ - \kappa_\text{\tiny{A}} \tau }}{ \kappa_\text{\tiny{A}} \tau } \right].
  \label{GainsAL}
\end{align}
It should be noted that the pre-factors of the admixed quadratures, $\hat{\bold{X}}_\text{\tiny{A}}^{0}$ and $\hat{\bold{P}}_\text{\tiny{L}}^0$ in Eqs.~(\ref{atom1},\ref{atom4}), are unequal ($\mathsf{G}_\text{\tiny{A}}\neq \mathsf{G}_\text{\tiny{L}}$).
Therefore, such a transformation is, in general, not symmetrical.
It is only in the limit of a perfect efficiency $\eta = 1$ and sufficiently long pulses $\kappa_\text{\tiny{A}} \tau \gg 1$ that we can characterise the gate with only a single gain parameter $\mathsf{G}$.
For a fixed $\tau\kappa_\text{\tiny{A}}$ this parameter is determined mostly by the coupling strength $g_\text{\tiny{A}}$.

Let us examine this non-ideal QND gate
and calculate the probability of detecting
two excitations at one output of the gate and zero at the other (the \emph{success} probability).
Such success appears already in a short-time evolution of ideal QND gate applied on two quanta.
This probability equals the HOM matrix element $\langle \text{HOM}| \rho\s{out}|\text{HOM}\rangle$,
where $\ket{\text{HOM}} = \left( \ket 2_\text{\tiny{L}} \ket 0_\text{\tiny{A}} - \ket 0_\text{\tiny{L}} \ket 2_\text{\tiny{A}} \right)/ \sqrt{2 }$.
The choice of such a definition of the HOM-state is worth explaining.
Our goal is to compare the HOM interference obtained via the QND gate  with the well known HOM effect obtained via the beam-splitter.
For a beam-splitter, this HOM state definition provides the maximum of the $\langle \text{HOM}| \rho\s{out}|\text{HOM}\rangle$.
Another definition, taken with a different sign, i.e. $\ket{\text{HOM}_+} = \left( \ket 2_\text{\tiny{L}} \ket 0_\text{\tiny{A}} + \ket 0_\text{\tiny{L}} \ket 2_\text{\tiny{A}} \right)/ \sqrt{2 }$, would lead to the zero HOM matrix element of the output state.
Same proved to be true for the QND transformation, and therefore, such an approach allows us to compare the HOM effect by the beam-splitter and the QND gate in the most appropriate way.

To evaluate the bunching of the output excitations and to determine whether it is caused by a truly non-classical interference of bosons, we define certain thresholds that correspond to the performance of classical coherent states after the same QND interaction.
We define the \emph{output threshold} as the value of this element for two arbitrary coherent states at the output of the gates.
Such interaction-independent threshold shows the maximal value of HOM element attainable by \emph{any} bipartite state with positive Glauber-Sudarshan $P$-function and as such is a fundamental boundary of nonclassicality of the output state of the gate (for details, see Methods and Supplementary).
In addition, we define the \emph{input threshold}, as the highest possible value of the HOM element after the same interaction for the case when both input modes were initially in phase-randomized coherent states, as a lower interaction-dependent nonclassicality threshold.
In order to devise a threshold for the output states produced from the input Fock states which do not posess well-defined phases and hence are incapable to exhibit the first-order interference, we wish to have a threshold based on classical states that are rid of the capability to exhibit the first-order interference as well.
Without this phase randomization classical states have been shown to exhibit a nearly-perfect visibility of HOM interference~\cite{kim_conditions_2013,sadana_near100_2019}.
It is phase randomization that allows to rule out the first-order interference from the input coherent states.
The threshold, thereby, shows the bunching achievable by classical states only due to the higher-order interference.
We assume that the HOM interference with nonclassical input states takes place if the corresponding input threshold is surpassed.

\begin{figure*}[ht]
\begin{center}
\includegraphics[width = \linewidth]{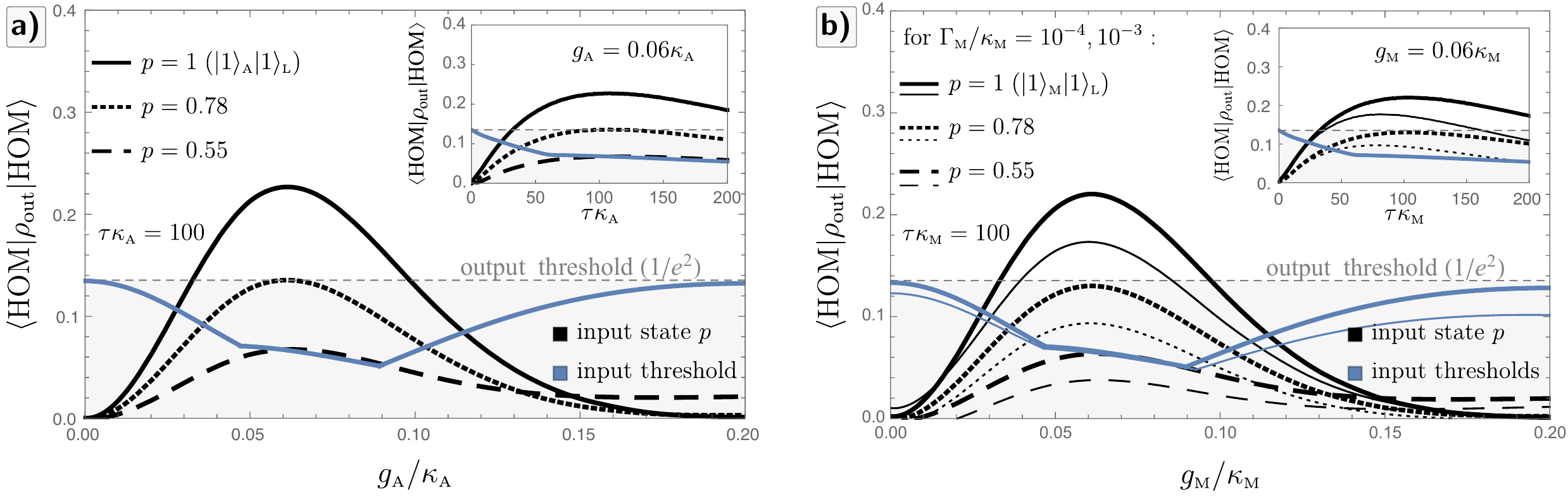}
\caption
{
Matrix element $\langle \text{HOM}| \rho\s{out}|\text{HOM}\rangle$ of the output states of the light-atom and optomechanical gates.
The element is plotted as a function of the coupling strength for the pulse duration $\tau\kappa_{\text{\tiny{A,M}}}=100$ and efficiency $\eta=0.9$.
At the input we consider the mixture ${ (p\;|1\rangle\langle1| + (1-p)\;|0\rangle\langle0| )_{\text{\tiny{A,M}}}\otimes(p\;|1\rangle\langle1| + (1-p)\;|0\rangle\langle0| )_{\text{\tiny{L}}}}$:
a)~Light-atom QND gate.
Dependence on the coupling strength $g_\text{\tiny{A}}$ for $p=1,\;0.78,\;0.55$.
A well pronounced maximum decreases with decreasing $p$.
The inset demonstrates the HOM element as the function of the pulse duration $\tau\kappa_\text{\tiny{A}}$.
b)~Optomechanical QND gate.
Dependence on the coupling strength $g_\text{\tiny{M}}$ for $p=1,\;0.78,\;0.55$ using two values of the rethermalization rate each provides its own input threshold (thick for $\Gamma_\text{\tiny{M}}=10^{-4}\kappa_\text{\tiny{M}}$ and thin for $\Gamma_\text{\tiny{M}}=10^{-3}\kappa_\text{\tiny{M}}$ ).
For both (a) and (b),
the dashed gray line is the \emph{output threshold} and
the blue curves of the corresponding thickness are the \emph{input thresholds} (phase randomized).
}
\label{Fig2}
\end{center}
\end{figure*}

\mbox{Figure~\ref{Fig2}(a)} demonstrates the dependence of the HOM element at the output of the QND gate on its coupling strength.
We consider a realistic incoherent mixture of vacuum and single-boson states at each input port of the gate,
assuming parameter $p$ as the fraction of the latter in the mixture, that is, the state $\rho = \rho\s{A} \otimes \rho\s{L}$ with $\rho_i = p \dyad 1 1 _i + ( 1 - p ) \dyad 0 0 _i$ for $i = \text{A,L}$ (see Sec.~\ref{methods} for details).
For simplicity, we assume equal contribution of excitations (that is, equal $p$) in each subsystem.
For $p=1$ both atomic and light modes are initially in a pure single-boson state $|1\rangle_\text{\tiny{A}}|1\rangle_\text{\tiny{L}}$.
Single boson (polariton) states are already achievable in the experiments for the different atomic systems~\cite{PhysRevA.89.033801,Cuevaseaao6814,Jackson2021}.

%%%%%%%%
%%%%%%%%

For a fixed pulse duration and efficiency, the HOM element, as a function of the coupling strength, has a well pronounced maximum that decreases with decreasing  $p$.
For $\tau\kappa_\text{\tiny{A}}=100$ and $\eta=0.9$ (i.e. parameters providing the gate performance close to the best possible)
the maximum of the HOM element located at $g_\text{\tiny{A}}=0.06\kappa_\text{\tiny{A}}$ decreases from $0.25$ at $p=1$  to  $1/e^2$ at $p \approx 0.78$, which corresponds to the output threshold.
At $p\approx0.55$ the HOM element already crosses the input threshold.
Note that the input threshold is determined only by the parameters of the gate, so $p$, as the parameter of the input state, does not affect it.
This threshold, shown by blue, is the same for the entire set of $p$.

For the input states with high $p$ ($p>0.8$), the HOM element surpasses the corresponding input threshold even for a very low efficiency $\eta=0.1$.
Thus the Hong-Ou-Mandel interference can be observed for a gate of a low quality if the input state was close to the pure one-boson state.
Moreover, for the high $p$ the HOM element satisfies a more stringent condition and lies above even the output threshold  for a wide range of efficiencies (up to $\eta>0.2$) if the pulse duration and coupling strength are optimized.
However, for the low $p$, the HOM element cannot surpass the input threshold even in the case of an ideal gate with $\eta=1$ and both optimal $\tau$ and $g_\text{\tiny{A}}$.

The pulse duration, as an argument of the interaction gains  $\mathsf{G}_\text{\tiny{A,L}}$, also deserves attention.
The dependence of the HOM element on $\tau$ for different $p$ accompanied with  their  common input threshold are shown on the inset of  \mbox{the Fig.~\ref{Fig2}(a)}.
These curves also have pronounced maxima, that is, there is an optimal pulse duration for each fixed coupling.

For this type of the gate, the longer the pulse, the higher the HOM element we can get and the smaller coupling is required, however for $\tau\kappa_\text{\tiny{A}}>100$ the advantage is already insignificant.

For any QND gate,
 the parity of the total number of excitations is preserved, and hence an odd number of excitations at the input will never turn into an even number at the output.
Thus, if $p =0$ for one subsystem and $p =1$ for the other (for the light-atom gate it corresponds to $|0\rangle_\text{\tiny{A}}|1\rangle_\text{\tiny{L}}$ or $|1\rangle_\text{\tiny{A}}|0\rangle_\text{\tiny{L}}$ inputs),
it will lead to  the zero HOM element of the output state,
i.e. no HOM effect.
In case of $p = 0$ for both subsystems, the input state is a pure vacuum state and the total number of excitations is even.
The HOM element of the output state is non-zero, but has a maximum that is lower but still well pronounced, so the bunching of excitations that are created during the interaction is present.

%----------------------------------------------------------------
\subsection{Optomechanical Hong-Ou-Mandel Interference}\label{secB}
%----------------------------------------------------------------

It is also possible to show the effect of bunching of optical photons and mechanical phonons in an optomechanical system.
Optomechanical QND gate between an incident light pulse and the mechanical oscillator can be realized using e.g. appropriately modulated classical drive~\cite{braginsky_quantum_1980,shomroni_optical_2019}.
To describe the mechanical part of the system we use quadratures $(\hat{\bold{X}}_\text{\tiny{M}},\hat{\bold{Y}}_\text{\tiny{M}})$ that refer to the dimensionless position and momentum of the mechanical oscillator.
In optomechanics, single phonons can be generated by optomechanical parametric down-conversion~\cite{hong_hanbury_2017} or swapped to the mechanical mode from light~\cite{rakhubovsky_photonphononphoton_2017,kiesewetter_pulsed_2017}.

Let us consider coupling of the same pair of quadratures and use the following effective linearized Hamiltonian for the optomechanical interaction $\hat{H}\s{\text{\tiny{LM}}} = \hbar g_\text{\tiny{M}} \hat{\bold{X}}_\text{\tiny{M}}\hat{p}_\text{c}$~\cite{Asp2014},
where $\hat{p}_\text{c}$ is the canonical quadrature of the intracavity light.
Same as with the atom-light interaction, the optomechanical interaction is single-mode, it couples only one mode of the cavity to a single mechanical mode.
The single-mode character of the coupling in practice requires negligible spatial and spectral overlap of other mechanical modes with the interacting one.
These conditions are standardly met with exceptional precision in most optomechanical experiments~\cite{aspelmeyer_cavity_2014}.
After the QND-type interaction with the coupling strength $g_\text{\tiny{M}}$, the quadratures of light and mechanics transform as:
\begin{align}
& \hat{\bold{X}}_\text{\tiny{M}} =\hat{\bold{X}}_\text{\tiny{M}}^0+ \hat{\mathbf{N}}_{\text{X}_\text{\tiny{M}}}, \\
& \hat{\bold{Y}}_\text{\tiny{M}} =\hat{\bold{Y}}_\text{\tiny{M}}^0-\mathsf{G}_\text{\tiny{M}}\hat{\bold{Y}}^{0}+ \hat{\mathbf{N}}_{\text{P}_\text{\tiny{M}}},\\
& \hat{\bold{X}}'{}\up{out}=\mathsf{T}_\text{\tiny{L}}\hat{\bold{X}}^{0}_\text{\tiny{L}}+\mathsf{G}_l\hat{\bold{X}}_\text{\tiny{M}}^0+ \hat{\mathbf{N}}_{\text{X}_\text{\tiny{L}}},\\
& \hat{\bold{Y}}'{}\up{out}=\mathsf{T}_\text{\tiny{L}}\hat{\bold{Y}}^{0}_\text{\tiny{L}}+ \hat{\mathbf{N}}_{\text{P}_\text{\tiny{L}}},
\end{align}
where canonical quadratures~$(\hat{\bold{X}}^{0},\hat{\bold{Y}}^{0})$ of the signal and the transfer factors~$\mathsf{T}_l$ are defined as in the previous subsection,
while the noises~$ \hat{\mathbf{N}}_\bullet{}$,
and their correlation relations (characterized by the physical parameters of the system)
are different.  Their exact definitions are cumbersome and thus are in the Supplementary Materials.
Along with the cavity linewidth~$\kappa_\text{\tiny{M}}$ and coupling rate~$g_\text{\tiny{M}}$, other important parameters of the optomechanical gate are
the gate efficiency~$\eta$ and
the mechanical damping coefficient~$\gamma_\text{\tiny{M}}$ that shows how good the mechanics is isolated from thermal bath
with average phonon number $n\s{th}$
(the two latter parameters are combined in the reheating rate $\Gamma_\text{\tiny{M}} = \gamma_\text{\tiny{M}} n\s{th}$).
Interaction gains~$\mathsf{G}_\text{\tiny{M,L}}$ coincide with the corresponding gains in the Eq.~\ref{GainsAL} with an evident subscript replacement~$\text{A} \mapsto \text{M}$.
Thus, as in the previous, atom-light case, this gate is asymmetric, i.e.~$\mathsf{G}_\text{\tiny{M}}\neq\mathsf{G}_\text{\tiny{L}} $,
and the main role of the gain is similarly determined by the coupling strength~$g_\text{\tiny{M}}$.

Despite very different physical nature,
both atom-light and optomechanical gates are described by very similar equations with the only apparent difference of the mechanical oscillator
being coupled to the environment with possibly very high occupation.
This makes the rethermalization of the mechanics the critical difference between the two gates.
Qualitatively, the behavior of the HOM element with respect to the efficiency and coupling strength is similar to the atom-light case, so we will focus on the rethermalization rate.

Expectedly, the rethermalization rate~$\Gamma_\text{\tiny{M}}$ is the most significant physical parameter, severely limiting the value of attainable~$\langle \text{HOM}|\rho\s{out}|\text{HOM}\rangle$:
the lower the rethermalization rate, the higher the HOM element.
\mbox{Figure~\ref{Fig2}(b)} shows the dependence of the HOM element and the corresponding thresholds on the coupling strength for the two rethermalization rates.

The rethermalization strongly affects the values of the HOM elements, but very slightly affects the input threshold.
To demonstrate it we chose $\Gamma_\text{\tiny{M}}=10^{-3}\kappa_\text{\tiny{M}}$, that is quite feasible (see, e.g., Refs~\cite{norte_mechanical_2016,tsaturyan_ultracoherent_2017,hoj_ultracoherent_2021}), and compared the result
with $\Gamma_\text{\tiny{M}}=10^{-4}\kappa_\text{\tiny{M}}$ that is attainable at the moment in the experiment. The HOM curves differ a lot, but not the thresholds.
For the smaller values of $\Gamma\s{\tiny{M}}$ the plots look very similar to the case in the previous section where there is no rethermalization.
That is, $\Gamma_\text{\tiny{M}}=10^{-4}\kappa_\text{\tiny{M}}$ is a type of a border and lower rethermalization does not significantly increase the maximum of the HOM element.

Note that, in contrast to the atom-light case, the pulse duration  has an optimal value
yielding the highest possible value of the HOM element,
e.g.  for the low rethermalization rate and high efficiency,~$g_\text{\tiny{M}}\approx0.06\kappa_\text{\tiny{M}}$ for $\tau\kappa_\text{\tiny{M}}=100$ are the optimal parameters to observe the Hong-Ou-Mandel interference.
The existence of this optimal pulse length is dictated by the non-negligible reheating rate that admixes to the mechanics thermal noise with variance increasing with the pulse duration.

%----------------------------------------------------------------
\subsection{Atom-Mechanical Hong-Ou-Mandel Interference}
%----------------------------------------------------------------

In this subsection, following our treatment in~\cite{Manukhova2020} we consider a hybrid QND gate between an atomic ensemble and a mechanical oscillator considered in a previous subsections.
Specifically, we investigate bunching of excitations in such system.

To establish the gate we connect the atomic and optomechanical cavities introduced in Sec.~\ref{secA} and~\ref{secB} in such a way that the light passes them sequentially, interacting first with atoms and then with mechanics.
At the moment, such systems have already been physically implemented.
For example, the works~\cite{Thomas2020} and~\cite{Karg2020} both  describe the systems
that theoretically allow observing the HOM effect if the parameters are properly coordinated.
In both works, the atoms are located in free space, which, however, does not affect the idea, since it is important to ensure the interaction of the QND type that could be done both ways, with or without the atomic cavity.
In addition, we emphasize that to observe the effect, it is critical to ensure a low rethermalization rate, therefore, the optomechanical cavity should be appropriately cooled.

We choose the phases of the drivings in a way that the effective Hamiltonians for the atom-light and mechanical-light interactions are
$\hat{H}\s{\text{\tiny{LA}}} = -\hbar g_\text{\tiny{A}} \hat{\mathbf P}_\text{\tiny{A}}\hat{x}_\text{c}$ and $\hat{H}\s{\text{\tiny{LM}}} = \hbar g_\text{\tiny{M}} \hat{\mathbf X}_\text{\tiny{M}}\hat{p}_\text{c}$, correspondingly.
Afterward the $\hat{X}$ quadrature of the pulse is homodyned and the output of the detection is used to displace the atoms in the phase space.
The choice of different types of Hamiltonians for the atomic-light and optomechanical parts of the scheme is dictated by the goal to couple the $P\s{\tiny{A}}$-quadrature of the atomic part with the  $X\s{\tiny{M}}$-quadrature of the mechanics, that is, to provide an effective Hamiltonian $H\propto X\s{\tiny{M}} P\s{\tiny{A}}$.
For the hybrid atom-mechanical gate, we also consider squeezing of the mediating light as a resource since for a QND coupling matter with light, squeezing can effectively enhance the interaction gain~\cite{filip_quantum_2009,rakhubovsky_squeezerbased_2016} which, in the gate, might enhance the HOM effect.
Strictly speaking, the squeezing could be beneficial also for the gates considered earlier, but for those cases the advantage is small, while for an atom-mechanical gate it is more noticeable.
Additionally, squeezing the input state for a HOM-like interferometry would cause an unwanted emphasis shift to the input state preparation which does not translate to the atom-mechanical case where squeezing of input state is dramatically more challenging.
Let us emphasize once again the important difference of this gate from those considered earlier~--- here, the light is only a mediator coupling two systems, it does not serve as a signal as it was for the atom-light and optomechanical ones.

The built QND gate relates the quantum state of the atoms and mechanics after the interactions with their initial states and the noises
and transforms the initial quadratures $\bold{r}\up{in}=(\hat{\bold{X}}_\text{\tiny{A}}^0,\hat{\bold{P}}_\text{\tiny{A}}^0,\hat{\bold{X}}_\text{\tiny{M}}^0,\hat{\bold{P}}_\text{\tiny{M}}^0)^T$
to the final quadratures $\bold{r}\up{out}\equiv\bold{r}=(x_\text{\tiny{A}},p_\text{\tiny{A}},x_\text{\tiny{M}},p_\text{\tiny{M}})^T$ as:
\begin{align}
 & x_\text{\tiny{A}}= \hat{\bold{X}}\up{out}_\text{\tiny{A}} =\coefficient{T}_\text{\tiny{A}}\hat{\bold{X}}_\text{\tiny{A}}^0-\coefficient{G}_\text{\tiny{A}}\hat{\bold{X}}_\text{\tiny{M}}^0+\hat{\mathbf{N}}_{\text{X}_\text{\tiny{A}}},\\
 & x_\text{\tiny{M}}=\hat{\bold{X}}\up{out}_\text{\tiny{M}} =\coefficient{T}_\text{\tiny{M}}\hat{\bold{X}}_\text{\tiny{M}}^0+\hat{\mathbf{N}}_{\text{X}_\text{\tiny{M}}}, \\
 & p_\text{\tiny{A}}=\hat{\bold{P}}\up{out}_\text{\tiny{A}} =\coefficient{T}_\text{\tiny{A}}\hat{\bold{P}}_\text{\tiny{A}}^0+\hat{\mathbf{N}}_{\text{P}_\text{\tiny{A}}},   \\
 & p_\text{\tiny{M}}=\hat{\bold{P}}\up{out}_\text{\tiny{M}}=\coefficient{T}_\text{\tiny{M}}\hat{\bold{P}}_\text{\tiny{M}}^0+\coefficient{G}_\text{\tiny{M}}\hat{\bold{P}}_\text{\tiny{A}}^0+\hat{\mathbf{N}}_{\text{P}_\text{\tiny{M}}},
  \label{gate}
\end{align}
where the controllable gains~$\coefficient{G}_{\text{\tiny{A,M}} }$, the transfer factors~$\coefficient{T}_{\text{\tiny{A,M}} }$ and the excess noises~$\hat{\mathbf{N}}_\bullet{}$ are complicated functions
of the interaction, loss and noise parameters of the system (see Supplementary for the full definitions).
For the atom-mechanical gate it is experimentally well justified to put $\coefficient{T}_\text{\tiny{A}}=\coefficient{T}_\text{\tiny{M}}=1$.
The system has several parameters affecting the process:
the pulse duration~$\tau$,
the coupling strengths~$g_\text{\tiny{A}}$ and $g_\text{\tiny{M}}$,
the initial squeezing of the mediator pulse~$S$,
the energy loss~$\eta$,
the optical damping rates~$\kappa_\text{\tiny{A}}$ and~$\kappa_\text{\tiny{M}}$ of the cavities (for simplicity here we take $\kappa_\text{\tiny{A}}=\kappa_\text{\tiny{M}}=\kappa$, but the difference can, in principle, serve as an additional degree of freedom),
and the rethermalization rate~$\Gamma_\text{\tiny{M}}$.
The feedforward procedure is carried out in a way that ensures $\coefficient{G}_\text{\tiny{A}}=\coefficient{G}_\text{\tiny{M}}$, i.e. this gate is symmetric unlike the atom-light and mechanical-light cases:
\begin{equation}
    \mathfrak{G}=
    g_\text{\tiny{M}} g_\text{\tiny{A}} \frac{ 2 \tau \sqrt{ \eta }}{ \kappa } \left[ 1 + e^{ - \kappa \tau }
    - \frac{ 2 }{ \kappa \tau } \left( 1 - e^{ - \kappa \tau } \right) \right].
\end{equation}
Note again that this interaction, same as the atom-optical and optomechanical, is a single-mode interaction that addresses only one atomic and one mechanical mode.
The input states to our HOM analogue are thus purely single-mode bosons attenuated by vacuum, in contrast to the usual multimode (spectrally, temporally etc.) character of the input states of optical HOM experiments.

\begin{figure*}[ht]
\begin{center}
\includegraphics[width = \linewidth]{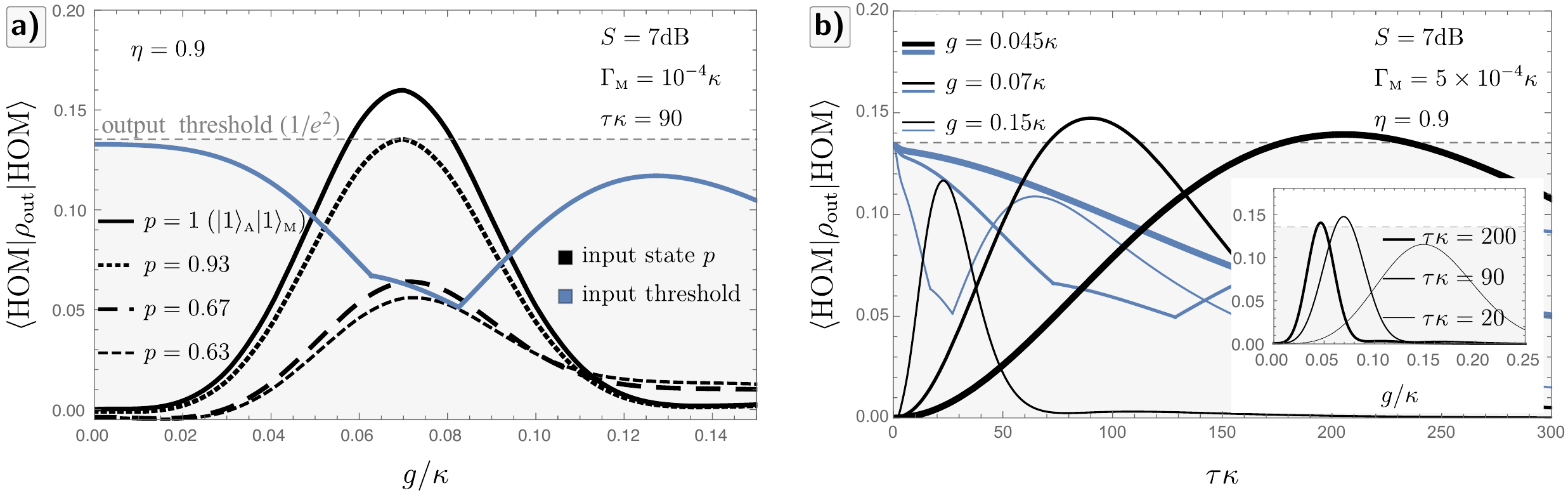}
\caption
{
Matrix element $\langle \text{HOM}| \rho\s{out}|\text{HOM}\rangle$ of the output state for the non-adiabatic atom-mechanical QND gate with
 the mixture $ (p\;|1\rangle\langle1| + (1-p)\;|0\rangle\langle0| )_\text{\tiny{A}}\otimes(p\;|1\rangle\langle1| + (1-p)\;|0\rangle\langle0| )_\text{\tiny{M}}$ at the input.
 a)~dependence on coupling strength for the different $p=1,\;0.93,\;0.67,\;0.63$.
 b)~dependence on pulse duration for the different $g$
 (thickness indicates the coupling strength~--- the thickest for $g=0.045\kappa$, the middle one for $g=0.07\kappa$ and the thinnest for $g=0.15\kappa$).
 The inset demonstrates dependence on coupling strength for the different pulse duration
 (thickness indicates the pulse duration~--- the thickest for $\tau\kappa=200$, the middle one for $\tau\kappa=90$ and the thinnest for $\tau\kappa=20$).
  For both (a) and (b), $\kappa_\text{\tiny{M}}=\kappa_\text{\tiny{A}}=\kappa,\;g_\text{\tiny{A}}=g_\text{\tiny{M}}=g,\; \eta=0.9,\;S=7$dB.
}
\label{Fig4}
\end{center}
\end{figure*}

Let us calculate the matrix element $\langle \text{HOM}| \rho\s{out}|\text{HOM}\rangle$ for the output state of the atom-mechanical QND gate
assuming single-boson states at each subsystem's input
(single-polariton for the atomic subsystem and single-phonon for the mechanics).
The gain $\coefficient{G}$ is defined by the pulse duration $\tau$ and by the product of the coupling strengths of the light-atom $g_\text{\tiny{A}}$ and optomechanical $g_\text{\tiny{M}}$ interactions.
\mbox{Figure~\ref{Fig4}(a)} demonstrates the dependence of the HOM elements and the corresponding input coherent thresholds on the value of the coupling rate for the different $p$,
assuming equal coupling $g_\text{\tiny{A}}=g_\text{\tiny{M}}=g$ for atoms and mechanics.
%%%%%%%%
%%%%%%%%
The gate parameters were chosen close to the optimal ones, i.e. providing the highest possible value of the HOM element, so the picture is similar to the previously considered gates.
However if the efficiency is too low, the HOM element cannot surpass the input threshold even for $p=1$ in contrast to the atom-light and mechanical-light gates.

The main part of  \mbox{the Fig.~\ref{Fig4}(b)} demonstrates the dependence of the HOM element on the pulse duration for the fixed $S$ and $g$
and the inset shows the dependence of this element on the coupling for the fixed $S$ and $\tau$.
These figures demonstrate that, similarly to the previously considered gates, to obtain the highest possible value of the HOM element one has to find the optimal values of both $g$ and $\tau$.
Note that squeezing $S$ itself is a parameter that has the optimal value,
but optimal squeezing is not very high regardless of all other parameters, and does not exceed $10$dB.
For $S=7$dB, high efficiency and low rethermalization the optimum is provided by $\tau\kappa\approx90,\;g\approx0.07\kappa$.

There is a monotonic dependence on the two remaining parameters, efficiency $\eta$ and the reheating rate $\Gamma\s{\text{\tiny{M}}}$.
Maximum of the HOM element is delivered by highest efficiency and lowest reheating rate.
There is always a threshold value for the rethermalization $\Gamma\s{\text{\tiny{M}}}$, that is for $\Gamma\s{\text{\tiny{M}}}/\kappa>0.01$ no effect can be observed even with perfect other parameters like $\eta=1$ (no optical losses) and $p=1$ (ideally prepared initial state).
Threshold for $\eta$ strongly depends on the values of the other parameters.

%%%%%%%%%%%%%%%%%%%%%%%%%%%%%%%%%%%%%%%
\section{Methods} \label{methods}
%%%%%%%%%%%%%%%%%%%%%%%%%%%%%%%%%%%%%%%

In this paper we investigate the possibility of observing an analogue of the HOM effect using a quantum nondemolition~(QND) gate by evaluating the HOM matrix element of the corresponding output quantum state.
We compare the performance of the gate with the one of a beam-splitter.
In our scheme, the input single bosons (phonons and polaritons) are excitations of strictly single-mode quantum oscillators ulnike multimode photons in the usual optical HOM experiment.
Therefore, our main focus is on the comparison of these two quantum maps (QND and beam-splitter).
In this section we elaborate on the definitions and the methods we use to perform the necessary computations.

A beam-splitter~(BS) transformation, defined by the unitary operator $ U\s{BS}=\exp{ \left[  \Theta \(a^\dag b-b^\dag a\) \right]}$,
describes an evolution of two quantum oscillators with annihilation operators, respectively, $a$ and $b$.
The only parameter of this transformation is the transmittance coefficient~$\mathsf{T}=\cos^2\Theta$.
A QND gate, defined by the unitary operator $U_\sfG= \exp\left[ \sfG (a+a^\dag)(b^\dag-b)/2 \right]$, describes another type of evolution of the two oscillators.
The gain~$\sfG$ is the only parameter characterizing the ideal QND gate transformation.

There is a significant difference between these two transformations.
The BS transformation is passive (energy conserving),
if initially there is exactly one excitation in each of the oscillators, at the output of a BS they can appear bunched in a single mode via the Hong-Ou-Mandel (HOM) effect.
To have only one excitation in a single mode is insufficient to observe bunching because of preservation of the total energy.
Unlike BS, the QND transformation is active, which means it is capable of changing the total number of excitations in the system (the energy of the system).
Despite the QND interaction can generate new quanta,
it is still possible to analyze whether the QND interaction is capable of generating the non-classical two-quanta superpositions going beyond any classical states serving as input to the QND interaction.
However, such analysis requires a general approach to the HOM interference beyond the simple case with the passive BS interaction.
The first step to extend HOM effect to active interaction has been presented in~\cite{Cerf33107}.

In our generalized description, the matrix element of the quantum state that equals the \emph{success} probability of detecting two-photon HOM entangled states (the HOM element) at the output of the unitary transformation $U$,
can be introduced as $|\langle \text{HOM} |U|\varphi\rangle\s{in}|^2$, where $|\varphi\rangle\s{in}$ is the initial state~\cite{Makinoe1501772}.
Here, the HOM state is defined as $\ket{ \text{HOM}}  = \left( \ket 0_\text{a} \ket 2_\text{b} - \ket 2_\text{a} \ket 0_\text{b} \right) /\sqrt{2}$.
It is well known that a BS provides an ideal photon bunching (the HOM effect).
This means that the success probability $|\langle \text{HOM} |U\s{BS}|\varphi\rangle\s{in}|^2$ at the output of the BS equals one.
This effect occurs when two identical quanta enter a balanced beam-splitter ($ \mathsf{T}=0.5$), one in each input port (the input state $|\varphi\rangle\s{in}=|1\rangle_\text{a}|1\rangle_\text{b}$).

In order to compare the two transformations in the context of the HOM effect, let us look at the matrix elements of the output state of each of the transformations.
For simplicity, first let us restrict the input $|\varphi\rangle\s{in}$ to the space of coherent superpositions of vacuum and one excitation of each mode, that is an arbitrary pure superposition of
$|1\rangle_\text{a}|1\rangle_\text{b},|0\rangle_\text{a}|1\rangle_\text{b},|1\rangle_\text{a}|0\rangle_\text{b}$ and $|0\rangle_\text{a}|0\rangle_\text{b}$ (for the case of full infinite space see Supplementary Materials).
Then, to obtain the desired matrix element we need:
\begin{align}
    \langle \text{HOM} |\;U\s{BS} & =-\sin{\(2\Theta\)}\;\langle1|\s{a}\langle1|\s{b},\\
    \langle \text{HOM} | \;U_\sfG & = \frac{4 \sfG (8 - \sfG^2)}{(4 + \sfG^2)^{5/2}}\;\langle1|\s{a}\langle1|\s{b}+\nn\\
                                  &+\frac{2 \sfG^2}{(4 + \sfG^2)^{3/2}}\;\langle0|\s{a}\langle0|\s{b} \label{Eq2}.
\end{align}
It clearly shows that the HOM matrix element provided by the inputs $|0\rangle\s{a}|1\rangle\s{b}$ and $|1\rangle\s{a}|0\rangle\s{b}$ is equal to zero for both BS and QND transformations.
The matrix element provided by $|0\rangle\s{a}|0\rangle\s{b}$ input is equal to zero in the case of a beam-splitter.
However, for the QND gate this element is a function of the gain $\sfG$ and equals zero only in the trivial case with $ \sfG=0$.
That is, by varying the gain of the QND gate, it is impossible to make the contribution of the input vacua $|0\rangle\s{a}|0\rangle\s{b}$ vanish, in order to render these two transformations fully analogous.
The active transformation can therefore generate a non-zero HOM element even from two ground (vacuum) states.

\begin{figure*}[ht]
\begin{center}
 \includegraphics[width =\linewidth]{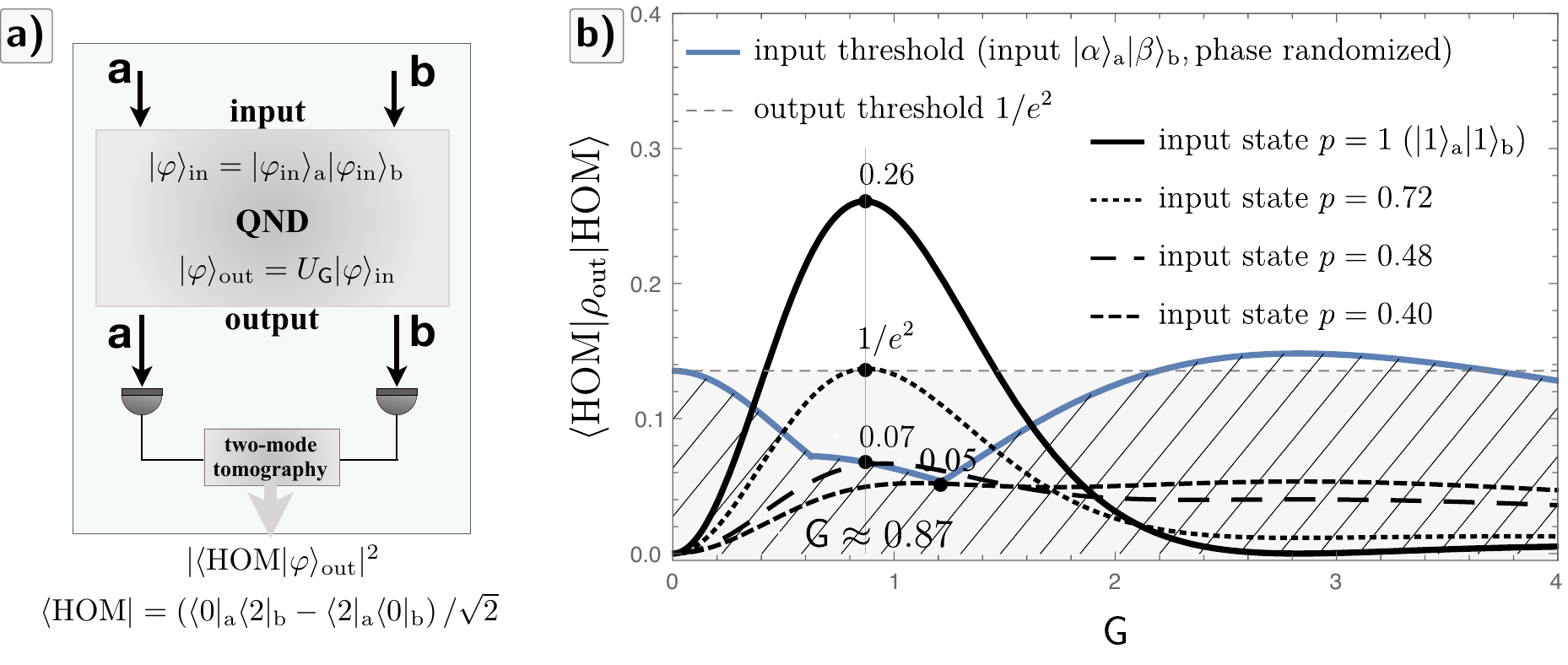}
\caption
{
Ideal QND interaction:
a) Scheme of the QND interaction. States at the input ports are independent in contrast to the output ones.
The output state is characterized by two-mode homodyne tomography~\cite{Makinoe1501772}.
b) $\langle \text{HOM}| \rho\s{out} |\text{HOM}\rangle$ matrix element of the output state for the ideal QND gate as a function of the gain $\mathsf{G}$ calculated for the different cases of the input:
quantum input $|1\rangle\s{a} |1\rangle\s{b}$ (solid black curves),
mixture input Eq.~(\ref{ME_superpositioninput_Eq1}) (dashed black curves, dashing scale indicates parameter $p$).
Dashed gray line is the \emph{output threshold}.
Blue curve is the \emph{input thresholds} (phase randomized) restricting area
that covers all the possible values of the matrix elements of the output state of the gate in the case of the random coherent input with averaged phases.
}
\label{FigSuperposition}
\end{center}
\end{figure*}

For the case of a QND gate with $|1\rangle\s{a}|1\rangle\s{b}$ at the input, one can observe
that for a certain region of the parameter $\sfG$, the probability of bunching of both excitations in one subsystem is higher than the probability of equal redistribution of the excitations between the subsystems.
Visually, it is characterized by the presence of the maximum of $\langle \text{HOM}| \rho\s{out}|\text{HOM}\rangle$ (see  \mbox{Fig.~\ref{FigSuperposition}(b)}) approximately equal to $0.26$ for $\sfG
\approx 0.87$ (as compared to $1$ for the BS with $\Theta=\pi/4$).
However, we should keep in mind that this correspondence to the case of a BS is not complete due to the non-zero contribution from the vacuum input that is non-zero for the case of the QND gate but does not exist in the case of a BS.

To overcome the issue of discriminating between the classical interference and non-classical bunching, we define two nonclassicality thresholds by evaluating the maximum of $\langle \text{HOM}|\rho\s{out}|\text{HOM}\rangle$ over (i) all superpositions of coherent states at the output of the QND interaction:
$\rho\s{out} = \rho\up{coh} %= \dyad{\alpha_a \beta_b}{\alpha_a \beta_b}
=| \alpha\rangle\s{a}|\beta\rangle\s{b}\langle\alpha|\s{a}\langle\beta|\s{b}$ and (ii) all superpositions of coherent states, phase averaged, before the QND interaction: $\rho\s{out} = \frac{ 1 }{ 4 \pi^2 } \iint d \varphi\s{a} d\varphi\s{b} U_\sfG \rho\up{coh} U_\sfG^\dag$.  Here we use the notation $\alpha = |\alpha| e^{ i \varphi\s{a}}$ and $\beta = |\beta| e^{ i \varphi\s{b}}$.
For the BS interaction, the two thresholds coincide.
%\textcolor{red}{Add motivation of the (ii) threshold.}

To obtain the first threshold let us take two random coherent states derived  in the Fock basis:
\begin{align}
& |\alpha\rangle=e^{-\frac{|\alpha|^2}{2}}\sum\limits_{n=0}^{\infty}\alpha^n\frac{|n\rangle}{\sqrt{n!}},\\
& |\beta\rangle=e^{-\frac{|\beta|^2}{2}}\sum\limits_{m=0}^{\infty}\beta^m\frac{|m\rangle}{\sqrt{m!}}.
\end{align}
The HOM element corresponding to this state is the following:
\begin{equation}
    h\s{out} = |\langle\text{HOM}| \alpha\rangle\s{a}|\beta\rangle\s{b}|^2=\frac{1}{4}e^{-|\alpha|^2-|\beta|^2}|\alpha^2-\beta^2|^2.
\end{equation}
It can be shown that $|\langle\text{HOM}| \alpha\rangle\s{a}|\beta\rangle\s{b}|^2 \leq 1/e^2$ on any coherent states, that is for any complex $\alpha$ and $\beta$.
This allows to prove that if the output state $\rho\s{out}$ is classical, i.e. is a mixture of coherent states, then $0\leq\langle \text{HOM}| \rho\s{out}|\text{HOM}\rangle\leq1/e^2$ (see Supplementary).
Thus, $1/e^2$ is the \emph{output threshold} for the HOM interference.
That is, when measuring the HOM element, a value greater than $1/e^2$ indicates the nonclassicality of the output state, as no classical state (no mixture of coherent states) is capable of providing such value.
This threshold is shown by a thin gray dashed line in the  \mbox{Fig.~\ref{FigSuperposition}(a)}.

To derive the \emph{input threshold}, let us use two arbitrary coherent states as the input states of the gate and calculate the maximal possible HOM element for the output state $ \rho\up{coh}\s{out}=U_\sfG \rho\up{coh}U_\sfG^\dag$.
Bunching of the excitations (maximum of the HOM element)  can be provided not only by the second-order interference (the non-classical effect we are interested in) but also by the classical first-order interference~\cite{Glauber,perina_quantum_2012}.
To avoid the input of the first-order interference we assume phase-randomized coherent input state which practically means averaging over the phases of the input coherent states.
This allows us to eliminate the interference enabled by the degree of freedom (phases), which only the coherent states have in contrast to the pure $|1\rangle\s{a}|1\rangle\s{b}$.
After the threshold is derived using the phase-randomized states, it indicates the bunching of excitations due to only the second-order interference, the only one available to the Fock states, so that the comparison is more adequate.
Thus the input threshold for the ideal gate is as follows:
\begin{align}
    h\s{in}  = \frac{1}{4\pi^2}\max\limits_{r\s{a},r\s{b}} & \iint d\varphi\s{a}\;d\varphi\s{b}
    \\
             & \times \Big|\langle \text{HOM} |U_\sfG|r\s{a} e^{i\varphi\s{a}}\rangle\s{a}|r\s{b} e^{i\varphi\s{b}}\rangle\s{b}\Big|^2.\nn
\end{align}
The dependence of the $\langle \text{HOM} | \rho\up{coh}\s{out}|\text{HOM} \rangle$ on the gain $\sfG$ for all the coherent states (see \mbox{Fig.~\ref{FigSuperposition}(b)}) is illustrated by the area restricted by an \emph{input threshold} (blue curve) that has a specific complex shape (explained in Supplementary).

In order to evaluate the robustness of the QND gate against photon loss, let us examine an incoherent mixture of vacuum and single-photon states at each input port of the gate.
At this point, the HOM effect is additionally influenced by the statistics of quantum states at the inputs.
Thus, we consider the following input state
\begin{align}
    \rho\s{in}^{(p)} =
    & \Big( p\s{a}\;|1\rangle\langle1| + (1-p\s{a})\;|0\rangle\langle0| \Big)\s{a}\nn\\
&    \otimes
    \Big(p\s{b}\;|1\rangle\langle1| + (1-p\s{b})\;|0\rangle\langle0| \Big)\s{b}, \label{ME_superpositioninput_Eq1}
\end{align}
where the parameters $p\s{a,b}$ characterize how much vacuum has been admixed to the single-photon state at the corresponding input ports, and calculate matrix elements for the output state of the gate.
Using Eq.~(\ref{Eq2}), we can obtain the HOM matrix element of the output state of the gate $\rho\s{out}^{(p)} \equiv U_\sfG \rho\s{in}^{(p)} U_\sfG^\dag$:
\begin{gather}
    \langle \text{HOM}| \rho\s{out}^{(p)} |\text{HOM}\rangle=
    p\s{a} p\s{b}\frac{16 \sfG^2 ( \sfG^2-8)^2}{(4 + \sfG^2)^5} \nn\\
    +(1-p\s{a})(1-p\s{b})\frac{4 \sfG^4}{(4 + \sfG^2)^3}. \label{MHOM_superpositioninput}
\end{gather}

This matrix element is symmetrical with respect to $p\s{a}$ and $p\s{b}$.
Surprisingly, the independent coherent superpositions $ \ket{\psi}\s{in} = (\sqrt{p\s{a}}\;|1\rangle + \sqrt{1-p\s{a}}\;|0\rangle)\s{a}\cdot(\sqrt{p\s{b}}\;|1\rangle + \sqrt{1-p\s{b}}\;|0\rangle)\s{b}$ at the input
give rise to exactly the same matrix element Eq.~(\ref{MHOM_superpositioninput}) as does the mixture Eq.~(\ref{ME_superpositioninput_Eq1}).

\mbox{Figure~\ref{FigSuperposition}(b)}  demonstrates $\langle \text{HOM}| \rho\s{out} |\text{HOM}\rangle$ depending on the gain, assuming $p\s{a}=p\s{b}=p$,
compared with the case of the pure input $|1\rangle\s{a}|1\rangle\s{b}$.
%%%%%%%%
%%%%%%%%
Expectedly, as the parameter $p$ decreases, the contribution from $|1\rangle\s{a}|1\rangle\s{b}$ term decreases, while the contribution from $|0\rangle\s{a}|0\rangle\s{b}$ term increases.
Visually, it is reflected in the gradual change of the curves' shape --
for relatively high $p$, the maximum first decreases, then smoothly shifts to the right.
Thus, the maximum of the HOM element decreases from $0.26$ at $p=1$ to $1/e^2$ at $p \approx 0.7$, which corresponds to the output threshold.
At $p\approx0.48$ it already crosses the input threshold, so for lower values of $p$ the HOM element lies below the input threshold at the gain $\mathsf{G}\approx0.87$.
For $p<0.40$ the HOM element lies below the input threshold for any gain.

In order to calculate the matrix elements for an ideal (without additional noise) transformation,
it is enough to know the form of the unitary transformation $U_\sfG$.
Thus the matrix element of the output can be calculated as:
\begin{align}
  &  \langle \text{HOM}| \rho\s{out} |\text{HOM}\rangle=|\langle \text{HOM} |U_\sfG|\varphi\rangle\s{in}|^2
    =\nn\\
  &  =\left|\langle \text{HOM} |U_\sfG \sum_{n,m } \dyad{n}{n}\s{a} \dyad{m}{m}\s{b} \ket{\varphi}\s{in} \right|^2,
\end{align}
This approach is convenient to use for calculation of the ME when the input state has an appropriate representation in the Fock-state basis $|n\rangle\s{a}|m\rangle\s{b}$.
We used this approach to obtain the output threshold and the HOM elements for the case of an ideal gate.

In the general case of a Gaussian transformation including noises it can be convenient to take a different approach described below.
We can use the Wigner function (WF) of the output state and define the matrix elements $\langle\psi| \rho\s{out} |\varphi\rangle$ of the output state as:
\begin{equation}
 \langle\psi| \rho\s{out} |\varphi\rangle=(4\pi)^2 \iiiint d\bold{r} \; W_{ |\varphi\rangle\langle\psi| }(\bold{r})\cdot W\s{out}(\bold{r}),\label{HOMMEWF}
\end{equation}
where $W\s{out}(\bold{r})$ is the WF of the output state,
and $W_{ |\varphi\rangle\langle\psi| }(\bold{r})$ is the WF corresponding to the operator $ |\varphi\rangle\langle\psi|$.
$W\s{out}$ is a function of the column-vector of quadratures $\bold{r}=\left(x\s{a},p\s{a},x\s{b},p\s{b}\right)^T$.
For the two mode case, the WF corresponding to the operator $\xi$, is defined as follows:
\begin{align}
    & W_\xi (\bold{r})=\frac{1}{(4\pi)^2}\iint dy\s{a}dy\s{b} \; e^{-{i(p\s{a}y\s{a}+p\s{b}y\s{b})}/{2}} \nn\\
& \times \langle x\s{a}+\frac{y\s{a}}{2}|\langle x\s{b}+\frac{y\s{b}}{2}|\xi|x\s{a}-\frac{y\s{a}}{2}\rangle |x\s{b}-\frac{y\s{b}}{2}\rangle.
\end{align}
Thus, the WF of the HOM projector $\dyad{\text{HOM}}{\text{HOM}}$ is the following:
\begin{align}
    W\s{HOM}& ( \bold{r})=
 \frac{1}{16\pi^2} \exp \left( -\frac{ p\s{a}^2+p\s{b}^2 + x\s{a}^2+x\s{b}^2 }{2} \right) \nn \\
 & \times ( ( p\s{a}+p\s{b})^2 + (x\s{a}+x\s{b})^2-2 )\nn\\
 & \times((p\s{a}-p\s{b})^2 + (x\s{a}-x\s{b})^2-2 ).\label{HOM}
\end{align}

Both described approaches are identical and, being applied to any ideal transformations, they give the same results.
However, for the non-ideal gate we can apply only the WF-based one since we have to take to account the noises.
Moreover, to obtain the WF of the output state in the case of a non-ideal gate, we have to use the language of the covariance matrices.
An arbitrary Gaussian transformation maps the quadratures of the oscillators as~\cite{weedbrook_gaussian_2012,Gaussian2015}:
\begin{align}
    & \bold{r}\up{out}=T_\sfG\bold{r}\up{in}+\vec{r}\up{N},\quad
    \text{where for a QND gate}\nn\\
    & T_\sfG=
    \begin{pmatrix}
        1 & 0 & \mathsf{G} & 0 \\
        0 & 1 & 0 & 0 \\
        0 & 0 & 1 & 0 \\
        0 & - \mathsf{G} & 0 & 1
    \end{pmatrix},
\end{align}
and $\vec r \up{N}$ is a vector of the quadratures of the added noise.
The covariance matrices are transformed according to
\begin{align}
&    V\s{out}=T_\sfG V\s{in}T_\sfG^T+V\s{N}, \qquad
    \text{where}\nn\\
&   [V_\bullet{}]_{ij}=\frac{1}{2}\langle \vec r ^\bullet{}_i \vec r ^\bullet{}_j + \vec r ^\bullet{}_j \vec r ^\bullet{}_i\rangle - \mean{\vec r ^\bullet{}_i} \mean{\vec r ^\bullet{}_j}\\
& \left(\bullet{}=\text{in,out,N} \right).\nn
\end{align}
This approach works the best with the Gaussian states, for which the WF can be written as
\begin{equation}
W(V,\bold{r},\bold{R})=\frac{\exp{\left(-\frac{1}{2}(\bold{r}- \bold{R})^T V^{-1} (\bold{r}- \bold{R})\right)}}{4\pi^2 \sqrt{\text{det}V}},
\end{equation}
where $\bold{R}\equiv  \left(X\s{a},Y\s{a},X\s{b},Y\s{b}\right)^T$ is the column-vector of means and $V$ is the covariance matrix.
Using an approximation~\cite{Gaussian2015} to represent a single-photon state as a combination of vacuum and a thermal state, we are able to use the Wigner-function based approach for non-Gaussian states such as $\ket 1$.
We used this approach to calculate the HOM elements and the input thresholds for the atom-light, optomechanical and atom-mechanical gates.

To calculate the input threshold for the non-ideal gate let us take two random coherent states as the inputs of the QND gate (for two coherent states, $V\s{in}=\mathbb{I}_{4\times4}$, a $4\times 4$ identity matrix):
\begin{align}
& W\s{in}(\bold{r},\bold{R})=W(V\s{out},\bold{r},\bold{R}),\\
& W\s{out}(\bold{r},\bold{R}, \mathsf{G})=W(V\s{out},\bold{r},T_\sfG\bold{R}).
\end{align}
Using Eqs.~(\ref{HOMMEWF}) and (\ref{HOM})  we can obtain the HOM element  $\langle \text{HOM}| \rho\s{out}\up{coh}|\text{HOM}\rangle=\mathsf{M}\s{HOM}(\bold{R}, \mathsf{G} )$ as a function of  $\mathsf{G}$ and $\bold{R}$.
We can obtain
the input threshold (phase randomized) as a function of $\mathsf{G}$:
\begin{align}
&    h\s{in} = \max\limits_{R\s{a},R\s{b}} \left[
\frac{1}{4\pi^2} \iint d\varphi\s{a}\;d\varphi\s{b}\;\;  \mathsf{M}\s{HOM}(\bold{R}, \mathsf{G} )\right],
\end{align}
where
\begin{align}
&\bold{R}=  \left(X\s{a},Y\s{a},X\s{b},Y\s{b}\right)^T=\\ &=\left(R\s{a} \cos[\varphi\s{a}],R\s{a} \sin[\varphi\s{a}],R\s{b} \cos[\varphi\s{b}],R\s{b} \sin[\varphi\s{b}]\right)^T,\nn
\end{align}
that can be used in the case of a non-ideal QND gate.
In our work, we investigated three types of non-ideal gates, for each of which we used the mixed state Eq.~(\ref{ME_superpositioninput_Eq1}) as the input.
Note that the output threshold is fixed, but the position of the input threshold strongly depends on the physical parameters of the gate.

%%%%%%%%%%%%%%%%%%%%%%%%%%%%%%%%%%%
\section{Discussion and Conclusion}
%%%%%%%%%%%%%%%%%%%%%%%%%%%%%%%%%%%

We have proposed and investigated an analogue of the celebrated Hong-Ou-Mandel bunching effect of single quanta excitations in a light-atom, an optomechanical and finally, a hybrid opto-atom-mechanical system.
In the paradigmatic configuration of the HOM experiment, two indistinguishable photons arrive simultaneously at two input channels of a beamsplitter, and both leave it together through one output port.
In order for the perfect coincidences at the output to take place, the incident photons have to be perfectly indistinguishable in all possible degrees of freedom.
This includes their spatio-temporal modes including arrival times, spectral and polarization modes etc.
In a realistic case, there is always a distinguishability parameter, such as a delay between the arrival times (the temporal mode mismatch), which allows two photons to emerge in different output channels.
The HOM effect is then directly witnessed by measuring the coincidences of photon counts between the output channels, which ideally vanishes at zero distinguishability (e.g., zero delay) thus leading to the famous HOM-dip.
The key properties of the optical HOM experiment are thus (i) the input states of the photons including their spatio-temporal, frequency and polarization modes, (ii) the input-output transformation, which is typically a symmetric beamsplitter, and (iii) the coincidence counting which serves the purpose of verification.

In contrast with the conventional HOM effect, we use a quantum non-demolition gate between the participating oscillators instead of the traditional beamsplitter.
This allows to consider analogy between the transmittance of the latter and the gain $\mathfrak G$ of the QND interaction, as well as all the parameters that influence the gain (light-matter coupling rates, mediating pulse duration and squeezing).
In our setting, we consider a realistic picture of the input states, where each input quantum oscillator (optical, atomic and mechanical) is described by a single quantum mode.
The single-mode picture of the matter modes is nevertheless fully feasible and is capable of perfectly accurate description of state-of-the-art experiments~\cite{aspelmeyer_cavity_2014,karg_lightmediated_2020}.
The single-mode picture rids us of the complications associated with multimode character of light field but allows to capture the crucial non-classical character of the excitations bunching.
Finally, while the conventional HOM allows a direct observation by the coincidence counting, in our system, the matter subsystems (atoms and mechanics) cannot be addressed directly.
Hence in a real experiment light should be used to perform the two-mode tomography~\cite{morin_remote_2014, makino_synchronization_2016, ulanov_losstolerant_2016, marek_direct_2017}.
As an example, it is possible to use a red-detuned drive of the optomechanical cavity on the lower mechanical sideband.
Such driving enables a state swap between the mechanics and the leaking light~\cite{aspelmeyer_cavity_2014,vanner_optomechanical_2015,rakhubovsky_photonphononphoton_2017} whose subsequent tomography allows to infer the initial mechanical state.
A QND coupling can be used in place of a beam-splitter in order to perform the tomography one quadrature at a time~\cite{lei_quantum_2016,shomroni_optical_2019}.
In the atomic subsystem, the spin waves can be read out with high efficiency using a quantum memory protocols based on counter-propagating quantum signal wave and strong classical reference wave through the atomic matter~\cite{Cao:20,PhysRevA.103.062426,Pu2017}.

Our main finding is that the atomic-mechanical HOM effect via realistic QND interaction with light reveals two-boson interference beyond the classical states.
In order to prove the non-classical character quanta bunching, we devise coherent-state-based thresholds for the output bipartite quantum state.
The output threshold, computed as the maximal HOM element achievable by two coherent states at the output, marks the HOM elements attainable by the classical states irrespectively to the interaction.
The input threshold, equal to the maximal HOM element possible to obtain at the output given two phase-randomized coherent states at the output, shows the bunching enabled by the interference of intensity of classical sources for the same interaction.
We prove that both thresholds can be overcome with feasible parameters of opto-atom-mechanical systems.

We found out that in the scheme, which can be controlled by the set of coupling rates, the mediator pulse duration, and squeezing, there are optimal values of each of these parameters to observe the HOM bunching.
These optimal control parameters are influenced by the value of the optical loss in the system and the heating rate due to the coupling of the mechanics to its environment.
Importantly, we have shown that the optimal parameters are either within the values implemented in the already reported experiments or within the reach.
For the choice of numerical parameters, we were inspired by Refs~\cite{Karg2020,Thomas2020}.
It should be noted that  for the HOM effect the important are not the absolute values of the parameters of the gate (with the exception of optical losses and thermal noises, which the smaller the better),
but the relations between them.

For the atom-light gate the only requirement to observe the HOM effect is to provide a strong QND interaction between a collective atomic spin and light~\cite{BOOK}.
Atoms, usually  a cloud of alkali metals as Rb or Ce with $10^7-10^{11}$ units, can be either cooled, or taken at the room temperature.
They can be placed into the cavity, but there is no need (if a cavity is used~--- adiabatic regime is preferable).
The coupling strength  $g_{\text{\tiny{A}}}$ obtained in the  experiments usually is about hundreds of Hz and can be varied by the number of atoms in a cloud or photons in the pump ~\cite{Vasilakis,AtomLightKohler2018,Novikova}.
For the HOM effect the important ratio is the pulse duration multiplied by the coupling strength.
To maximize the HOM effect it should be $g_{\text{\tiny{A}}}\tau\approx0.87$ for the free space case (and  $g_{\text{\tiny{A}}}\tau\approx5$  with $g_{\text{\tiny{A}}}\approx0.05\kappa_{\text{\tiny{A}}}$ if cavity is used), which allows to use of a wide range of the pulse durations.

To obtain the HOM effect for the optomechanical gate one has to provide a QND interaction between a mechanical oscillator and light with a sufficient coupling strength $g_{\text{\tiny{M}}}$.
Normally $g_{\text{\tiny{M}}}$ is about tenths of the optical decay rate~$\kappa_{\text{\tiny{M}}}$~\cite{Delic} that is enough to achieve the result.
The HOM effect occurs when the interaction is in the adiabatic regime.
Similar to the  atom-light gate the maximum of the HOM element is provided at the ratio $g_{\text{\tiny{M}}}\tau\approx5$ with $g_{\text{\tiny{M}}}\approx0.05\kappa_{\text{\tiny{M}}}$.
For this type of the gate it is extremely important to cool the mechanics since the thermal noises can totally destroy the effect.
For the rethermalization rate $\Gamma\s{\text{\tiny{M}}} / \kappa\s{\text{\tiny{M}}}> 10^{-2}$ the effect vanishes.
In recent experiments it is already possible to decrease this value up to the $\Gamma_{\text{\tiny{M}}} = 10^{-4}\kappa_{\text{\tiny{M}}}$~\cite{tsaturyan_ultracoherent_2017,hoj_ultracoherent_2021}
that for the HOM effect can be considered as an ideal cooling since a lower rethermalization would not give a noticeable advantage.

The atom-mechanical gate combines recommendations for the both previous gates.
In this work,
%\textcolor{red}{for simplicity and unification},
we assumed the atomic ensemble be placed in a cavity of the same optical decay rate as for the optomechanical cavity and investigated the situation when the coupling constants are equal.
Both these assumptions are not necessary to observe the effect.

It is required to ensure that the light that has interacted with the atomic subsystem completely enters the optomechanical cavity (to increase the efficiency of the gate),
but even strong optical losses still do not lead to the complete disappearance of the maximum of the HOM element.
This means for this effect (evaluated by the input coherent threshold) problems associated with distortion of the temporal profile of the pulse during the interactions are not so significant.
Moreover, the effective value of the loss can be decreased by a proper engineering of the driving field temporal shapes~\cite{Kuz’min2015}.
In addition, the overall gain of the interaction can be slightly increased by performing an optimal mode-matching of the homodyne local oscillator to the light that leaks from the optomechanical cavity.
If atoms are in a free space the time durations of about $1$ms  that are usually used are suitable.
For the cavity configuration with the same parameters for the both parts of the scheme  (assumed in our work) we need to be careful with the optical decay rate.
We need to decrease the pulse duration from $1$ms to $\tau=0.14$ms (it is possible, still much higher than the atom transient time and the oscillator period), keeping the coupling constant as $g_{\text{\tiny{A}}}=2\pi\times 7$kHz~\cite{Vasilakis}.
According to~\cite{Delic}  for the optomechanical part $\kappa_{\text{\tiny{M}}}\approx2\pi\times100$kHz ($\tau\kappa_{\text{\tiny{M}}}=90$ for $\tau=0.14$ms), while the coupling strength is still about tenths of the optical decay rate.
Thus, relations $\tau\kappa=90$ providing  $g_{\text{\tiny{A}}}=g_{\text{\tiny{M}}}=0.07\kappa$ are experimentally achievable, but it is better to keep atoms in a free space and do not chase to make two subsystems identical.
Recent experiments~\cite{Karg2020,Thomas2020} demonstrate the possibility to obtain the gate between mechanical and a spin oscillators with the achievable coupling of the order $1-10$kHz.

The atom-mechanical gate  is symmetric.
For this type of the gate it is possible to increase the value of the maximum of the HOM element, i.e. improve the visibility of the HOM effect, by using a squeezed pulse as a light mediator.
Squeezing should be optimized in accordance with other gate parameters, but anyway it should not be strong: $5-7$dB is good enough, strong squeezing would destroy the effect.

Thus, we can safely say that at the moment the physical capabilities of the experiment allow observing the HOM effect for all three types of gates.
The bunching of phonons and polaritons whose success depends on the experimental parameters, can be used for a quantum-enhanced estimation of coupling between atoms and mechanics in systems like reported in~\cite{Karg2020,Thomas2020}.
Multiplexing of the matter systems can potentally address dual-rail encoding in the output light and allow generation of polarization-entangled light states.
Besides this interesting perspectives, a HOM-like experiment in the spirit of proposed here will be advanced test of the hybrid pulsed gates opening joint experiments with non-Gaussian states of atoms and mechanical oscillators and, in future, nonlinear hybrid gates using atomic and mechanical nonlinearities.

\section{Data availability}
%%%%%%%%%%%%%

The datasets generated and analysed during the current study are available from the corresponding author on reasonable request.

%%%%%%%%%%%%%%%%
\section{Acknowledgements}
%%%%%%%%%%%%%%

We all acknowledge the support of project 20-16577S of the Czech Science Foundation.
A.D.M. has been also supported by the project CZ.02.1.01/0.0/0.0/16\_026/0008460 
of the Czech Ministry of Education, Youth and Sport (MEYS).
A.D.M and R.F. also acknowledge the support of by national funding from
MEYS and the European Union’s Horizon 2020 (20142020) research
and innovation framework programme under grant agreement program under
Grant No. 731473 (project 8C20002 ShoQC).
Project ShoQC has received funding from the
QuantERA ERA-NET Cofund in Quantum Technologies
implemented within the European Union’s Horizon 2020 program.

%%%%%%%%%%%%%%
\section{Competing interests}
%%%%%%%%%%%%%%

The authors declare no competing interests.

%%%%%%%%%%%%%%%
\section{Author Contributions}
%%%%%%%%%%%%%%%

R.F. developed theoretical idea and supervised the project. A.D.M. performed calculations and made the figures with input from A.A.R.
All authors jointly contributed to analysis and preparation of the manuscript.

\medskip

%Sets the bibliography style to UNSRT and imports the
%bibliography file "samples.bib".
% \bibliographystyle{unsrt}
% \bibliography{sample}
{\sloppy
\printbibliography
}

\onecolumn

\appendix

%%%%%%%%%%%%%%%%%%%%
%%%%%%%%%%%%%%%%%%%%
\section{Model of the atomic-opto-mechanical gate}\L{app1}
%%%%%%%%%%%%%%%%%%%%
%%%%%%%%%%%%%%%%%%%%

In this section we derive the input-output relations that characterize the atom-optical, opto-mechanical and atom-mechanical interactions which take place in our proposed setup.

\begin{figure}[ht]
\begin{center}
\includegraphics[width=0.8\linewidth]{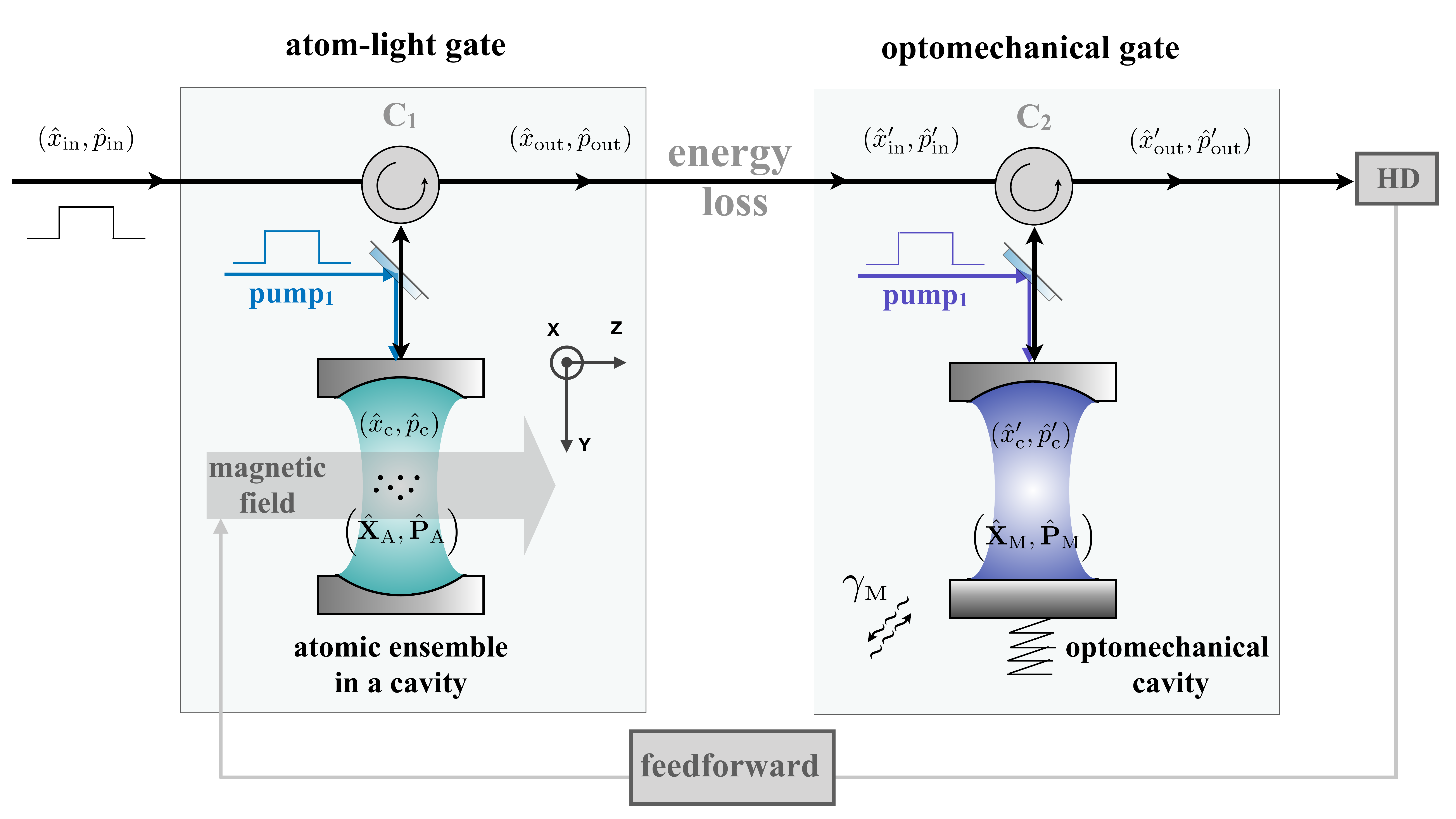}
\caption{QND gate between an atomic ensemble and a mechanical oscillator:
a quantum light pulse with a rectangular temporal profile first passes the atomic ensemble in a cavity
and then the optomechanical cavity via circulators $\text{C}_{1,2}$ and then goes to the homodyne detector (HD).
Within the cavities the optical pulse is coupled to atoms and mechanics respectively via QND interactions enabled by strong classical pumps.
The homodyne detection data are used to control the optical feedforward procedure  after the detection to shift the atomic quadratures.
Canonical variables $(\hat{X}\s{\tiny{A}},\hat{P}\s{\tiny{A}})$, $(\hat{X}\s{\tiny{M}},\hat{P}\s{\tiny{M}})$, $\(\hat{x}\s{c},\hat{p}\s{c}\)$,  and  $\(\hat{x}'\s{c},\hat{p}'\s{c}\)$ are the quadratures of the collective atomic spin,
mechanical oscillator and intracavity modes respectively;
non-canonical variables $\(\hat{x}\s{in},\hat{p}\s{in}\)$,   $\(\hat{x}\s{out},\hat{p}\s{out}\)$, $\(\hat{x}'\s{in},\hat{p}'\s{in}\)$ and  $\(\hat{x}'\s{out},\hat{p}'\s{out}\)$
are the quadratures of the light field  outside the cavities in free space at the corresponding parts of the scheme.
The homodyne measurement and magnetic feedforward control via magnetic field phase shifter are optimized to perform the QND interaction and the squeezed light is used to achieve large entangling power.}
\label{Fig1sup}
\end{center}
\end{figure}

The schematic diagram of the setup that allows realization of the gate is shown in the Fig.~\ref{Fig1sup}.
Note that such an atomic-mechanical gate consists of two parts,
which in turn are the QND gates coupling the optical mode with the atomic (for the atom-light gate) or mechanical (for the optomechanical gate) modes.
Let us describe the model of each in detail.

%%%%%%%%%%%%%%%%%%%%
\subsection{Atom-light QND gate}\L{app3i}
%%%%%%%%%%%%%%%%%%%%

A quantum pulse in a free space, described by the quadratures $( \hat{x}\s{in}(t),\hat{p}\s{in}(t))$, enters the cavity (See Fig.~\ref{Fig1}, the atom-light gate part) that contains a cloud of alkali-metal atoms.
To describe the atomic subsystem we consider the state of an ensemble of atoms at room temperature, each having two stable ground states.
We assume a strong magnetic driving along the $Z$-axis for the atomic ensemble that allows us to apply the Holstein-Primakoff transformation
and consider normalized collective spins~$(\hat{X}_\text{A},\hat{P}_\text{A})$  as very long-lived canonical atomic variables.
Duration of the initial pulse is $\tau$.
Optical damping rate of the cavity is $\kappa_a$.
The pulse is accompanied by the strong classical driving that ensures the QND type interaction with the Hamiltonian $\hat{H}\s{LA}  = \hbar g\s{\tiny{A}} \hat{X}\s{\tiny{A}}\hat{p}\s{c}$.
For the atomic part the coupling constant is the following~\cite{BOOK}:
$g\s{\tiny{A}}=3\Gamma\sigma \sqrt{N\s{ph}}\sqrt{N\s{at}}/(2\tau\Delta A)$, where
$\Gamma$ -- total spontaneous decay rate of the upper state,
$\sigma$ -- resonant photon absorption cross section ($\sigma =\lambda^2/2\pi$),
$\tau$ -- pulse duration,
$\Delta$ -- the Raman detuning,
$A$ -- the beam cross section,
$N\s{ph}$, $N\s{at}$ -- number of photons in the driving pulse and the number of atoms in the atomic ensemble.

The intra cavity field described by the canonical quadratures $( \hat{x}\s{c},\hat{p}\s{c})$ evolves in accordance with this Hamiltonian.
After the interaction field leaves the cavity, the field at the cavity output is described by the quadrature pair  $(\hat{x}\s{out}(t),\hat{p}\s{out}(t))$.
Now the light and atoms are coupled.
At this stage we take the losses into account so we introduce final field $(\hat{x}\s{detect}(t),\hat{p}\s{detect}(t))$ that is the $(\hat{x}\s{out}(t),\hat{p}\s{out}(t))$-field but with the admixed vacuum.
We have to keep in mind that losses are presented at any stage of the gate,
so the losses characterized by a single parameter $\eta$ (efficiency of the gate) are effective and describe all possible losses in the system.
We introduce the canonical quadratures for the initial and final states of the light pulse as $(\hat{\bold{X}}\s{\tiny{L}}^{0},\hat{\bold{Y}}\s{\tiny{L}}^{0})$ and  $( \hat{\bold{X}}{}\up{out}, \hat{\bold{Y}}{}\up{out})$ as the spectra at the zero frequency (formal definition is in Eqns.~(\ref{eq:pulse:quada},\ref{eq:pulse:quadb}) below).
Thus, the atom-light gate transforms the initial vector $(\hat{X}\s{\tiny{A}}(0),\hat{P}\s{\tiny{A}}(0),\hat{\bold{X}}\s{\tiny{L}}^{0},\hat{\bold{Y}}\s{\tiny{L}}^{0})$ to the final vector $(\hat{X}\s{\tiny{A}}(\tau),\hat{P}\s{\tiny{A}}(\tau),\hat{\bold{X}}{}\up{out},\hat{\bold{Y}}{}\up{out})$.

Mathematically, the whole process is as follows.
The Heisenberg-Langevin equations set is:
\begin{align}
\begin{cases}
 \dot{\hat x}\s{c}(t)=-\kappa\s{\tiny{A}} \hat{x}\s{c}(t)+\sqrt{2\kappa\s{\tiny{A}}}\hat{x}\s{in}(t)+g\s{\tiny{A}} \hat{X}\s{\tiny{A}}(t),\\
 \dot{ \hat{p}}\s{c}(t) = -\kappa\s{\tiny{A}} \hat{p}\s{c}(t)+\sqrt{2\kappa\s{\tiny{A}}}\hat{p}\s{in}(t),\\
 \dot{ \hat{X}}\s{\tiny{A}}(t) =0,\\
 \dot{ \hat{P}}\s{\tiny{A}}(t) =- g\s{\tiny{A}} \hat{p}\s{c}(t).
\end{cases}
\end{align}
The solution of the set is:
\begin{align}
\begin{cases}
 \hat{X}\s{\tiny{A}}(t)=\hat{X}\s{\tiny{A}}(0)\\
 \hat{P}\s{\tiny{A}}(t)=\hat{P}\s{\tiny{A}}(0)- \sqrt{2\kappa\s{\tiny{A}}}\(\Theta_{\kappa\s{\tiny{A}}} (t)\ast\hat{p}\s{in}(t)\)- \hat{p}\s{c}(0)\Theta_{\kappa\s{\tiny{A}}} (t)\\
 \hat{x}\s{c}(t)  =\sqrt{2\kappa\s{\tiny{A}}}(e^{-\kappa\s{\tiny{A}} t}\ast \hat{x}\s{in}(t))+\hat{X}\s{\tiny{A}}(0)\Theta_{\kappa\s{\tiny{A}}} (t)+ \hat{x}\s{c}(0)e^{-\kappa\s{\tiny{A}} t}\\
 \hat{p}\s{c}(t)=\sqrt{2\kappa\s{\tiny{A}}}\(e^{-\kappa\s{\tiny{A}} t}\ast \hat{p}\s{in}(t)\)+\hat{p}\s{c}(0)e^{-\kappa\s{\tiny{A}} t}
\end{cases},
\;\;\; \Theta_{\kappa\s{\tiny{A}}} (t)=g\s{\tiny{A}} \frac{1-e^{-\kappa\s{\tiny{A}} t} }{ \kappa\s{\tiny{A}}}.
\end{align}
Here, the $\ast$-symbol  is a convolution, i.e. $f(t)\ast g(t)=\int_0^t f(t-t')g(t')dt'$.

We use the input-output relations to obtain  the field at the cavity output (outside the cavity) and take the optical losses into account:
\begin{align}
\begin{cases}
&\hat{x}\s{detect}(t)=\sqrt{\eta}\;\hat{x}\s{out}(t)+\sqrt{1-\eta}\;\hat{x}\s{vac}(t)\\
&\hat{p}\s{detect}(t)=\sqrt{\eta}\;\hat{p}\s{out}(t)+\sqrt{1-\eta}\;\hat{p}\s{vac}(t)
\end{cases},
\qquad\text{where}\qquad
\begin{cases}
&\hat{x}\s{out}(t)=\sqrt{2\kappa_m}\hat x\s{c}(t)-\hat x{}\s{in}(t)\\
&\hat{p}\s{out}(t)=\sqrt{2\kappa_m}\hat p\s{c}(t)-\hat p{}\s{in}(t)
\end{cases}.
\end{align}
We introduce the canonical variables for the initial and final light pulses as:
\begin{align}
    \label{eq:pulse:quada}
& \hat{\bold{X}}\s{\tiny{L}}^{0}= \frac{1}{\sqrt{\tau}} \int_0^\tau     \hat{x}\s{in}(t) dt,            && \hat{ \bold{Y}}\s{\tiny{L}}^{0}= \frac{1}{\sqrt{\tau}} \int_0^\tau     \hat{p}\s{in}(t)    dt, \\
    \label{eq:pulse:quadb}
& \hat{\bold{X}}{}\up{out}=\frac{1}{\sqrt{\tau}}\int_0^\tau \hat{x}\s{detect}(t)dt,     && \hat{\bold{Y}}{}\up{out}=\frac{1}{\sqrt{\tau}}\int_0^\tau \hat{p}\s{detect}(t)dt.
\end{align}
Thus , we obtain the following atom-light gate ($\hat {X}\s{\tiny{A}}(0)\equiv\hat{\bold{X}}\s{\tiny{A}}^0,\hat{P}\s{\tiny{A}}(0)\equiv\hat{\bold{Y}}\s{\tiny{A}}^0, \hat{X}\s{\tiny{A}}(\tau) \equiv \hat{\bold{X}}\s{\tiny{A}}, \hat{P}\s{\tiny{A}}(\tau)\equiv \hat{\bold{Y}}\s{\tiny{A}} $):
\begin{align}
&  \hat{\bold{X}}\s{\tiny{A}} =\hat{\bold{X}}\s{\tiny{A}}^0+ \hat{\mathbf{N}}_{\text{X}\s{\tiny{A}}},                                                    && \hat{\bold{X}}{}\up{out}=\mathsf{T}\s{\tiny{L}}\hat{\bold{X}}^{0}\s{\tiny{L}}+\mathsf{G}\s{\tiny{L}}\hat{\bold{X}}\s{\tiny{A}}^0+ \hat{\mathbf{N}}_{\text{X}\s{\tiny{L}}},\\
&  \hat{\bold{Y}}\s{\tiny{A}} =\hat{\bold{Y}}\s{\tiny{A}}^0-\mathsf{G}\s{\tiny{A}}\hat{\bold{Y}}\s{\tiny{L}}^{0}+ \hat{\mathbf{N}}_{\text{P}\s{\tiny{A}}}, && \hat{\bold{Y}}{}\up{out}=\mathsf{T}\s{\tiny{L}}\hat{\bold{Y}}^{0}\s{\tiny{L}}+ \hat{\mathbf{N}}_{\text{P}\s{\tiny{L}}},
\end{align}
transforming the initial vector $(\hat{\bold{X}}\s{\tiny{A}}^0,\hat{\bold{Y}}\s{\tiny{A}}^0,\hat{\bold{X}}\s{\tiny{L}}^{0},\hat{\bold{Y}}\s{\tiny{L}}^{0})$ to the final vector $( \hat{\bold{X}}\s{\tiny{A}}, \hat{\bold{Y}}\s{\tiny{A}},\hat{\bold{X}}{}\up{out},\hat{\bold{Y}}{}\up{out})$.
Here, the gains $\mathsf{G}\s{\tiny{A,L}}$, transfer factors $\mathsf{T}\s{\tiny{L}}$ and noises $\hat{\mathbf{N}}_{\text{X}\s{\tiny{A,L}},\text{P}\s{\tiny{A,L}}} $ are:
\begin{align}
&   \mathsf{G}\s{\tiny{A}}=g\s{\tiny{A}}\sqrt{\frac{2\tau}{\kappa\s{\tiny{A}}}},
  \qquad
  \mathsf{G}\s{\tiny{L}} =
  g\s{\tiny{A}} \sqrt{ \frac{ 2 \tau }{ \kappa\s{\tiny{A}} }} \times \sqrt{ \eta } \left[ 1 - \frac{ 1 - e^{ - \kappa\s{\tiny{A}} \tau }}{ \kappa\s{\tiny{A}} \tau } \right],\\
& \mathsf{T}\s{\tiny{L}}=\sqrt{\eta}\(\coeffL-1\),
\end{align}
\begin{align}
&\hat{\mathbf{N}}_{\text{X}\s{\tiny{A}}}=0,\\
&\hat{\mathbf{N}}_{\text{P}\s{\tiny{A}}}=- \hat{p}\s{c}(0)\Theta_{\kappa\s{\tiny{A}}} (t)+\frac{  \mathsf{G}\s{\tiny{A}}\coeffK_1}{\sqrt{\tau}}\hat{\bold{Y}}_{0\kappa\s{\tiny{A}}},\\
&\hat{\mathbf{N}}_{\text{X}\s{\tiny{L}}}=\sqrt{1-\eta}\;\hat{\bold{x}}\s{vac}+\frac{\sqrt{2\eta}(1 - e^{-\kappa\s{\tiny{A}} \tau})}{ \sqrt{\kappa\s{\tiny{A}}\tau}}\hat{x}\s{c}(0) +\sqrt{\eta}\; \coeffL\coeffL_1\hat{\bold{X}}_{0\text{f}_1},\\
&\hat{\mathbf{N}}_{\text{P}\s{\tiny{L}}}=\sqrt{1-\eta}\;\hat{\bold{p}}\s{vac}
+\frac{\sqrt{2\eta}(1 - e^{-\kappa\s{\tiny{A}} \tau})}{ \sqrt{\kappa\s{\tiny{A}}\tau}}\hat{p}\s{c}(0)+\sqrt{\eta}\; \coeffL\coeffL_1\hat{\bold{Y}}_{0\text{f}_1}.
\end{align}
We use the following canonical variables
\begin{align}
& \hat{\bold{Y}}_{0\kappa\s{\tiny{A}}}= \frac{1}{\coeffK_1} \int_0^\tau   dt e^{-\kappa\s{\tiny{A}}(\tau-t)}\hat{p}\s{in}(t),\\
& \hat{\bold{Y}}_{0\text{f}_1}=\frac{1}{\coeffL_1} \int_0^\tau   dt \(\frac{f_1(t)} {\coeffL}- \frac{1}{\sqrt{\tau}}\) \hat{p}\s{in}(t), && \hat{\bold{X}}_{0\text{f}_1}=\frac{1}{\coeffL_1} \int_0^\tau   dt \(\frac{f_1(t)} {\coeffL}- \frac{1}{\sqrt{\tau}}\) \hat{x}\s{in}(t),
\end{align}
with the following correlation relations (all other unspecified combinations correlate to zero)
\begin{align}
&\langle \hat{\bold{Y}}\s{\tiny{L}}^{0} \hat{\bold{Y}}\s{\tiny{L}}^{0}\rangle =\langle  \hat{\bold{Y}}_{0\kappa\s{\tiny{A}}} \hat{\bold{Y}}_{0\kappa\s{\tiny{A}}}\rangle=\langle \hat{ \bold{Y}}_{0\text{f}_1} \hat{\bold{Y}}_{0\text{f}_1}\rangle=\langle \hat{\bold{X}}\s{\tiny{L}}^{0} \hat{\bold{X}}\s{\tiny{L}}^{0}\rangle =\langle \hat{\bold{X}}_{0\text{f}_1} \hat{\bold{X}}_{0\text{f}_1}\rangle=1,\\
&\langle\hat{ \bold{Y}}_{0\kappa\s{\tiny{A}}} \hat{\bold{Y}}_{0}\rangle =\frac{\coeffK_\text{f}}{\coeffK_1},
\;\langle \hat{\bold{Y}}_{0\text{f}_1} \hat{\bold{Y}}\s{\tiny{L}}^{0}\rangle =\langle \hat{\bold{X}}_{0\text{f}_1} \hat{\bold{X}}\s{\tiny{L}}^{0}\rangle=\frac{\coeffK_{\text{f}_1}}{\coeffL_1} ,
\;\langle \hat{\bold{Y}}_{0\text{f}_1} \hat{\bold{Y}}_{0\kappa\s{\tiny{A}}}\rangle =\frac{\coeffK_{\text{ff}_1}}{\coeffL_1\coeffK_1},
\end{align}
where the constants are determined as follows:
\begin{align}
 & \coeffK_1=\sqrt{ \int_0^\tau   dt \(e^{-\kappa\s{\tiny{A}}(\tau-t)}\)^2  },
 \qquad
 \coeffL=\sqrt{\int_0^\tau  f_1^2(t)dt},
 \qquad
 \coeffL_1=\sqrt{ \int_0^\tau   dt \(\frac{f_1(t)} {\coeffL}- \frac{1}{\sqrt{\tau}}\)^2  }, \nn\\
 & \text{where }f_1(t)=\frac{ 2(1 - e^{-\kappa\s{\tiny{A}}(\tau- t)})}{\sqrt{\tau}},\\
 &
 \coeffK\s{f}=\frac{1}{\sqrt{\tau}} \int_0^\tau   dt e^{-\kappa\s{\tiny{A}}(\tau-t)} , \;
 \coeffK_{\text{f}_1}=\frac{1}{\sqrt{\tau}} \int_0^\tau   dt \(\frac{f_1(t)} {\coeffL}- \frac{1}{\sqrt{\tau}}\),\;
 \coeffK_{\text{ff}_1}=\int_0^\tau   dt e^{-\kappa_a(\tau-t)}\(\frac{f_1(t)} {\coeffL}- \frac{1}{\sqrt{\tau}}\).\nn
\end{align}
%%%%%%%%%%%%%%%%%%%%
\subsection{Optomechanical QND gate}\L{app3}
%%%%%%%%%%%%%%%%%%%%

A quantum pulse described by the quadratures $( \hat{x}'\s{in}(t),\hat{p}'\s{in}(t))$ enters the optomechanical cavity (See Fig.~\ref{Fig1}, the optomechanical gate part).
To describe the mechanical part  of the system we use quadratures $(\hat{X}\s{\tiny{M}},\hat{P}\s{\tiny{M}})$ that refer to the dimensionless position and momentum of the mechanical oscillator.
Duration of the initial pulse is $\tau$.
Optical damping rate of the cavity is $\kappa\s{\tiny{M}}$, the rethermalization rate is $\Gamma\s{\tiny{M}}$.
The pulse is accompanied by the strong classical driving that ensures the QND type interaction with the Hamiltonian $\hat{H}\s{\tiny{LM}}  = \hbar g\s{\tiny{M}} \hat{X}\s{\tiny{M}}\hat{p}'\s{c}$.
The intra cavity field described by the canonical quadratures $( \hat{x}'\s{c},\hat{p}'\s{c})$ evolves in accordance with the Hamiltonian.
After the interaction field leaves the cavity, the field at the cavity output is described by the quadrature pair  $(\hat{x}'\s{out}(t),\hat{p}'\s{out}(t))$.
Now the light and mechanics are coupled
(and we can evaluate the HOM-element).
At this stage we take the losses into account so we introduce final field $(\hat{x}'\s{detect}(t),\hat{p}'\s{detect}(t))$ that is the $(\hat{x}'\s{out}(t),\hat{p}'\s{out}(t))$-field but with the admixed vacuum.
We have to keep in mind that losses are presented at any stage of the gate,
so the losses characterized by a single parameter $\eta$ (efficiency of the gate) are effective and describe all possible losses in the system.
We introduce the canonical quadratures for the initial and final  light pulses as $(\hat{\bold{X}}\s{\tiny{L}}^{0},\hat{\bold{Y}}\s{\tiny{L}}^{0})$ and  $( \hat{\bold{X}}'{}\up{out}, \hat{\bold{Y}}'{}\up{out})$ as the spectra at the zero frequency.
Thus, the optomechanical gate transforms the initial vector $(\hat{X}\s{\tiny{M}}(0),\hat{P}\s{\tiny{M}}(0),\hat{\bold{X}}\s{\tiny{L}}^{0},\hat{\bold{Y}}\s{\tiny{L}}^{0})$ to the final vector $(\hat{X}\s{\tiny{M}}(\tau),\hat{P}\s{\tiny{M}}(\tau),\hat{\bold{X}}'{}\up{out},\hat{\bold{Y}}'{}\up{out})$.

Mathematically, the whole process is as follows.
The Heisenberg-Langevin equations are:
\begin{align}
\begin{cases}
 \dot{\hat x}'\s{c}(t)=-\kappa\s{\tiny{M}} \hat{x}'\s{c}(t)+\sqrt{2\kappa\s{\tiny{M}}}\hat{x}'\s{in}(t)+g\s{\tiny{M}} \hat{X}\s{\tiny{M}}(t),\\
 \dot{ \hat{p}}'\s{c}(t) = -\kappa\s{\tiny{M}} \hat{p}'\s{c}(t)+\sqrt{2\kappa\s{\tiny{M}}}\hat{p}'\s{in}(t),\\
 \dot{ \hat{X}}\s{\tiny{M}}(t) =\hat{\zeta}_{\text{X}\s{\tiny{M}}},\\
 \dot{ \hat{P}}\s{\tiny{M}}(t) =\hat{\zeta}_{\text{P}\s{\tiny{M}}}- g\s{\tiny{M}} \hat{p}'\s{c}(t).
\end{cases}
\end{align}
where $\hat{\zeta}_{\text{X}\s{\tiny{M}},\text{P}\s{\tiny{M}}}$ are the operators for the mechanical noises
with the following correlation relations $\langle\hat{\zeta}_{\text{X}\s{\tiny{M}},\text{P}\s{\tiny{M}}}(t)\hat{\zeta}_{\text{X}\s{\tiny{M}},\text{P}\s{\tiny{M}}}(t')\rangle=\gamma\s{\tiny{M}} (2n\s{th}+1)\delta(t-t')\approx 2\Gamma\s{\tiny{M}}\delta(t-t')$.
The solution of the set is:
\begin{align}
&\begin{cases}
 \hat{X}\s{\tiny{M}}(t)=\sqrt{\tau2\Gamma\s{\tiny{M}}}\;\bm{\hat{\zeta}}^{\text{X}\s{\tiny{M}}}+\hat{X}\s{\tiny{M}}(0)\\
 \hat{P}\s{\tiny{M}}(t)=\sqrt{\tau2\Gamma\s{\tiny{M}}}\;\bm{\hat{\zeta}}^{\text{P}\s{\tiny{M}}}+\hat{P}\s{\tiny{M}}(0)- \sqrt{2\kappa\s{\tiny{M}}}\(\Theta_{\kappa\s{\tiny{M}}} (t)\ast\hat{p}'\s{in}(t)\)- \hat{p}'\s{c}(0)\Theta_{\kappa\s{\tiny{M}}} (t)\\
 \hat{x}'\s{c}(t)  =\sqrt{2\kappa\s{\tiny{M}}}(e^{-\kappa\s{\tiny{M}} t}\ast \hat{x}'\s{in}(t))+\(\Theta_{\kappa\s{\tiny{M}}} (t)\ast \hat{\zeta}_{\text{X}\s{\tiny{M}}}(t)\)+\hat{X}\s{\tiny{M}}(0)\Theta_{\kappa\s{\tiny{M}}} (t)+ \hat{x}'\s{c}(0)e^{-\kappa\s{\tiny{M}} t}\\
 \hat{p}'\s{c}(t)=\sqrt{2\kappa\s{\tiny{M}}}\(e^{-\kappa\s{\tiny{M}} t}\ast \hat{p}'\s{in}(t)\)+\hat{p}'\s{c}(0)e^{-\kappa\s{\tiny{M}} t}
\end{cases},
\\
& \Theta_{\kappa\s{\tiny{M}}} (t)=g\s{\tiny{M}} \frac{1-e^{-\kappa\s{\tiny{M}} t} }{ \kappa\s{\tiny{M}}}.\nn
\end{align}
%
% Here, the $\ast$-symbol  is a convolution, i.e. $f(t)\ast g(t)=\int_0^t f(t-t')g(t')dt'$.

We use the input-output relations to obtain  the field at the cavity output (outside the cavity) and take the optical losses into account:
\begin{align}
\begin{cases}
&\hat{x}'\s{detect}(t)=\sqrt{\eta}\;\hat{x}'\s{out}(t)+\sqrt{1-\eta}\;\hat{x}\s{vac}(t)\\
&\hat{p}'\s{detect}(t)=\sqrt{\eta}\;\hat{p}'\s{out}(t)+\sqrt{1-\eta}\;\hat{p}\s{vac}(t)
\end{cases},
\qquad\text{where}\qquad
\begin{cases}
&\hat{x}'\s{out}(t)=\sqrt{2\kappa\s{\tiny{M}} }\hat x'\s{c}(t)-\hat x'{}\s{in}(t)\\
&\hat{p}'\s{out}(t)=\sqrt{2\kappa\s{\tiny{M}} }\hat p'\s{c}(t)-\hat p'{}\s{in}(t)
\end{cases}.
\end{align}
We introduce the canonical variables for the initial and final light pulses as:
\begin{align}
& \hat{\bold{X}}\s{\tiny{L}}^{0}= \frac{1}{\sqrt{\tau}} \int_0^\tau     \hat{x}'\s{in}(t) dt,            && \hat{ \bold{Y}}\s{\tiny{L}}^{0}= \frac{1}{\sqrt{\tau}} \int_0^\tau     \hat{p}'\s{in}(t)    dt, \\
& \hat{\bold{X}}'{}\up{out}=\frac{1}{\sqrt{\tau}}\int_0^\tau \hat{x}'\s{detect}(t)dt,     && \hat{\bold{Y}}'{}\up{out}=\frac{1}{\sqrt{\tau}}\int_0^\tau \hat{p}'\s{detect}(t)dt.
\end{align}
Thus , we obtain the following optomechanical gate ($\hat {X}\s{\tiny{M}}(0)\equiv\hat{\bold{X}}\s{\tiny{M}}^0,\hat{P}\s{\tiny{M}}(0)\equiv\hat{\bold{Y}}\s{\tiny{M}}^0, \hat{X}\s{\tiny{M}}(\tau) \equiv \hat{\bold{X}}\s{\tiny{M}}, \hat{P}\s{\tiny{M}}(\tau)\equiv \hat{\bold{Y}}\s{\tiny{M}} $):
\begin{align}
&  \hat{\bold{X}}\s{\tiny{M}} =\hat{\bold{X}}\s{\tiny{M}}^0+ \hat{\mathbf{N}}_{\text{X}\s{\tiny{M}}},                                                    && \hat{\bold{X}}'{}\up{out}=\mathsf{T}\s{\tiny{L}}\hat{\bold{X}}^{0}\s{\tiny{L}}+\mathsf{G}\s{\tiny{L}}\hat{\bold{X}}\s{\tiny{M}}^0+ \hat{\mathbf{N}}_{\text{X}\s{\tiny{L}}},\\
&  \hat{\bold{Y}}\s{\tiny{M}} =\hat{\bold{Y}}\s{\tiny{M}}^0-\mathsf{G}\s{\tiny{M}}\hat{\bold{Y}}\s{\tiny{L}}^{0}+ \hat{\mathbf{N}}_{\text{P}\s{\tiny{M}}}, && \hat{\bold{Y}}'{}\up{out}=\mathsf{T}\s{\tiny{L}}\hat{\bold{Y}}^{0}\s{\tiny{L}}+ \hat{\mathbf{N}}_{\text{P}\s{\tiny{L}}},
\end{align}
transforming the initial vector $(\hat{\bold{X}}\s{\tiny{M}}^0,\hat{\bold{Y}}\s{\tiny{M}}^0,\hat{\bold{X}}\s{\tiny{L}}^{0},\hat{\bold{Y}}\s{\tiny{L}}^{0})$ to the final vector $( \hat{\bold{X}}\s{\tiny{M}}, \hat{\bold{Y}}\s{\tiny{M}},\hat{\bold{X}}'{}\up{out},\hat{\bold{Y}}'{}\up{out})$.
Here, the gains $\mathsf{G}\s{\tiny{M,L}}$, transfer factors $\mathsf{T}\s{\tiny{L}}$ and noises $\hat{\mathbf{N}}_{\text{X}\s{\tiny{M,L}},\text{P}\s{\tiny{M,L}}} $ are:
\begin{align}
&   \mathsf{G}\s{\tiny{M}}=g\s{\tiny{M}}\sqrt{\frac{2\tau}{\kappa\s{\tiny{M}}}},
  \qquad
  \mathsf{G}\s{\tiny{L}} =
  g\s{\tiny{M}} \sqrt{ \frac{ 2 \tau }{ \kappa\s{\tiny{M}} }} \times \sqrt{ \eta } \left[ 1 - \frac{ 1 - e^{ - \kappa\s{\tiny{M}} \tau }}{ \kappa\s{\tiny{M}} \tau } \right],\\
& \mathsf{T}\s{\tiny{L}}=\sqrt{\eta}\(\coeffL-1\),
\end{align}
%
%%%%%%%
\begin{align}
&\hat{\mathbf{N}}_{\text{X}\s{\tiny{M}}}=\sqrt{\tau2\Gamma\s{\tiny{M}}}\;\bm{\hat{\zeta}}^{\text{X}\s{\tiny{M}}},\\
&\hat{\mathbf{N}}_{\text{P}\s{\tiny{M}}}=\sqrt{\tau2\Gamma\s{\tiny{M}}}\;\bm{\hat{\zeta}}^{\text{P}\s{\tiny{M}}}- \hat{p}'\s{c}(0)\Theta_{\kappa\s{\tiny{M}}} (t)
+\frac{  \mathsf{G}\s{\tiny{M}}\coeffK_1}{\sqrt{\tau}}
\hat{\bold{Y}}_{0\kappa\s{\tiny{M}}},\\
&\hat{\mathbf{N}}_{\text{X}\s{\tiny{L}}}=\sqrt{1-\eta}\;\hat{\bold{x}}\s{vac}+
\frac{\sqrt{2\eta}(1 - e^{-\kappa\s{\tiny{M}} \tau})}{ \sqrt{\kappa\s{\tiny{M}}\tau}}\hat{x}'\s{c}(0) +\sqrt{\eta}\; \coeffL\coeffL_1\hat{\bold{X}}_{0\text{f}_1}+
\sqrt{\eta}\sqrt{2\Gamma\s{\tiny{M}}}\coeffM
\bm{\hat{\zeta}}^{\text{X}\s{\tiny{M}}}_{\text{f}_2},\\
&\hat{\mathbf{N}}_{\text{P}\s{\tiny{L}}}=\sqrt{1-\eta}\;\hat{\bold{p}}\s{vac}
+\frac{\sqrt{2\eta}(1 - e^{-\kappa\s{\tiny{M}} \tau})}{ \sqrt{\kappa\s{\tiny{M}}\tau}}\hat{p}'\s{c}(0)
+\sqrt{\eta}\; \coeffL\coeffL_1\hat{\bold{Y}}_{0\text{f}_1}.
\end{align}

We use the following canonical variables
\begin{align}
& \hat{\bold{Y}}_{0\kappa\s{\tiny{M}}}= \frac{1}{\coeffK_1} \int_0^\tau   dt e^{-\kappa\s{\tiny{M}}(\tau-t)}\hat{p}'\s{in}(t),\\
& \hat{\bold{Y}}_{0\text{f}_1}=\frac{1}{\coeffL_1} \int_0^\tau   dt \(\frac{f_1(t)} {\coeffL}- \frac{1}{\sqrt{\tau}}\) \hat{p}'\s{in}(t), && \hat{\bold{X}}_{0\text{f}_1}=\frac{1}{\coeffL_1} \int_0^\tau   dt \(\frac{f_1(t)} {\coeffL}- \frac{1}{\sqrt{\tau}}\) \hat{x}'\s{in}(t),\\
& \bm{\hat{\zeta}}^{\text{X}\s{\tiny{M}}}_{f_2}=\frac{1}{ \sqrt{2\Gamma\s{\tiny{M}}}\coeffM}\int_0^\tau  dt \(f_3(t) \hat{\zeta}_{\text{X}\s{\tiny{M}}}(t)\),\\
& \bm{\hat{\zeta}}^{\text{X}\s{\tiny{M}}}=\frac{1}{ \sqrt{2\Gamma\s{\tiny{M}}\tau}}\int_0^\tau  dt \hat{\zeta}_{\text{X}\s{\tiny{M}}}(t).
\end{align}
with the following correlation relations (all other unspecified combinations correlate to zero)
\begin{align}
&\langle \hat{\bold{Y}}\s{\tiny{L}}^{0} \hat{\bold{Y}}\s{\tiny{L}}^{0}\rangle =\langle  \hat{\bold{Y}}_{0\kappa\s{\tiny{M}}} \hat{\bold{Y}}_{0\kappa\s{\tiny{M}}}\rangle=\langle \hat{ \bold{Y}}_{0\text{f}_1} \hat{\bold{Y}}_{0\text{f}_1}\rangle=\langle \hat{\bold{X}}\s{\tiny{L}}^{0} \hat{\bold{X}}\s{\tiny{L}}^{0}\rangle =\langle \hat{\bold{X}}_{0\text{f}_1} \hat{\bold{X}}_{0\text{f}_1}\rangle=
 \langle\bm{\hat{\zeta}}^{\text{X}\s{\tiny{M}}}\bm{\hat{\zeta}}^{\text{X}\s{\tiny{M}}}\rangle= \langle\bm{\hat{\zeta}}^{\text{X}\s{\tiny{M}}}_{\text{f}_2}\bm{\hat{\zeta}}^{\text{X}\s{\tiny{M}}}_{\text{f}_2}\rangle=1,\\
&\langle\hat{ \bold{Y}}_{0\kappa\s{\tiny{M}}} \hat{\bold{Y}}_{0}\rangle =\frac{\coeffK_\text{f}}{\coeffK_1},
\;\langle \hat{\bold{Y}}_{0\text{f}_1} \hat{\bold{Y}}\s{\tiny{L}}^{0}\rangle =\langle \hat{\bold{X}}_{0\text{f}_1} \hat{\bold{X}}\s{\tiny{L}}^{0}\rangle=\frac{\coeffK_{\text{f}_1}}{\coeffL_1} ,
\;\langle \hat{\bold{Y}}_{0\text{f}_1} \hat{\bold{Y}}_{0\kappa\s{\tiny{M}}}\rangle =\frac{\coeffK_{\text{ff}_1}}{\coeffL_1\coeffK_1},
\; \langle\bm{\hat{\zeta}}^{\text{X}\s{\tiny{M}}}_{\text{f}_2}\bm{\hat{\zeta}}^{\text{X}\s{\tiny{M}}}\rangle=\frac{\coeffM_1}{ \sqrt{\tau} \coeffM},
\end{align}
where the constants are determined as follows:
\begin{align}
 & \coeffK_1=\sqrt{ \int_0^\tau   dt \(e^{-\kappa\s{\tiny{M}}(\tau-t)}\)^2  },
 \coeffL=\sqrt{\int_0^\tau  f_1^2(t)dt},
  \coeffL_1=\sqrt{ \int_0^\tau   dt \(\frac{f_1(t)} {\coeffL}- \frac{1}{\sqrt{\tau}}\)^2  },\\
  &
\coeffM=\sqrt{ \int_0^\tau   dt f^2_3(t) },\coeffM_1=\int_0^\tau  dt f_3(t),\nn\\
& \text{where }f_1(t)=\frac{ 2(1 - e^{-\kappa\s{\tiny{M}}(\tau- t)})}{\sqrt{\tau}},\;  f_3(t)=\frac{\sqrt{2\kappa\s{\tiny{M}}}}{\sqrt{\tau}} \frac{g\s{\tiny{M}} (\kappa\s{\tiny{M}} (\tau- t)-1 + e^{-\kappa\s{\tiny{M}} (\tau- t)} )}{\kappa\s{\tiny{M}}^2},\nn\\
&
\coeffK_\text{f}=\frac{1}{\sqrt{\tau}} \int_0^\tau   dt e^{-\kappa\s{\tiny{M}}(\tau-t)} , \;
\coeffK_{\text{f}_1}=\frac{1}{\sqrt{\tau}} \int_0^\tau   dt \(\frac{f_1(t)} {\coeffL}- \frac{1}{\sqrt{\tau}}\),\;
\coeffK_{\text{ff}_1}=\int_0^\tau   dt e^{-\kappa\s{\tiny{M}}(\tau-t)}\(\frac{f_1(t)} {\coeffL}- \frac{1}{\sqrt{\tau}}\).\nn
 \end{align}

%%%%%%%%%%%%%%%%%%
%%%%%%%%%%%%%%%%%%%%
\subsection{Atom-mechanical QND gate}
%%%%%%%%%%%%%%%%%%%%
%%%%%%%%%%%%%%%%%%%%

To establish the gate we connect the atomic and optomechanical cavities  in such a way that the light passes them sequentially, interacting first with atoms and then with mechanics.
Thus, as the input light for the optomechanical part we take the output light of the atom-light part.

A quantum pulse described by the quadratures $\( \hat{x}\s{in}(t),\hat{p}\s{in}(t)\)$ enters the cavity (See Fig.~\ref{Fig1}) that contains a cloud of alkali-metal atoms.
To describe the atomic subsystem we consider the state of an ensemble of atoms at room temperature, each having two stable ground states.
We assume a strong magnetic driving along the Z-axis for the atomic ensemble that allows us to apply the Holstein-Primakoff transformation
and consider normalized collective spins~$(\hat{X}\s{\tiny{A}},\hat{P}\s{\tiny{A}})$  as very long-lived canonical atomic variables.
Duration of the initial pulse is $\tau$.
Optical damping rate of the cavity is $\kappa\s{\tiny{A}}$.
The pulse is accompanied by the strong classical driving that ensures the QND type interaction with the Hamiltonian $\hat{H}\s{LA}  = -\hbar g\s{\tiny{A}} \hat{P}\s{\tiny{A}}\hat{x}\s{c}$.
The intra cavity field described by the canonical quadratures $\( \hat{x}\s{c},\hat{p}\s{c}\)$ evolves in accordance with this Hamiltonian.
After the interaction, field leaves the cavity, the field at the cavity output is described by the quadrature pair  $\(\hat{x}\s{out}(t),\hat{p}\s{out}(t)\)$.
Now the light and atoms are coupled.

At this stage we take the losses into account so we introduce the field  $\( \hat{x}'\s{in}(t),\hat{p}'\s{in}(t)\)$ that is the $\(\hat{x}\s{out}(t),\hat{p}\s{out}(t)\)$-field but with the admixed vacuum.
We have to keep in mind that losses are presented at any stage of the gate,
so the losses characterized by a single parameter $\eta$ (efficiency of the gate) are effective and describe all possible losses in the system.

A quantum pulse described by the quadratures $\( \hat{x}'\s{in}(t),\hat{p}'\s{in}(t)\)$ enters the optomechanical cavity.
To describe the mechanical part  of the system we use quadratures $(\hat{X}\s{\tiny{M}},\hat{P}\s{\tiny{M}})$ that refer to the dimensionless position and momentum of the mechanical oscillator.
Optical damping rate of the cavity is $\kappa\s{\tiny{M}}$, the rethermalization rate is $\Gamma\s{\tiny{M}}$.
The pulse is accompanied by the strong classical driving that ensures the QND type interaction with the Hamiltonian $\hat{H}\s{LM}  = \hbar g\s{\tiny{M}} \hat{X}\s{\tiny{M}}\hat{p}'\s{c}$.
The intra cavity field described by the canonical quadratures $\( \hat{x}'\s{c},\hat{p}'\s{c}\)$ evolves in accordance with the Hamiltonian.
After the interaction field leaves the cavity, the field at the cavity output is described by the quadrature pair  $\(\hat{x}'\s{out}(t),\hat{p}'\s{out}(t)\)$.
Now the atoms and mechanics are coupled.

Afterward the pulse is homodyned and the output of the detection is used to displace the atoms in the phase space.
Here, we consider squeezed light, since for such a gate, squeezing allows to increase the HOM element value.

Mathematically, the whole process is as follows.
The Heisenberg-Langevin equations are:
\BY
&&\dot{\hat x}\s{c}(t)=-\kappa\s{\tiny{A}} \hat{x}\s{c}+\sqrt{2\kappa\s{\tiny{A}}}\hat{x}\s{in}\\
&&\dot{ \hat{p}}\s{c} (t)= g\s{\tiny{A}} \hat{P}\s{\tiny{A}}(t)-\kappa\s{\tiny{A}} \hat{p}\s{c}+\sqrt{2\kappa\s{\tiny{A}}}\hat{p}\s{in}\\
&&\dot{ \hat{X}}\s{\tiny{A}} (t)= - g\s{\tiny{A}}\hat{x}\s{c}(t) \\
&&\dot{\hat P}\s{\tiny{A}}(t) =0.
\EY
The solution of the set is the following:
\BY
&&\hat{X}\s{\tiny{A}}(t)=\hat{X}\s{\tiny{A}}(0)- \sqrt{2\kappa\s{\tiny{A}}}\(\Theta_{\kappa\s{\tiny{A}}}  (t)\ast\hat{x}\s{in}(t)\)- \hat{x}\s{c}(0)\Theta_{\kappa\s{\tiny{A}}}  (t),\\
&&\hat{P}\s{\tiny{A}}(t)=\hat{P}\s{\tiny{A}}(0),\\
&&\hat{x}\s{c}(t)=\sqrt{2\kappa\s{\tiny{A}}}\(e^{-\kappa\s{\tiny{A}} t}\ast\hat{x}\s{in}(t)\)+\hat{x}\s{c}(0)e^{-\kappa\s{\tiny{A}} t},\\
&& \hat{p}\s{c}(t)=\sqrt{2\kappa\s{\tiny{A}}}\(e^{-\kappa\s{\tiny{A}} t}\ast\hat{p}\s{in}(t)\)+ \hat{P}\s{\tiny{A}}(0)\Theta_{\kappa\s{\tiny{A}}}  (t)+ \hat{p}\s{c}(0)e^{-\kappa\s{\tiny{A}} t},
\EY
where  $\Theta_{\kappa\s{\tiny{A}}} (t)$ is:
\BY
\Theta_{\kappa\s{\tiny{A}}}  (t)=g\s{\tiny{A}} \frac{1-e^{-\kappa\s{\tiny{A}} t} }{ \kappa\s{\tiny{A}}}.
\EY
We use the following input-output relation:
\BY
&&\hat Q\up{out}(t)=\sqrt{2\kappa\s{\tiny{A}}}\hat Q(t)-\hat Q\up{in}(t), \qquad \hat{Q}=\hat{x}\s{c},\hat{p}\s{c}.
\EY
Thus, the field at the cavity output (outside the cavity):
\BY
&&\hat{x}\s{out}(t)=\(\(2\kappa\s{\tiny{A}} e^{-\kappa\s{\tiny{A}} t}-\delta(t)\)\ast\hat{x}\s{in}(t)\)+\sqrt{2\kappa\s{\tiny{A}}}\hat{x}\s{c}(0)e^{-\kappa\s{\tiny{A}} t},\\
&&\hat{p}\s{out}(t)=\( \(2\kappa\s{\tiny{A}} e^{-\kappa\s{\tiny{A}} t}-\delta(t)\)\ast\hat{p}\s{in}(t)\)+ \sqrt{2\kappa\s{\tiny{A}}}\hat{P}\s{\tiny{A}}(0)\Theta_{\kappa\s{\tiny{A}}}  (t)+ \sqrt{2\kappa\s{\tiny{A}}}\hat{p}\s{c}(0)e^{-\kappa\s{\tiny{A}} t}.
\EY

Next is the optomechanical interaction.
The Heisenberg-Langevin set is:
\BY
&& \dot{\hat x}'\s{c}(t)=-\kappa\s{\tiny{M}} \hat{x}'\s{c}(t)+\sqrt{2\kappa\s{\tiny{M}}}\hat{x}'\s{in}(t)+g\s{\tiny{M}} \hat{X}\s{\tiny{M}}(t)\\
&&\dot{ \hat{p}}'\s{c}(t) = -\kappa\s{\tiny{M}} \hat{p}'\s{c}(t)+\sqrt{2\kappa\s{\tiny{M}}}\hat{p}'\s{in}(t)\\
&& \dot{ \hat{X}}\s{\tiny{M}}(t) =\hat{\zeta}_{\text{X}\s{\tiny{M}}}\\
&&\dot{ \hat{P}}\s{\tiny{M}}(t) =\hat{\zeta}_{\text{P}\s{\tiny{M}}}- g\s{\tiny{M}} \hat{p}'\s{c}(t)
\EY
with correlation $\langle\hat{\zeta}_{\text{X}\s{\tiny{M}},\text{P}\s{\tiny{M}}}(t)\hat{\zeta}_{\text{X}\s{\tiny{M}},\text{P}\s{\tiny{M}}}(t')\rangle \approx 2\Gamma\s{\tiny{M}}\delta(t-t')$.\\
\\
The solution is
\BY
&&\hat{X}\s{\tiny{M}}(t)=\sqrt{\tau2\Gamma\s{\tiny{M}}}\;\bm{\hat{\zeta}}^{\text{X}\s{\tiny{M}}}+\hat{X}\s{\tiny{M}}(0)\\
&&\hat{P}\s{\tiny{M}}(t)=\sqrt{\tau2\Gamma\s{\tiny{M}}}\;\bm{\hat{\zeta}}^{\text{P}\s{\tiny{M}}}+\hat{P}\s{\tiny{M}}(0)- \sqrt{2\kappa\s{\tiny{M}}}\(\Theta_{\kappa\s{\tiny{M}}} (t)\ast\hat{p}'\s{in}(t)\)- \hat{p}'\s{c}(0)\Theta_{\kappa\s{\tiny{M}}} (t)\\
&& \hat{x}'\s{c}(t)  =\sqrt{2\kappa\s{\tiny{M}}}(e^{-\kappa\s{\tiny{M}} t}\ast \hat{x}'\s{in}(t))+\(\Theta_{\kappa\s{\tiny{M}}} (t)\ast \hat{\zeta}_{\text{X}\s{\tiny{M}}}(t)\)+\hat{X}\s{\tiny{M}}(0)\Theta_{\kappa\s{\tiny{M}}} (t)+ \hat{x}'\s{c}(0)e^{-\kappa\s{\tiny{M}} t}\\
&&\hat{p}'\s{c}(t)=\sqrt{2\kappa\s{\tiny{M}}}\(e^{-\kappa\s{\tiny{M}} t}\ast \hat{p}'\s{in}(t)\)+\hat{p}'\s{c}(0)e^{-\kappa\s{\tiny{M}} t}
\EY
where  $\Theta_{\kappa\s{\tiny{M}}} (t)$ is:
\BY
\Theta_{\kappa\s{\tiny{M}}} (t)=g\s{\tiny{M}} \frac{1-e^{-\kappa\s{\tiny{M}} t} }{ \kappa\s{\tiny{M}}}.
\EY
We use the following input-output relation:
\BY
&&\hat Q'{}\up{out}(t)=\sqrt{2\kappa\s{\tiny{M}}}\hat Q'(t)-\hat Q'{}\up{in}(t), \qquad \hat{Q}'=\hat{x}'\s{c},\hat{p}'\s{c}.
\EY
Thus, the field at the cavity output (outside the cavity):
\BY
&&\hat{x}'\s{out}(t)=\nn\\
&&=\(\(2\kappa\s{\tiny{M}} e^{-\kappa\s{\tiny{M}} t}-\delta(t)\)\ast\hat{x}'\s{in}(t)\)+\sqrt{2\kappa\s{\tiny{M}}}\(\(\Theta_{\kappa\s{\tiny{M}}} (t)\ast \hat{\zeta}_{\text{X}\s{\tiny{M}}}(t)\)+\hat{X}\s{\tiny{M}}(0)\Theta_{\kappa\s{\tiny{M}}} (t)+ \hat{x}'\s{c}(0)e^{-\kappa\s{\tiny{M}} t}\),\nn\\
&&\hat{p}'\s{out}(t)=\( \(2\kappa\s{\tiny{M}} e^{-\kappa\s{\tiny{M}} t}-\delta(t)\)\ast\hat{p}'\s{in}(t)\)+ \sqrt{2\kappa\s{\tiny{M}}}\hat{p}'\s{c}(0)e^{-\kappa\s{\tiny{M}} t}.\nn
\EY

Let us take the optical \textbf{loss} into account as follows:

\BY
&&\hat{x}'\s{in}(t)=\sqrt{\eta}\;\hat{x}\s{out}(t)+\sqrt{1-\eta}\;\hat{x}\s{vac}(t)\\
&&\hat{p}'\s{in}(t)=\sqrt{\eta}\;\hat{p}\s{out}(t)+\sqrt{1-\eta}\;\hat{p}\s{vac}(t)
\EY
\\

Now we have to \textbf{detect} the light outside the optomechanical cavity.
For simplicity let us take $\kappa\s{\tiny{A}}=\kappa\s{\tiny{M}}=\kappa$.
We have to detect the X-quadrature of the output light:
\BY
&&\hat{x}'\s{out}(t)=\sqrt{\eta} \(  \(4\kappa^2te^{-\kappa t}-4\kappa e^{-\kappa t}+\delta(t)\) \ast\hat{x}\s{in}(t)\)+\sqrt{2\kappa}\hat{X}\s{\tiny{M}}(0)\Theta_{\kappa\s{\tiny{M}}} (t)+\text{Noise}.
\EY
We assume a rectangular pulse. Thus
\BY
 &&\hat {\bold{X}}'{}\up{out}=\frac{1}{\sqrt{\tau}}\int_0^\tau \hat{x}'\s{out}(t)dt.
 \EY

To obtain a symmetric gate we need to \textbf{shift} one of the quadratures.
Thus, after detection we shift of the atomic quadrature by the feedforward procedure with  $\bold{K}\s{f}$ coefficient:
\BY
\hat{X}\s{\tiny{A}}(\tau) \rightarrow \hat{X}\s{\tiny{A}}(\tau)+\bold{K}\s{f} \hat {\bold{X}}'{}\up{out},\;\bold{K}\s{f}=\frac{\sqrt{2\eta \tau}g\s{\tiny{M}}\(e^{-\kappa \tau} ( \kappa \tau+2) + \kappa \tau - 2\)}{\sqrt{\kappa}(\kappa \tau- 1 + e^{-\kappa \tau})}.
\EY

The \textbf{symmetric QND Gate} is characterized by the equations that read:
\BY
\begin{cases}
&\hat{X}\s{\tiny{A}}(\tau)=\hat {X}\s{\tiny{A}}(0)+\mathfrak{G}\hat{X}\s{\tiny{M}}(0)+\mathfrak{N}_{\text{X}\s{\tiny{A}}},\\
&\hat{P}\s{\tiny{A}}(\tau)=\hat{P}\s{\tiny{A}}(0),\\
&\hat{X}\s{\tiny{M}}(\tau)=\hat{X}\s{\tiny{M}}(0)+\mathfrak{N}_{\text{X}\s{\tiny{M}}},\\
&\hat{P}\s{\tiny{M}}(\tau)=\hat{P}\s{\tiny{M}}(0)-  \mathfrak{G}\hat{P}\s{\tiny{A}}(0)+\mathfrak{N}_{\text{P}\s{\tiny{M}}},
\end{cases},\text{where}\qquad
\mathfrak{G}=g\s{\tiny{A}}g\s{\tiny{M}}\sqrt{\eta}\frac{2}{\kappa^2}\(e^{-\kappa \tau} ( \kappa \tau+2) + \kappa \tau - 2\)
\EY
Here, the noises are defined as:
\BY
&&\mathfrak{N}_{\text{P}\s{\tiny{A}}}=0,\\
&&\mathfrak{N}_{\text{X}\s{\tiny{M}}}=\sqrt{\tau2\Gamma\s{\tiny{M}}}\;\bm{\hat{\zeta}}^{\text{X}\s{\tiny{M}}},\nn\\
&&\mathfrak{N}_{\text{X}\s{\tiny{A}}}=
- \frac{g\s{\tiny{A}}\sqrt{2}}{ \bold{K}_{2}\sqrt{\kappa}}\hat {\bold{X}}\up{in}+\frac{\bold{K}\s{f}}{\bold{K}_{5}} \sqrt{\frac{\eta}{\tau}} \hat {\bold{X}}\up{in}\s{f} +\frac{\bold{K}\s{f}}{\bold{K}_{1} }\sqrt{\frac{1-\eta}{\tau}}\hat {\bold{X}}\up{vac}+\frac{g\s{\tiny{M}}\bold{K}\s{f}}{\kappa\bold{K}_{3}}\sqrt{\frac{4\Gamma\s{\tiny{M}}}{\kappa \tau}}\bm{\hat{\zeta}}^{\text{X}\s{\tiny{M}}}\s{f}+\nn\\
&&+\(\bold{K}\s{f}\sqrt{\frac{2\eta}{\kappa \tau}}( 1 -e^{-\kappa \tau}  (2 \kappa \tau+1))-g\s{\tiny{A}}(1-e^{-\kappa \tau}) \) \hat{x}\s{c}(0) + \bold{K}\s{f}\sqrt{\frac{2}{\kappa \tau} }   (1 - e^{-\kappa \tau})  \hat{x}'\s{c}(0)\nn\\
&&\mathfrak{N}_{\text{P}\s{\tiny{M}}}=-g\s{\tiny{M}} (1-e^{-\kappa \tau}) \hat{p}'\s{c}(0)+\sqrt{\tau2\Gamma\s{\tiny{M}}}\;\bm{\hat{\zeta}}^{\text{P}\s{\tiny{M}}}
- \sqrt{2\kappa\eta}\; \frac{g\s{\tiny{M}}}{\kappa\bold{K}_{6}}
\hat {\bold{P}}\up{in} -\nn\\
&&
- 2 \sqrt{\eta}g\s{\tiny{M}} \frac{(1 - e^{-\kappa \tau} (1 + \kappa \tau))}{\kappa}  \hat{p}\s{c}(0)
- \frac{g\s{\tiny{M}}}{\bold{K}_{2}}\sqrt{2\kappa(1-\eta)}
\hat {\bold{P}}\up{vac}.\nn
\EY

\BY
&&\bm{\hat{\zeta}}^{\text{X}\s{\tiny{M}},\text{P}\s{\tiny{M}}}=\frac{1}{\sqrt{\tau2\Gamma\s{\tiny{M}}}}\int_0^\tau dt \hat{\zeta}_{\text{X}\s{\tiny{M}},\text{P}\s{\tiny{M}}}(t),
\\
&&
 \langle \bm{\hat{\zeta}}^{\text{X}\s{\tiny{M}}} \bm{\hat{\zeta}}^{\text{X}\s{\tiny{M}}} \rangle=\bm{\hat{\zeta}}^{\text{P}\s{\tiny{M}}} \bm{\hat{\zeta}}^{\text{P}\s{\tiny{M}}} \rangle=1,\;
 \langle \bm{\hat{\zeta}}^{\text{X}\s{\tiny{M}}} \bm{\hat{\zeta}}^{\text{P}\s{\tiny{M}}} \rangle=\bm{\hat{\zeta}}^{\text{P}\s{\tiny{M}}} \bm{\hat{\zeta}}^{\text{X}\s{\tiny{M}}} \rangle=0\\
 &&\bm{\hat{\zeta}}^{\text{X}\s{\tiny{M}}}\s{f}= \frac{\bold{K}_{3} }{\sqrt{2\Gamma\s{\tiny{M}}}}\(  (\kappa \tau -1 + e^{-\kappa \tau}) \ast \hat{\zeta}_{\text{X}\s{\tiny{M}}}(\tau)\),
 \qquad\langle \bm{\hat{\zeta}}^{\text{X}\s{\tiny{M}}}\s{f} \bm{\hat{\zeta}}^{\text{X}\s{\tiny{M}}} \rangle=\frac{ \bold{K}_{3}  \bold{K}_{4}}{\sqrt{\tau}} ,
\\
&&  \bold{K}_{3} =\sqrt{\frac{6 \kappa}{3 + 2 \kappa \tau (3 + \kappa \tau ( \kappa \tau-3 ))) - 3 e^{-2 \kappa \tau} - 12 e^{-\kappa \tau} \kappa \tau}}\\
&& \bold{K}_{4}=\frac{2 (1 -  \kappa  \tau) - 2 e^{- \kappa  \tau} +  \kappa^2 \tau^2}{2 \kappa }
\EY
\BY
&&\hat {\bold{X}}\up{vac}=\bold{K}_{1} \( \(1-2 e^{-\kappa \tau}\)\ast \hat{x}\s{vac}(\tau)  \), \qquad \bold{K}_{1}= \sqrt{\frac {\kappa}{\kappa \tau-2 + 4 e^{-\kappa \tau} - 2 e^{-2 \kappa \tau} }}\\
&&\hat {\bold{P}}\up{vac}=\bold{K}_{2}\(\(1-e^{-\kappa \tau }\) \ast \hat{p}\s{vac}(\tau)    \), \qquad \bold{K}_{2}=\sqrt{ \frac{2 \kappa}{  4 e^{-\kappa \tau} + 2 \kappa \tau - 3 - e^{-2 \kappa \tau}}}\\
&&\hat {\bold{P}}\up{in}=\bold{K}_{6}\((1 - e^{-\kappa \tau} (2 \kappa \tau + 1))\ast\hat{p}\s{in}(\tau)\),  \\
&& \bold{K}_{6}=\sqrt{\frac{2 \kappa}{(2 \kappa \tau - 7) + 4 e^{-\kappa \tau} (2 \kappa \tau + 3) -
 e^{-2 \kappa \tau} (5 + 4 \kappa \tau (2 + \kappa \tau))}}
\EY
\BY
 && \hat {\bold{X}}\up{in}=  \bold{K}_{2}  \(\(1-e^{-\kappa \tau}\)\ast \hat{x}\s{in}(\tau) \)\\
 && \hat {\bold{X}}\up{in}\s{f}=  \bold{K}_{5}  \(\(1-4\kappa \tau e^{-\kappa \tau}\)\ast \hat{x}\s{in}(\tau) \),\\
 && \bold{K}_{5}=\sqrt{ \frac{\kappa}{  (\kappa \tau - 4) + 8 e^{-\kappa \tau} (1 + \kappa \tau) - 4 e^{-2 \kappa \tau} (1 + 2 \kappa \tau (1 + \kappa \tau))}}\\
 &&\bold{K}_{7}=\frac{2 e^{-\kappa \tau} (3 + 2 \kappa \tau) + (\kappa \tau - 4) - 2 e^{-2 \kappa \tau} (1 + 2 \kappa \tau)}{ \kappa},  \qquad\langle  \hat {\bold{X}}\up{in}  \hat {\bold{X}}\up{in}\s{f} \rangle=\bold{K}_{2} \bold{K}_{5} \bold{K}_{7}
\EY

The physical parameters of the systems are
the pulse duration $\tau$,
the optical decay rates of the cavity $\kappa\s{\tiny{A,M}}$,
the coupling strengths $g\s{\tiny{A,M}}$,
the gate efficiency $\eta$,
the mechanical  damping coefficient $\gamma\s{\tiny{M}}$ that shows how good the mechanics is isolated from the thermal bath
with average phonon number $n\s{th}$
(the two latter parameters are combined in the reheating rate  $\Gamma\s{\tiny{M}} = \gamma\s{\tiny{M}} (2n\s{th} + 1)$).

Figures show the dependencies of the HOM element on different parameters.
We assume that Hong-Ou-Mandel interference takes place if the HOM-element of the output state of the gate lies above its corresponding input threshold.

%%%%%%%%
\begin{figure}[ht]
\begin{center}
       \begin{minipage}[ht]{0.45\linewidth}
    \center{a\\ \includegraphics[width = \linewidth]{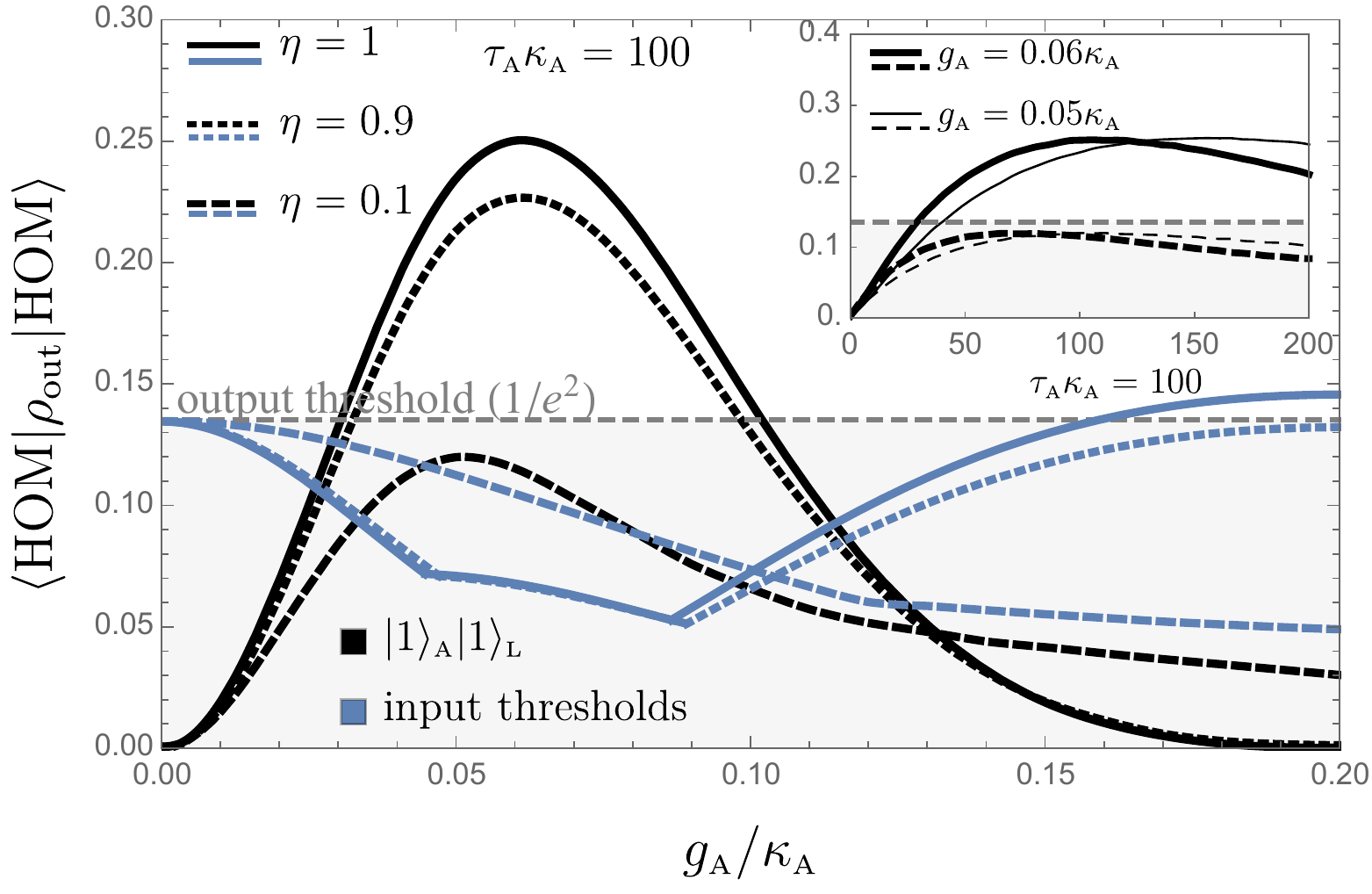}}
   \end{minipage}
      \begin{minipage}[ht]{0.45\linewidth}
    \center{b\\ \includegraphics[width = \linewidth]{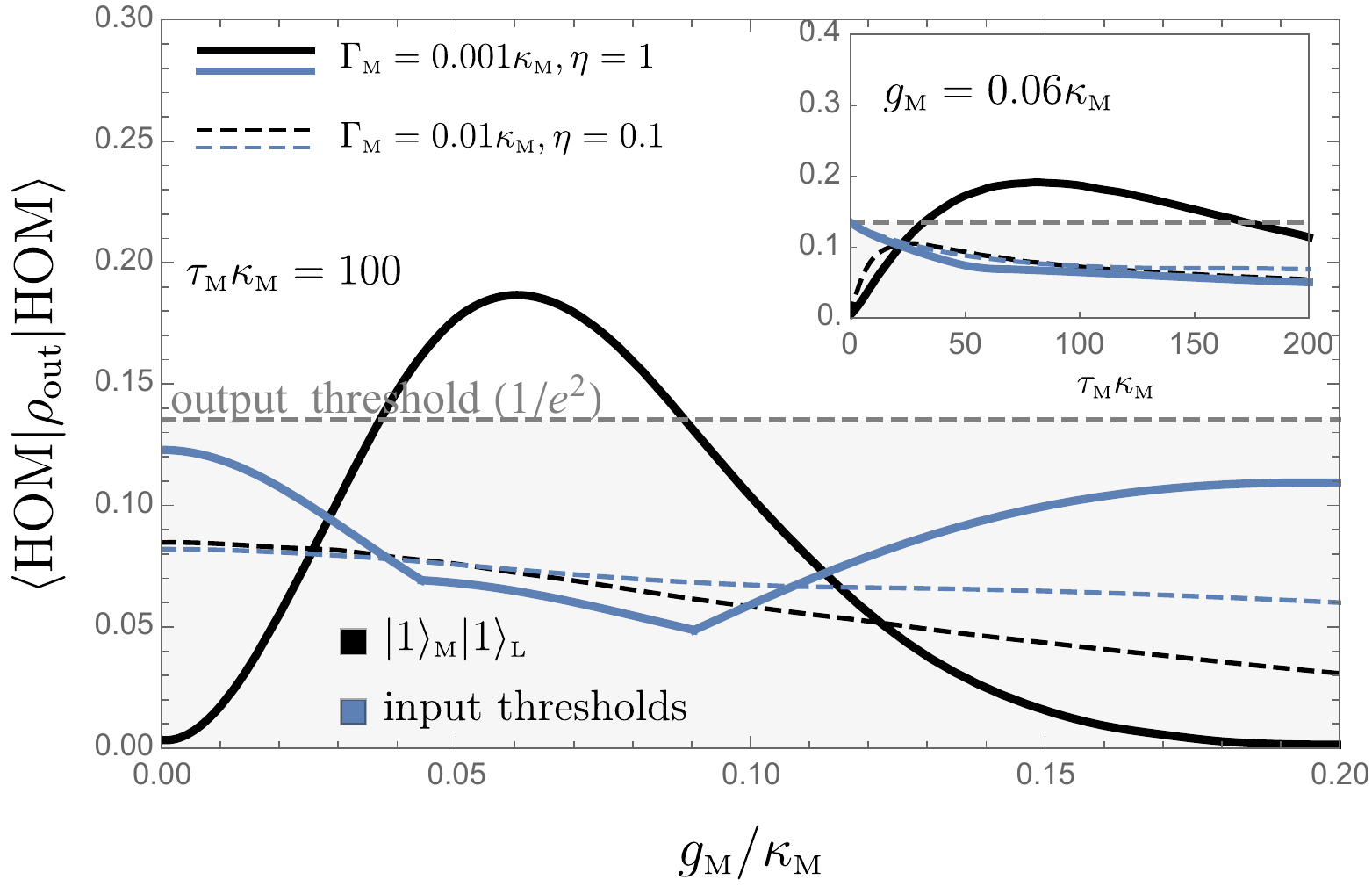}}
   \end{minipage}
\caption
{
Matrix element $\langle \text{HOM}| \rho\s{out}|\text{HOM}\rangle$ of the output state of the gate as a function of the coupling strength for the pulse length  $\tau\kappa\s{\tiny{A}}=100$:
a)~Light-atom QND gate with the independent single-photon (light) and single-polariton (atom) states at the input.
Dependence on coupling strength $g\s{\tiny{A}}$ with efficiencies $\eta=1,\;0.9,\;0.1$.
Well pronounced maximum shifts to the left with decreasing  efficiency.
The inset demonstrates  HOM-element as the function of the pulse duration $\tau\kappa\s{\tiny{A}}$.
b)~Light-mechanical QND gate with the independent single-photon (light) and single-phonon (mechanics) states at the input.
Dependence on coupling strength $g\s{\tiny{M}}$ for the rethermalization rates $\Gamma\s{\tiny{M}}=0.01\kappa\s{\tiny{M}},\;0.001\kappa\s{\tiny{M}}$ for the different efficiencies ($\eta=1,\;0.1$).
Note, $\Gamma\s{\tiny{M}}=0.01\kappa\s{\tiny{M}}$ is already too high for $\langle \text{HOM}| \rho\s{out}|\text{HOM}\rangle$ to surpass the output classical threshold, but even for $\eta=0.1$  the HOM-element lies quite close to the input classical threshold.
For both (a) and (b),
the dashed gray line is the {\em output threshold} and
the blue curves of the corresponding thickness and dashing are the {\em input  thresholds} (phase randomized).
}
%\label{Fig2sup}
\end{center}
\end{figure}
%%%%%%%%
%%%%%%%%
\begin{figure}[ht]
\begin{center}
     \begin{minipage}[ht]{0.45\linewidth}
    \center{a\\ \includegraphics[width = 0.98\linewidth]{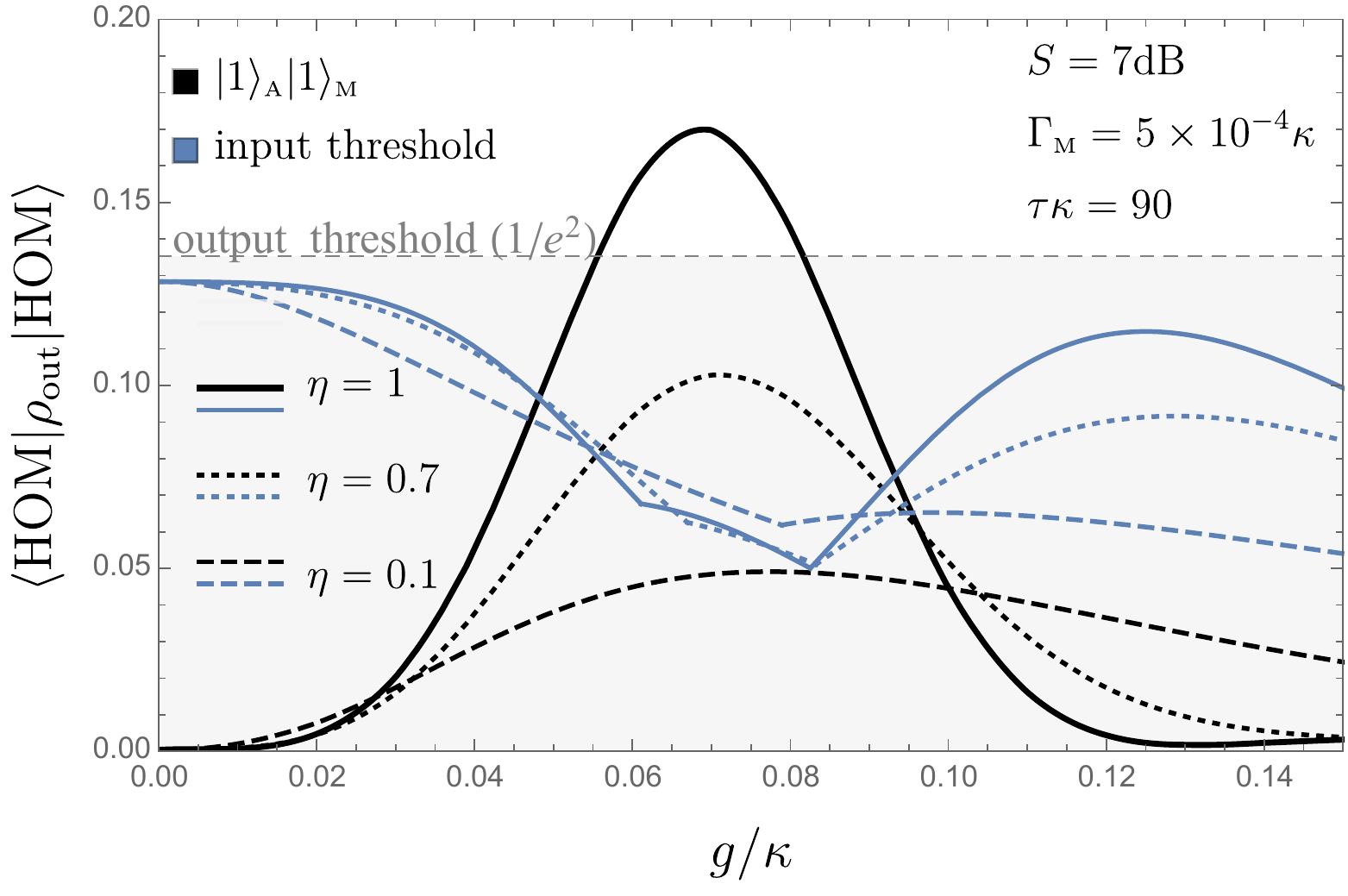}}
  \end{minipage}
    \begin{minipage}[ht]{0.45\linewidth}
    \center{b\\ \includegraphics[width = \linewidth]{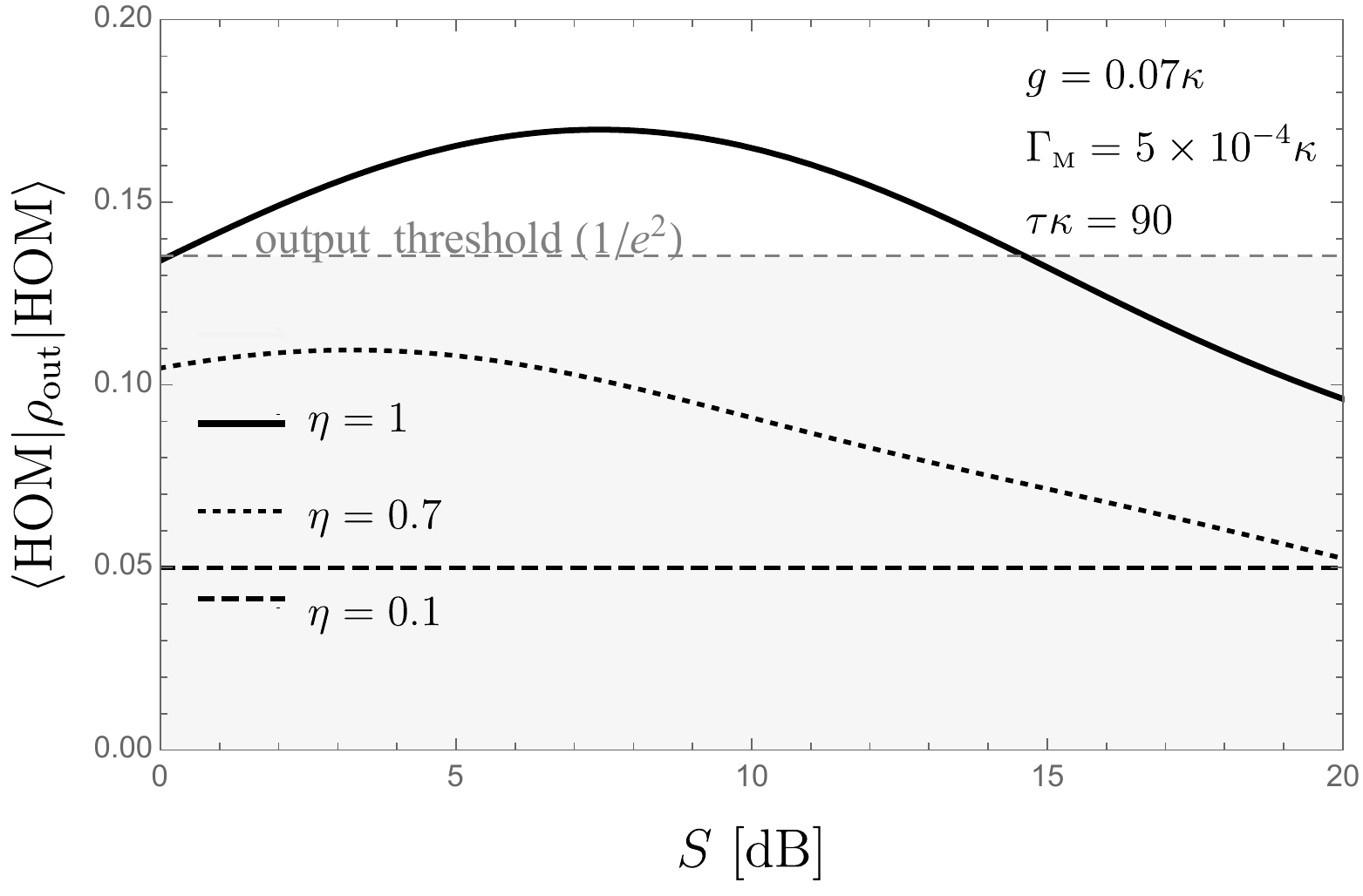}}
  \end{minipage}
\caption
{
Matrix element $\langle \text{HOM}| \rho\s{out}|\text{HOM}\rangle$ of the output state for the non-adiabatic atom-mechanical  QND gate with the independent single-boson states at the input:
 a)~dependence on coupling strength for the different efficiencies $\eta=1,\;0.8,\;0.1$.
 Blue curves of the same dashing indicate the corresponding input thresholds.
 Efficiency $\eta=0.1$ is too small to allow HOM-element to surpass the input thresholds.
  b)~dependence on the squeezing $S$ of the light pulse for the different efficiencies $\eta=1,\;0.8,\;0.1$.
  For both (a) and (b), $\kappa\s{\tiny{A}}=\kappa\s{\tiny{M}}=\kappa,\;g\s{\tiny{A}}=g\s{\tiny{M}}=g$.
}
%\label{Fig4}
\end{center}
\end{figure}
%%%%%%%%

%%%%%%%%%%%%%%%%%%%%%%%%%%%%%%%%%%%%%%%%%%%%%
%%%%%%%%%%%%%%%%%%%%%%%%%%%%%%%%%%%%%%%%%%
\section{Ideal QND gate. Nonclassicality thresholds}
%%%%%%%%%%%%%%%%%%%%%%%%%%%%%%%%%%%%%%%%%%%%%
%%%%%%%%%%%%%%%%%%%%%%%%%%%%%%%%%%%%%%%%%%%%%

Here, we first describe the ideal case of the Quantum NonDemolition (QND) transformation and compare it with the beam splitter (BS) transformation.
We then proceed to introduce the matrix elements of the output quantum state, that correspond to the bunching of excitations, and describe how to calculate them in different cases.

%%%%%%%%%%%%%%%%%%%%%%%%%%%%%%%%%%%%%%%%%%
\subsection{Comparison of a beam splitter with a QND transformation}
%%%%%%%%%%%%%%%%%%%%%%%%%%%%%%%%%%%%%%%%%%%%%

A beam splitter~(BS) transformation, characterized by the Hamiltonian $ H\s{BS}=\Theta \hbar i\(a^\dag b-b^\dag a\)$, describes an evolution of two quantum oscillators $\text{a}$ and $\text{b}$.
The only parameter of this transformation is the transmittance coefficient~$\mathsf{T}=\cos^2\Theta$.

A quantum Non-Demolition~(QND) gate, characterized by the Hamiltonian $H_\sfG=\sfG\hbar i(a+a^\dag)(b^\dag-b)/2$, describes another type of evolution of the two oscillators.
Gain~$\sfG$ is the only parameter characterizing the ideal QND gate transformation.

Hamiltonians and unitary transformations corresponding to the BS and QND transformations are the following:
\begin{align}
&H\s{BS}=\Theta \hbar i\(a^\dag b-b^\dag a\), &&U\s{BS}=\sum\limits_{n=0}^\infty \frac{\Theta^n (a^\dag b-b^\dag a)^n }{n!}),\\
&H_\mathsf{G}=\frac{\mathsf{G}}{2}\hbar i(a+a^\dag)(b^\dag-b), &&U_\mathsf{G}=\sum\limits_{n=0}^\infty \frac{g^n(a^\dag+a)^n(b^\dag-b)^n}{n!}.
\end{align}
For $U_\sfG$ we can use the Zassenhaus formula and derive it in the normal order:
\begin{align}
& U_\mathsf{G}=\sum\limits_{n=0}^\infty  g^n n! \sum\limits_{i,j,l,m=0}^\infty \frac{(a^\dag)^i a^j}{i!j!\(\frac{n-i-j}{2}\)!2^{\frac{n-i-j}{2}}} \frac{(-1)^m(b^\dag)^l b^m}{l!m!\(\frac{n-l-m}{2}\)!(-2)^{\frac{n-l-m}{2}}},
\; \frac{n-l-m}{2},\;\frac{n-i-j}{2} \in  \mathbb{N}_0,\nn
\end{align}
where $\mathbb{N}_0$ is the set of natural numbers including $0$.
 \begin{align}
& e^{t(a^\dag+a)}=e^{t a^\dag}e^{ta }e^{-\frac{t^2}{2}[a^\dag,a]}=e^{t a^\dag}e^{ ta}e^{\frac{t^2}{2}},\qquad e^{t(b^\dag-b)}=e^{t b^\dag}e^{-tb}e^{-\frac{t^2}{2}[b^\dag,-b]}=e^{t b^\dag}e^{-tb}e^{-\frac{t^2}{2}}\nn\\
& \sum\limits_{n=0}^\infty \frac{(a^\dag+a)^n }{n!}t^n=\sum\limits_{i,j,k=0}^\infty \frac{(a^\dag)^i a^j}{i!j!k!2^k}t^{i+j+2k},\qquad
\sum\limits_{n=0}^\infty \frac{(b^\dag-b)^n }{n!}t^n=\sum\limits_{l,m,k=0}^\infty \frac{(-1)^m(b^\dag)^l b^m}{l!m!k!(-2)^k}t^{l+m+2k},\\
& \frac{(a^\dag+a)^n }{n!}=\sum\limits_{i,j,k=0}^\infty \frac{\delta_{n,i+j+2k}}{i!j!k!2^k}(a^\dag)^i a^j=\sum\limits_{i,j=0}^\infty \frac{(a^\dag)^i a^j}{i!j!\(\frac{n-i-j}{2}\)!2^{\frac{n-i-j}{2}}}\\
& \frac{(b^\dag-b)^n }{n!}=\sum\limits_{l,m,k=0}^\infty \frac{(-1)^m\cdot\delta_{n,l+m+2k}}{l!m!k!(-2)^k}(b^\dag)^l b^m=\sum\limits_{l,m=0}^\infty \frac{(-1)^m(b^\dag)^l b^m}{l!m!\(\frac{n-l-m}{2}\)!(-2)^{\frac{n-l-m}{2}}}
 \end{align}

In the article we are interested in  $\langle \text{HOM} |U\s{G,BS}|\varphi\rangle\s{in}$,
where $ \langle \text{HOM} |=(1/\sqrt{2})\( \langle0|\s{a}\langle2|\s{b}- \langle2|\s{a}\langle0|\s{b}\)$.
Then we need $\langle \text{HOM} |U\s{G,BS}$.
Note that even $ \langle1|\s{a}(a^\dag)^{2}=0$ (and $ \langle2|\s{a}(a^\dag)^{3}=0$).
Then, using $ \langle0|\s{a}(a)^k =\sqrt{k!} \langle k|\s{a},\;\langle1|\s{a}(a)^k =\sqrt{(k+1)!} \langle k+1|\s{a},\;\langle2|\s{a}(a)^k =\sqrt{(k+2)!/2} \langle k+2|\s{a}$, we can easily obtain:
\\
\\
I. For the QND transformation:
\begin{align}
	 & \langle \text{HOM} | \;U_\sfG =
	\frac{1}{\sqrt{2}}\( \langle0|\s{a}\langle2|\s{b}- \langle2|\s{a}\langle0|\s{b}\) \;U_\sfG =                                                                                                                               \\
	 & =    \sum\limits_{n,m,j=0}^\infty  \frac{(-1)^mg^n n!}{\sqrt{j!m!}}
	(
	\frac{\sqrt{(m+2)(m+1)}\langle m+2|\s{b}\langle j |\s{a}+\sqrt{(j+2)(j+1)}\langle m|\s{b}\langle j+2 |\s{a}}{2\cdot \(\frac{n-j}{2}\)! \cdot 2^{\frac{n-j}{2} } \cdot  \(\frac{n-m}{2}\)! \cdot(-2)^{\frac{n-m}{2} } }-\nn \\
	 & -\frac{\sqrt{j+1} \langle m|\s{b}\langle j+1 |\s{a} }{\(\frac{n-1-j}{2}\)! \cdot 2^{\frac{n-1-j}{2} } \cdot  \(\frac{n-m}{2}\)! \cdot(-2)^{\frac{n-m}{2} } }
	-\frac{\langle m|\s{b}\langle j |\s{a} }{2\cdot \(\frac{n-2-j}{2}\)! \cdot 2^{\frac{n-2-j}{2} } \cdot  \(\frac{n-m}{2}\)! \cdot(-2)^{\frac{n-m}{2} } }+\nn                                                                 \\
	 & +\frac{\sqrt{m+1} \langle m+1|\s{b}\langle j |\s{a}}{\(\frac{n-j}{2}\)! \cdot 2^{\frac{n-j}{2} } \cdot  \(\frac{n-1-m}{2}\)! \cdot(-2)^{\frac{n-1-m}{2} } }
	+\frac{\langle m|\s{b}\langle j |\s{a}}{2\cdot\(\frac{n-j}{2}\)! \cdot 2^{\frac{n-j}{2} } \cdot  \(\frac{n-2-m}{2}\)! \cdot(-2)^{\frac{n-2-m}{2} }}) \nn.
\end{align}
Let us use $|\varphi\rangle\s{in}= |1\rangle\s{b}|1\rangle\s{a}$ and obtain matrix element $| \langle  \text{HOM} | \;U_\sfG  \; |1\rangle\s{b}|1\rangle\s{a}|^2$:
\begin{align}
	 & | \langle \text{HOM} | \;U_\sfG  \; |1\rangle\s{b}|1\rangle\s{a}|^2=|\sum\limits_{k=0}^\infty
	\frac{ (-1)\up{k}g^{2k+1} (2k+1)!}{(k!)^2 2^{2k-1}}+\sum\limits_{k=1}^\infty
	\frac{ (-1)\up{k}g^{2k+1} (2k+1)!}{k!(k-1)! 2^{2k-1}}  |^2=\frac{16  \mathsf{G}^2 (-8 +  \mathsf{G}^2)^2}{(4 +  \mathsf{G}^2)^5},\nn\\
	&( g=\mathsf{G}/2).
\end{align}
II. For the BS transformation:
\begin{align}
	 & \langle \text{HOM} |\;U\s{BS} =\frac{1}{\sqrt{2}}\(\langle2|\s{b}\langle0|\s{a}-\langle0|\s{b}\langle2|\s{a}\)U=\nn                                                                                                                                                                 \\
	 & =\frac{1}{\sqrt{2}}\sum\limits_{n=0}^\infty \(\frac{(-1)^{n+1}\Theta^{2n+1}(2\sqrt{2})^{n+1}(\sqrt{2})\up{n}\langle1|\s{a}\langle1|\s{b}}{(2n+1)!}+\frac{(-1)^n\Theta^{2n}(2\sqrt{2})\up{n}(\sqrt{2})\up{n}\(\langle2|\s{b}\langle0|\s{a}-\langle0|\s{b}\langle2|\s{a}\)}{(2n)!}\).
\end{align}
Let us check the result by obtaining  $| \langle \text{HOM}  |\;U\s{BS}  \; |1\rangle\s{b}|1\rangle\s{a}|^2$:
\begin{align}
	 & | \langle \text{HOM} |\;U\s{BS}  \; |1\rangle\s{b}|1\rangle\s{a}|^2
	=|\frac{1}{\sqrt{2}}\sum\limits_{n=0}^\infty \frac{(-1)^{n+1}\Theta^{2n+1}(2\sqrt{2})^{n+1}(\sqrt{2})\up{n}}{(2n+1)!}|^2=4\mathsf{T}( 1- \mathsf{T} ), \qquad (\cos^2 \Theta=\mathsf{T}).
\end{align}
If we restrict ourselves by the limited input subspace, including only  $|1\rangle\s{a}|1\rangle\s{b},|0\rangle\s{a}|1\rangle\s{b},|1\rangle\s{a}|0\rangle\s{b}$ and $|0\rangle\s{a}|0\rangle\s{b}$, the formulas can be simplified:
\begin{align}
&  \langle \text{HOM} |\;U\s{BS} =
\sum\limits_{n=0}^\infty \frac{(-1)^{n+1}\(2\Theta\)^{2n+1}}{(2n+1)!}\langle1|_a\langle1|_b=-\sin\(2\Theta\)\;\langle1|\s{a}\langle1|\s{b},\\
&  \langle \text{HOM} | \;U_\sfG
= \sum\limits_{k=0}^\infty \frac{ (-1)^k (\mathsf{G}/2)^{2 k + 1} (2 k + 1)! (1 + k)}{ (k!)^2 2^{2 k - 1}}\;\langle1|_a\langle1|_b+
\frac{(-1)^k (\mathsf{G}/2)^{2 k + 2} (2 k + 2)!}{ (k!) (k + 1)! 2^{2 k + 1}}\;\langle0|\s{a}\langle0|\s{b}=\nn\\
&= \frac{4 \mathsf{G} (8 - \mathsf{G}^2)}{(4 + \mathsf{G}^2)^{5/2}}\;\langle1|\s{a}\langle1|\s{b}+\frac{2 \mathsf{G}^2}{(4 + \mathsf{G}^2)^{3/2}}\;\langle0|\s{a}\langle0|\s{b}.
\end{align}
%
%%%%%%%%%%%%%%
\begin{table} [htbp]
\centering
\parbox{15cm}{\caption{Matrix elements of the output state corresponding the different cases of the input for the BS and QND gate transformations}\label{table2}}
\begin{center}
\begin{tabularx}{0.5\textwidth}{|*{5}{Y|}}
\hline
\multirow{2}{*}{input}
  &\multicolumn{2}{c|}{ $\langle \text{HOM}| \rho\s{out} |\text{HOM}\rangle$}\\
\cline{2-3}
             & BS      & QND gate \\
\hline
$|1\rangle\s{a}|1\rangle\s{b}$ & $4 \mathsf{T}( 1-\mathsf{T})$ & $\frac{16 \mathsf{G}^2 ( \mathsf{G}^2-8)^2}{(4 + \mathsf{G}^2)^5}$ \\
 % & & \\
    \hline
  %     & &\\
   $|1\rangle\s{a}|0\rangle\s{b}$ & $0$ & $0$\\
 %     & &\\
    \hline
 %      & &\\
  $|0\rangle\s{a}|1\rangle\s{b}$ & $0$ & $0$\\
 %    & &\\
    \hline
 %     & &\\
   $|0\rangle\s{a}|0\rangle\s{b}$ & $0$ & $ \frac{4 \mathsf{G}^4}{(4 + \mathsf{G}^2)^3}$\\
 %    & &\\
    \hline
\end{tabularx}
\end{center}
\end{table}
%%%%%%%%%%%%%%
In the case of the general input BS transformation provides non-zero matrix elements exclusively for $|\varphi\rangle\s{in}=|1\rangle\s{a}|1\rangle\s{b},|0\rangle\s{a}|2\rangle\s{b},|2\rangle\s{a}|0\rangle\s{b}$.
However,  the gate  transformation generates and annihilates excitations in pairs.
That means that, for instance, if at the gate input there are states with an even number of bosons (as  $  |0\rangle\s{a}|0\rangle\s{b},|1\rangle\s{a}|1\rangle\s{b},|1\rangle\s{a}|3\rangle\s{b} $ etc.),
then $\langle \text{HOM}| \rho\s{out}  |\text{HOM}\rangle$ will be nonzero (some functions of the gain).
If at the input of the gate there is a state with an odd number of excitations (as $|0\rangle\s{a}|1\rangle\s{b},|1\rangle\s{a}|0\rangle\s{b},|1\rangle\s{a}|2\rangle\s{b},|2\rangle\s{a}|1\rangle\s{b} $ etc.),
then the matrix elements of the output state will be zero.

There is a significant difference between these two transformations.
The BS transformation is passive, it neither creates nor annihilates excitations in a system of two harmonic oscillators.
If initially there are exactly one excitation in each of the oscillators, at the output of a BS they can appear bunched in a single mode via the Hong-Ou-Mandel (HOM) effect.
It is not sufficient to have just one excitation in a single mode to observe bunching.
Unlike BS, the QND transformation is active, which means it can possibly change the total number of excitations in the system (the energy of the system).
The creation of quanta by QND interaction can produce effects that resemble the HOM interference.
Although such case can be confused with the HOM interference effect, it is still possible to analyze whether the QND interaction is capable of generating the non-classical two-quanta superpositions going beyond any classical states serving as input to the QND interaction.
However, such analysis requires a general approach to the HOM interference beyond the simple case with the passive BS interaction.

In this generalized description, the matrix element of the output state $\rho\s{out}$, describing the {\em success} probability of detection of two-photon HOM-entangled states (the HOM element),
can be introduced as $|\langle \text{HOM} |U|\varphi\rangle\s{in}|^2$, where $|\varphi\rangle\s{in}$ is an initial state, and $U$ is a unitary transformation.
Here, the HOM-state is determined as $\langle \text{HOM} | =\(\langle2|\s{b}\langle0|\s{a}-\langle0|\s{b}\langle2|\s{a}\)/\sqrt{2}$.
It is well known that a BS provides an ideal photon bunching (the HOM effect).
This means that the success probability $|\langle \text{HOM} |U\s{BS}|\varphi\rangle\s{in}|^2$ at the output of the BS equals one.
This effect occurs when two identical quanta enter a balanced beam splitter ($ \mathsf{T}=0.5$), one in each input port (the input state $|\varphi\rangle\s{in}=|1\rangle\s{a}|1\rangle\s{b}$).

In order to compare the two transformations in the context of the HOM effect, let us look at the matrix elements of the output state of each of the transformations.
For simplicity, first let us restrict the subspace of the input and assume that
 $|\varphi\rangle\s{in}$ belongs to the space of coherent superpositions of vacuum and one excitation of each mode, that is an arbitrary pure superposition of
 $|1\rangle\s{a}|1\rangle\s{b},|0\rangle\s{a}|1\rangle\s{b},|1\rangle\s{a}|0\rangle\s{b}$ and $|0\rangle\s{a}|0\rangle\s{b}$.
Then, to obtain the desired matrix element we need:
\begin{align}
\langle \text{HOM} |\;U\s{BS} =-\sin\(2\Theta\)\;\langle1|\s{a}\langle1|\s{b},
\; \langle \text{HOM} | \;U_\sfG = \frac{4 \sfG (8 - \sfG^2)}{(4 + \sfG^2)^{5/2}}\;\langle1|\s{a}\langle1|\s{b}+\frac{2 \sfG^2}{(4 + \sfG^2)^{3/2}}\;\langle0|\s{a}\langle0|\s{b} .%\label{Eq2}.
\end{align}
It clearly shows that the HOM matrix element provided by the inputs $|0\rangle\s{a}|1\rangle\s{b}$ and $|1\rangle\s{a}|0\rangle\s{b}$ is equal to zero for both BS and QND transformations.
Matrix element provided by $|0\rangle\s{a}|0\rangle\s{b}$ input is equal to zero in the case of a beamsplitter.
However, for the QND gate this element is a function of the gain $\sfG$ and equals zero only in the trivial case with $ \sfG=0$.
That is, by varying the gain of the QND gate, it is impossible to make the contribution of the input vacua $|0\rangle\s{a}|0\rangle\s{b}$ vanish, in order to render these two transformations fully analogous.

For the case of a QND gate with $|1\rangle\s{a}|1\rangle\s{b}$ at the input, one can observe
that for a certain region of the parameter $\sfG$, the probability of bunching of both excitations in one subsystem is higher than the probability of equal redistribution of the excitations between the subsystems.
Visually, it is characterized by the presence of the maximum of $\langle \text{HOM}| \rho\s{out}|\text{HOM}\rangle$ (see Fig.~\ref{FigSuperposition}(b) )approximately equal to $0.26$ for $\sfG=\sqrt{11 - \sqrt{105}}
\approx 0.87$ (as compared to $1$ for the BS with $\Theta=\pi/4$).
However, we should keep in mind that this correspondence to the case of a BS is not complete due to the non-zero contribution from the vacuum input for the gate case that does not exist in the case of a BS.
%%%%%%%%%%%%%%%%%%%%%%%%%%%%%%%%%%
\subsection{BS and QND transformations using quadratures.  }\L{app01}
%%%%%%%%%%%%%%%%%%%%%%%%%%%%%%%%%%

Both Quantum Non-Demolition (QND) gate and beam splitter BS, transform quadratures of two quantum oscillators $\text{a}$ and $\text{b}$, correlating the quadratures of the two oscillators.
The initial quadratures $\bold{r}\up{in}=(\hat{X}\s{a}(0),\hat{P}\s{a}(0),\hat{X}\s{b}(0),\hat{P}\s{b}(0))^T$ after transformation should relate to the final quadratures $\bold{r}\up{out}\equiv\bold{r}=(x\s{a},p\s{a},x\s{b},p\s{b})^T$ as
\begin{align}
& \bold{r}\up{out}=T\s{G,BS}\bold{r}\up{in}+N,\;
\text{where}\\
&
T\s{G}=\left(
    \begin{array}{cccc}
        \coefficient{T}\s{a} &  0  & \coefficient{G}\s{a} & 0 \\
        0  &   \coefficient{T}\s{a}  & 0 & 0 \\
        0  &  0  & \coefficient{T}\s{b} & 0 \\
        0 &  \coefficient{G}\s{b}  & 0 &  \coefficient{T}\s{b}
    \end{array}
    \right),\;
    T\s{BS}=\left(
    \begin{array}{cccc}
        \sqrt{\mathsf{T}}  &   0      &\sqrt{1-\mathsf{T}} &0 \\
        0       &  \sqrt{\mathsf{T}} &0&\sqrt{1-\mathsf{T}}  \\
        -\sqrt{1-\mathsf{T}}        & 0 &  \sqrt{\mathsf{T}}&     0 \\
        0       & -\sqrt{1-\mathsf{T}}  &  0       &      \sqrt{\mathsf{T}} \\
    \end{array}
\right).
\label{transformation}
\end{align}
where $\mathsf{T}$  is the transmittance coefficient (BS transformation);
$\coefficient{G}\s{a,b}$ are the controllable gains of the built QND gate,
$\coefficient{T}\s{a,b}$ are the transfer factors  (QND transformation).
$N\equiv(\coefficient{N}_{\text{X}\s{a}}, \coefficient{N}_{\text{P}\s{a}} , \coefficient{N}_{\text{X}\s{b}} , \coefficient{N}_{\text{P}\s{b}})^T$ describe the excess noises, thus
for the ideal transformation they should be negligible ($\coefficient{N}_{\text{X}\s{a},\text{P}\s{a},\text{X}\s{b},\text{P}\s{b}}\rightarrow 0$) (and for QND transfer factors equal one ($\coefficient{T}\s{a,b}=1$),
while the gains are of the same magnitude but opposite sign ($\mathsf{G}=\coefficient{G}\s{a}=-\coefficient{G}\s{b}$).
Thus, gain $\mathsf{G}$ is the only parameter characterizing the ideal QND gate transformation, while
the BS transformation also is characterized by a single parameter $\mathsf{T}$.

%%%%%%%%%%%%%%%%%%%%%%%%%%%%%%%%%%%%%%%%%%%%%%%%%%%%%%%%%%%%%%%
\subsection{Wigner function of the state}
%%%%%%%%%%%%%%%%%%%%%%%%%%%%%%%%%%%%%%%%%%%%%%%%%%%%%%%%%%%%%%%

The Wigner functions (WF) can be used to calculate the HOM element.
Let us demonstrate how to obtain the WF corresponding to an arbitrary  operator.
First, let us remind that the wave-function of the $n$-th excited level (Fock state $\ket n$) can be derived using Hermite polynomials as:
\begin{align}
\psi_n(x)=\langle x|n\rangle=\frac{1}{\sqrt{2^{n+1/2}\pi^{1/2}n!}}H_n\(\frac{x}{\sqrt{2}}\)\exp\(-\frac{x^2}{4}\),
\end{align}
\begin{align}
&H_0(z)=1, \\
&H_1(z)=2z, \\
&H_2(z)=4z^2-2=2(2z^2-1), ...
\end{align}
The WF of the operator $|\varphi\rangle\langle\psi|$ in the $s$-dimensional space is as follows:
\begin{align}
W_{|\varphi\rangle\langle\psi|}(\bold{x},\bold{p})=\frac{1}{(4\pi)^s}\int\limits_{-\infty}^{\infty} d\bold{y} \; e^{-\frac{i\bold{p}\bold{y}}{2}}\langle \bold{x}+\frac{\bold{y}}{2}|\varphi\rangle\langle\psi|\bold{x}-\frac{\bold{y}}{2}\rangle,
\;\;\text{where}\; \bold{q}=(q_1,q_2,...q_s)\;\; \text{for}\;\;\mathbf{q=x,y,p}.
\end{align}
Here,
\begin{align}
&\langle x+\frac{y}{2}|2\rangle=\psi_{|2\rangle}(x+\frac{y}{2})=
\frac{1}{\sqrt{2}\sqrt[4]{2\pi}}\(\(x+\frac{y}{2}\)^2-1\)\exp\(-\frac{(x+\frac{y}{2})^2}{4}\)\\
&\langle x+\frac{y}{2}|1\rangle=\psi_{|1\rangle}(x+\frac{y}{2})=
\frac{1}{\sqrt[4]{2\pi}}\(x+\frac{y}{2}\)\exp\(-\frac{(x+\frac{y}{2})^2}{4}\)\\
&\langle x+\frac{y}{2}|0\rangle =\psi_{|0\rangle}(x+\frac{y}{2})=
\frac{1}{\sqrt[4]{2\pi}}\exp\(-\frac{(x+\frac{y}{2})^2}{4}\)\\
&\langle0|x-\frac{y}{2}\rangle =\psi^\ast_{|0\rangle}(x-\frac{y}{2})=
\frac{1}{\sqrt[4]{2\pi}}\exp\(-\frac{(x-\frac{y}{2})^2}{4}\)\qquad \text{etc}.
\end{align}

In the article, we work in the $4$-dimensional space.
Thus, for the operator
$A=|\text{HOM}\rangle \langle \text{HOM}|$,
associated with the state $|\text{HOM}\rangle=\(|0_2\;2_1\rangle - |2_2\;0_1\rangle\)/\sqrt{2}$, the WF can be calculated as:
\begin{align}
&W_{\text{HOM}}(x_1,p_1,x_2,p_2)=\frac{1}{(4\pi)^2}\iint dy_1dy_2 \; e^{-\frac{i(p_1y_1+p_2y_2)}{2}}f\s{HOM}(x_1,y_1,x_2,y_2)\nn\\
&=\frac{1}{16\pi^2} \exp \( -\frac{p_1^2 + p_2^2 + x_1^2 +  x_2^2}{2}\)
     ((p_1 - p_2)^2 + (x_1 - x_2)^2-2 ) ((p_1 +
     p_2)^2 + (x_1 + x_2)^2-2 ).
\end{align}
where
\begin{align}
& f_{\text{HOM}}(x_1,y_1,x_2,y_2)=
\langle x_1+\frac{y_1}{2} | \langle x_2+\frac{y_2}{2} |A \;|x_2-\frac{y_2}{2}  \rangle | x_1-\frac{y_1}{2} \rangle=\nn\\
&=\frac{1}{2}\(\langle x_1+\frac{y_1}{2} ||2_1\rangle \langle x_2+\frac{y_2}{2} ||0_2\rangle - \langle x_2+\frac{y_2}{2} ||2_2\rangle\;\langle x_1+\frac{y_1}{2} || 0_1\rangle  \)\otimes\nn\\
&\otimes
\( \langle0_1| | x_1-\frac{y_1}{2} \rangle\langle2_2||x_2-\frac{y_2}{2}  \rangle - \langle2_1||  x_1-\frac{y_1}{2} \rangle\langle0_2| |x_2-\frac{y_2}{2}  \rangle \)=\nn\\
&=\frac{1}{8\pi} \exp\(-\frac{(x^2_2+\frac{y^2_2}{4})+(x^2_1+\frac{y^2_1}{4})}{2}\)
\( \(x_1+\frac{y_1}{2}\)^2-\(x_2+\frac{y_2}{2}\)^2 \)
\( \(x_1-\frac{y_1}{2}\)^2-\(x_2-\frac{y_2}{2}\)^2 \).
\end{align}

Analogically, for the operator $A'=|1_21_1\rangle\langle1_11_2|$:
\begin{align}
& f_{|11\rangle\langle11|}(x_1,y_1,x_2,y_2)=
\frac{1}{2\pi}\(x_1^2-\(\frac{y_1}{2}\)^2\)\(x_2^2-\(\frac{y_2}{2}\)^2\) \exp\(-\frac{x_1^2+(\frac{y_1}{2})^2+x_2^2+(\frac{y_2}{2})^2}{2}\),\\
&W_{|11\rangle\langle11|}(x_1,p_1,x_2,p_2)=  \frac{1}{4\pi^2} \exp\(- \frac{p_1^2 + p_2^2 + x_1^2 + x_2^2}{2}\)  (-1 + p_1^2 + x_1^2) (-1 + p_2^2 + x_2^2)).
\end{align}
%%%%%%%%%%%%%%%%%%%%%%%%%%%%%%%%%%%%%%%%%%%%%%%%%%%%%%%%%%%%%%%%%%%%%

%%%%%%%%%%%%%%%%%%%%%%%%%%%%%%%%%%%%%%%%%%%%%%%%%%%%%%%%%%%%%%%
\subsection{Calculation of the HOM matrix element}\L{app2}
%%%%%%%%%%%%%%%%%%%%%%%%%%%%%%%%%%%%%%%%%%%%%%%%%%%%%%%%%%%%%%%

In order to calculate the matrix elements for an ideal (without additional noise) transformation,
it is enough to know the form of the unitary transformation $U$.
Thus the matrix element of the output can be calculated as:
\begin{equation}
  \sfM\s{HOM}=\langle \text{HOM}| \rho\s{out} |\text{HOM}\rangle=|\langle \text{HOM} |U|\varphi\rangle\s{in}|^2.%=|\langle \text{HOM} |U|n\rangle\s{a}|m\rangle\s{b}|^2,
\end{equation}

To evaluate the robustness of the QND gate against photon loss, we examine an incoherent mixture of vacuum and single-photon states at each input port of the gate
\begin{equation}
  \rho\s{in}^{(p)} = (p\s{a}\;|1\rangle\langle1| + (1-p\s{a})\;|0\rangle\langle0| )\s{a}\cdot(p\s{b}\;|1\rangle\langle1| + (1-p\s{b})\;|0\rangle\langle0| )\s{b}, %\label{ME_superpositioninput_Eq1}
\end{equation}
where the parameter $p\s{a,b}$ characterizes how much vacuum has been admixed to the single-photon state at the input ports, and calculate matrix elements for the output state of the gate.
Using Eq.~(\ref{Eq2}), we can obtain the HOM matrix element of the output state of the gate:
\begin{equation}
  \langle \text{HOM}| \rho\s{out}^{(p)}|\text{HOM}\rangle= p\s{a} p\s{b}\frac{16 \sfG^2 ( \sfG^2-8)^2}{(4 + \sfG^2)^5}+(1-p\s{a})(1-p\s{b})\frac{4 \sfG^4}{(4 + \sfG^2)^3}. %\label{MHOM_superpositioninput}
\end{equation}

This matrix element is symmetrical with respect to $p\s{a}$ and $p\s{b}$.
Surprisingly, the independent coherent superpositions $(\sqrt{p\s{a}}\;|1\rangle + \sqrt{1-p\s{a}}\;|0\rangle)\s{a}\cdot(\sqrt{p\s{b}}\;|1\rangle + \sqrt{1-p\s{b}}\;|0\rangle)\s{b}$ at the input
give rise to the same matrix element Eq.~(\ref{MHOM_superpositioninput}) as the mixture Eq.~(\ref{ME_superpositioninput_Eq1}).

This approach is good to calculate the ME when the input state is a $|n\rangle\s{a}|m\rangle\s{b}$-boson state or some combination of them.
Sometimes it is more convenient to take a different approach described below.
We can use the Wigner function (WF) and the matrix elements $\langle\psi| \rho\s{out} |\varphi\rangle$ of the output state can be defined as:
\begin{equation}
  \sfM_{ |\varphi\rangle\langle\psi| }=\langle\psi| \rho\s{out} |\varphi\rangle=(4\pi)^2 \iiiint d\bold{r} \; W_{ |\varphi\rangle\langle\psi| }(\bold{r})\cdot W\s{out}(\bold{r}),
  \label{MatrixEl}
\end{equation}
where $W\s{out}(\bold{r})$ is the WF of the output state, $\bold{r}=\bold{r}\up{out}=(x\s{a},p\s{a},x\s{b},p\s{b})^T$.
$W_{ |\varphi\rangle\langle\psi| }(\bold{r})$ corresponds to the projector
$ |\varphi\rangle\langle\psi|$, e.g. to calculate $\sfM\s{HOM}$ we need the WF of the HOM operator:
\begin{equation}
  W\s{HOM}( \bold{r})=\frac{1}{16\pi^2}
  \exp \( -\frac{ p\s{a}^2+p\s{b}^2 + x\s{a}^2+x\s{b}^2 }{2} \)
  ( ( p\s{a}+p\s{b})^2 + (x\s{a}+x\s{b})^2-2 ) ((p\s{a}-p\s{b})^2 + (x\s{a}-x\s{b})^2-2 ).%\label{HOM}
\end{equation}

The WF approach is convenient when we know the exact WF of the input state and how the transformation changes the quadratures of the oscillators.
%i.e. for the ideal transformation $\bold{r}\up{out}=T_ {\mathsf{G},\text{BS}}\bold{r}\up{in}$ (no noise).
This approach is also suitable for the case of $|n\rangle\s{a}|m\rangle\s{b}$ input.
We demonstrate it for the gate transformation with $|1\rangle\s{a}|1\rangle\s{b}$ input.
Let us assume that both quantum oscillators were initially in single-boson states.
Since they are independent, the exact Wigner function of the initial state of the system can be obtained by multiplying of the two single-photon state Wigner functions:
\begin{align}
  & W\s{in}(\bold{r})=W\s{a}|_{|1\rangle}(\bold{r}\s{a})W\s{b}|_{|1\rangle}(\bold{r}\s{b}), \;\; \text{where} \;\; W\s{a,b }|_{|1\rangle}(\bold{r}\s{a,b })=\frac{1}{2\pi} \(\bold{r}\s{a,b }^2-1\) \exp\(-\frac{\bold{r}\s{a,b }^2}{2}\),\\
  & \bold{r}\s{a,b }=(x\s{a,b },p\s{a,b })^T.
\end{align}
The WF of the output state of the system for the ideal gate:
\begin{align}
  & W\s{out}(\bold{r})
   =\frac{1}{4\pi^2}( ( p\s{b}+\mathsf{G} p\s{a})^2 + x\s{b}^2-1 ) (p\s{a}^2 + (x\s{a} - \mathsf{G} x\s{b})^2-1 )\times \nn\\
  & \times\exp \( -\frac{ (p\s{b} +\mathsf{G} p\s{a})^2 + x\s{b}^2 +p\s{a}^2 + (x\s{a} - \mathsf{G} x\s{b})^2 }{2} \).
\end{align}
Using Eq.(\ref{MatrixEl}), we will obtain the same result as in Eq.( \ref{MHOM_superpositioninput}) for $p\s{a}=p\s{b}=1$.
Both described approaches are identical and, being applied to any ideal transformations, they give the same results.

For the non-ideal case we need to take to account the noises,
which is possible using the language of covariance matrices.
A Gaussian quantum state (such as vacuum, coherent, squeezed or thermal states) can be fully described by the first and second statistical moments, that is, a vector of means and a covariance matrix.
The Wigner function of an arbitrary Gaussian state in $\mathbb {R}_s$-dimension can be represented using the covariance matrices $V$ as:
\begin{align}
& W(\bold{r})=\text{W}(V,\bold{r})=\frac{1}{2\pi \sqrt{\det V}} \exp\(-\frac{1}{2}(\bold{r}-\bar{\bold{r}})^T V^{-1} (\bold{r}-\bar{\bold{r}})\),\label{WFformula}\\
& \bold{r}=\(x_1,p_1,...,x_s,p_s\),\; \bar{\bold{r}}=\(\langle X_1 \rangle, \langle P_1 \rangle,...,\langle X_s \rangle, \langle P_s \rangle \).\nn
\end{align}

After a nonideal transformation $T$ of the vector $\bold{r}$, the evolution of the covariance matrices and the WF are as follows:
\begin{align}
& W\s{in,out}(\bold{r})=\text{W}(\bold{r},V\s{in,out}),
\qquad\text{where}\;
V\up{out}=T_\sfG V\up{in}T_\sfG^T+V\s{N},\\
&  [V_N]_{ij}=\frac{1}{2}\langle N_i N_j + N_j N_i\rangle \quad(i,j=1,...,s).
\end{align}
Thus, we can use Eq.(\ref{MatrixEl}) and obtain the HOM matrix element of the output state for the nonideal transformation.

However, neither single-photon nor HOM states are not Gaussian.
Nevertheless, the single-photon state can be approximated by the superposition of the thermal $W\s{th}(\bold{r})$ and vacuum $W_{|0\rangle}(\bold{r})$ states as:
\begin{align}
W|_{|1\rangle}(\bold{r})\approx \frac{1}{n}\((n+1)W\s{th}(\bold{r})- W_{|0\rangle}(\bold{r})\), \qquad n\ll1. \label{approximation}
\end{align}
Here $W\s{th}$ is the Wigner function of a thermal state with mean occupation $n$, that is a Gaussian state with zero means and covariance matrix equal to $V\s{th} = ( 2n + 1 ) \mathbb{I}_{2\times 2}.$

Thermal and vacuum states are Gaussian, and, in turn, they can be represented using Eq.~(\ref{WFformula}) with $\bar{\bold{r}}=0$ and $s=4$.
Thus, to describe the initial state at the input of the gate between a-oscillator and b-oscillator we can use the following approximated WF:
\begin{align}
&W\s{in}(\bold{r})=W\s{a}|_{|1\rangle}(\bold{r}\s{a})W\s{b}|_{|1\rangle}(\bold{r}\s{b})\approx \frac{1}{n}\((n+1)W\s{th}(\bold{r}\s{a})- W_{|0\rangle}(\bold{r}\s{a})\) \frac{1}{n}\((n+1)W\s{th}(\bold{r}\s{b})- W_{|0\rangle}(\bold{r}\s{b})\)=\nn\\
&=\frac{1}{n^2}\((n+1)^2\text{W}(V\up{in}_{\text{th}\s{a},\text{th}\s{b}}, \bold{r})- (n+1)\text{W}(V\up{in}_{\text{vac}\s{a},\text{th}\s{b}},\bold{r})-(n+1)\text{W}(V\up{in}_{\text{th}\s{a},\text{vac}\s{b}},\bold{r})+\text{W}(V\up{in}_{\text{vac}\s{a},\text{vac}\s{b}},\bold{r})\)=\nn\\
&=\frac{1}{n^2}\sum\limits_{k,l=0,1}(-(n+1))^{k+l}\text{W}(V\up{in}\s{k,l}, \bold{r}).
\end{align}
Here, the corresponding covariance matrices are
\begin{align}
& V\up{in}_{1,1}=(2n+1) \mathbb{I}_4, \;
 V\up{in}_{0,0}=\mathbb{I}_4, \;
  V\up{in}_{1,0}=
\begin{pmatrix}
  2n+1 & 0 & 0 & 0 \\
  0 & 2n+1 & 0 & 0 \\
  0 & 0 & 1 & 0 \\
  0 & 0 & 0 & 1
  \end{pmatrix}, \;
  V\up{in}_{0,1}=
\left(
  \begin{array}{cccc}
  1 & 0 & 0 & 0 \\
  0 & 1 & 0 & 0 \\
  0 & 0 & 2n+1 & 0 \\
  0 & 0 & 0 & 2n+1
  \end{array}
  \right),\nn\\
& V\up{out}_{k,l }=T_\mathsf{G}V\up{in}_{k,l }T_\mathsf{G}^T+V_\text{N}, \qquad [V_\text{N}]_{ij}=\frac{1}{2}\langle N_i N_j + N_j N_i\rangle,
\end{align}
where $\mathbb{I}_4$ is the identity matrix of size $4$, $k=0,1;\;l=0,1$ and $i,j=1,...,4$.

After the gate transformation of the vector $\bold{r}$,  the WF of the state at the output of the gate:
\begin{align}
&W\s{out}(\bold{r})=\frac{1}{n^2}\sum\limits_{k,l=0,1}(-(n+1))^{k+l}\text{W}(V\up{out}_{k,l}, \bold{r}).
\end{align}

Let us derive  the approximated WF of the HOM state.
Since $W\s{HOM}( \bold{r})$ is the WF of the state that would be at the output of the 1:1 beamsplitter  if at the input  there were two single-photon states,
we use the same approach as we used to obtain $W\s{out}(\bold{r})$ and get
\begin{align}
& W\s{HOM}( \bold{r})=
\frac{1}{n^2}\sum\limits_{k,l=0,1}(-(n+1))^{k+l}W(V'_{k,l}, \bold{r}),\\
&
V'_{k,l}=T\s{bs}V\up{in}_{k,l}T\s{bs}^T, \qquad T\s{bs}= \frac{\sqrt{2}}{2}
\left(
  \begin{array}{cccc}
  1 & 0 & 1 & 0 \\
  0 & 1 & 0 & 1 \\
  - 1 & 0 & 1 & 0 \\
  0 & - 1 & 0 & 1
  \end{array}
\right).
\end{align}
To calculate the HOM matrix element  let us use the  rule $\int d \bold{r} \;  \exp\{-\frac{1}{2}\bold{r}^TV^{-1}\bold{r}\}=4\pi^2\sqrt{ \text{det}[V]}$ that works for every symmetric positive-definite matrix $V$.
Then, keeping in mind $\bold{r} ^T(A+B)\bold{r} =\bold{r} ^TA\bold{r} +\bold{r} ^TB\bold{r}$,  we can calculate the HOM matrix element as:

\begin{align}
& \mathsf{M}\s{HOM}=\frac{ 4}{n^4}\sum\limits_{k,l,m,d=0,1}(-(n+1))^{k+l+m+d} \cdot F(V'_{k,l },V\up{out}_{m,d }),\qquad  F(A,B)=\(\sqrt{\det [A+B]}\)^{-1}.\label{Approx}\\
\end{align}
%

%%%%%%%%%%%%%
Using the same approach we can derive the HOM element for the case of the the mixture state input:
\begin{align}
& \mathsf{M}\s{HOM}=\frac{ 4}{n^4}p^2\sum\limits_{k,l,m,d=0,1}(-(n+1))^{k+l+m+d} \cdot F(V'_{k,l },V\up{out}_{m,d })+\nn\\
&+\frac{ 4}{n^3}p(1-p)\sum\limits_{k,l,m=0,1}(-(n+1))^{k+l+m} \cdot F(V'_{k,l },V\up{out}_{m,0 })+\nn\\
&+\frac{ 4}{n^3}p(1-p)\sum\limits_{k,l,d=0,1}(-(n+1))^{k+l+d} \cdot F(V'_{k,l },V\up{out}_{0,d })+\nn\\
&+\frac{ 4}{n^2}(1-p)^2\sum\limits_{k,l=0,1}(-(n+1))^{k+l} \cdot F(V'_{k,l },V\up{out}_{0,0 }).
\end{align}

%%%%%%%%%%%%%%%%%%%%%%%%%%%%%%%%%%%%%%%%%%%%%%%%%%%%%%%%%%%%%%%%%%

%%%%%%%%%%%%%%%%%%%%%%%%%%%%%%%%%%%%%%%%%%%%%%%%%%%%%%%%%%%%%%%
\subsection{Input and output thresholds (nonclassicality borders)}
%%%%%%%%%%%%%%%%%%%%%%%%%%%%%%%%%%%%%%%%%%%%%%%%%%%%%%%%%%%%%%%

We define two nonclassicality thresholds by evaluating the maximum of $\langle \text{HOM}|\rho\s{out}|\text{HOM}\rangle$ over (i) all superpositions of the coherent states at the output of the QND interaction: $\rho\s{out} = \rho\s{coh} = \dyad{\alpha\s{a} \beta\s{b}}{\alpha\s{a} \beta\s{b}}$ and (ii) before the QND interaction: $\rho\s{out} = U\s{QND} \rho\s{coh} U\s{QND}^\dag$.
For the BS interaction, such thresholds coincide.

If the output state $\rho\s{out}$ is classical, i.e. is a mixture
of coherent states, then it turns out that $0\leq\langle \text{HOM}| \rho\s{out}|\text{HOM}\rangle\leq1/e^2$.
Thus, $1/e^2$ is the {\em output threshold} for the HOM interference.
That is, when measuring the HOM element, if we get a value greater than $1/e^2$, then the measured state for sure is a non-classical one.
This threshold is shown by a thin gray dashed line in the Fig.~\ref{FigSuperposition}(a).

To derive the \emph{input threshold}, let us use two random coherent states as the input states of the gate and calculate the HOM element for the output state $ \rho\up{coh}\s{out}=U \rho\up{coh}U^\dag$.
We assume phase-randomized input state which means that phases of the input coherent states are averaged.
If we examine the dependence of the $\langle \text{HOM} | \rho\up{coh}\s{out}|\text{HOM} \rangle$ on the gain $\sfG$ for all the coherent states (see Fig.~\ref{FigSuperposition}(b)) we will obtain
the area restricted by an {\em input threshold} (blue curve) that has a specific complex shape.

To calculate the input threshold (phase randomized)
let us take two random coherent states as the inputs of the QND gate.
The WF of the input state (two independent coherent states) is the following
\begin{align}
W_{in}(\bold{r},\bold{R})=\frac{1}{4\pi^2 \sqrt{\text{det}V_{in}}} \exp\(-\frac{1}{2}(\bold{r}- \bold{R})^T V_{in}^{-1} (\bold{r}- \bold{R})\).
\end{align}
Here $\bold{r}=\(x\s{a},p\s{a},x\s{b},p\s{b}\)^T$, $\bold{R}\equiv  \(X\s{a},Y\s{a},X\s{b},Y\s{b}\)^T$ is the colomn-vector of means and
$V_{in}=\mathbb{I}_{4\times4}$ is the covariance matrix  of the initial state (coherent).
The vector of means changes as $\bold{R}\rightarrow T_\mathsf{G}\bold{R}$.
Thus, the WF of the output state is as follows:
\begin{align}
W_{out}(\bold{r},\bold{R}, \mathsf{G})=\frac{1}{4\pi^2 \sqrt{\text{det}V\s{out}}} \exp\(-\frac{1}{2}(\bold{r}-T_\mathsf{G}\bold{R})^T V\s{out}^{-1} (\bold{r}-T_\mathsf{G}\bold{R})\).
\end{align}
Using this WF we can obtain the HOM element as
\begin{align}
& \mathsf{M}\s{HOM}(\bold{R}, \mathsf{G} )= \langle \text{HOM}| \rho\s{out}\text{HOM}\rangle\equiv
(4\pi)^2 \iiiint d\bold{r}\; W\s{HOM}(\bold{r})\cdot W\s{out}(\bold{r},\bold{R}, \mathsf{G}).
\end{align}

If, calculating the matrix elements for $\rho^{coh}_{out}$ over all the possible coherent states averaged over phases, then the range of possible values for the matrix elements will significantly change.
Assuming $\bold{R}=  \(X\s{a},Y\s{a},X\s{b},Y\s{b}\)^T= \(R\s{a} \cos[\varphi\s{a}],R\s{a} \sin[\varphi\s{a}],R\s{b} \cos[\varphi\s{b}],R\s{b} \sin[\varphi\s{b}]\)^T$ we can obtain

\begin{align}
& \mathsf{M}\up{av}\s{HOM}(R\s{a}, R\s{b}, \mathsf{G})=
\frac{1}{4\pi^2} \iint d\varphi\s{a}\;d\varphi\s{b}\;\;  \mathsf{M}\s{HOM}(\bold{R}, \mathsf{G} )
\end{align}

and investigate it over all possible $R\s{a}, R\s{b}$  for the certain $\mathsf{G}$.

%%%%%%%%
\begin{figure}[ht]
\begin{center}
\begin{minipage}[ht]{0.45\linewidth}
  \center{a\\ \includegraphics[width =\linewidth]{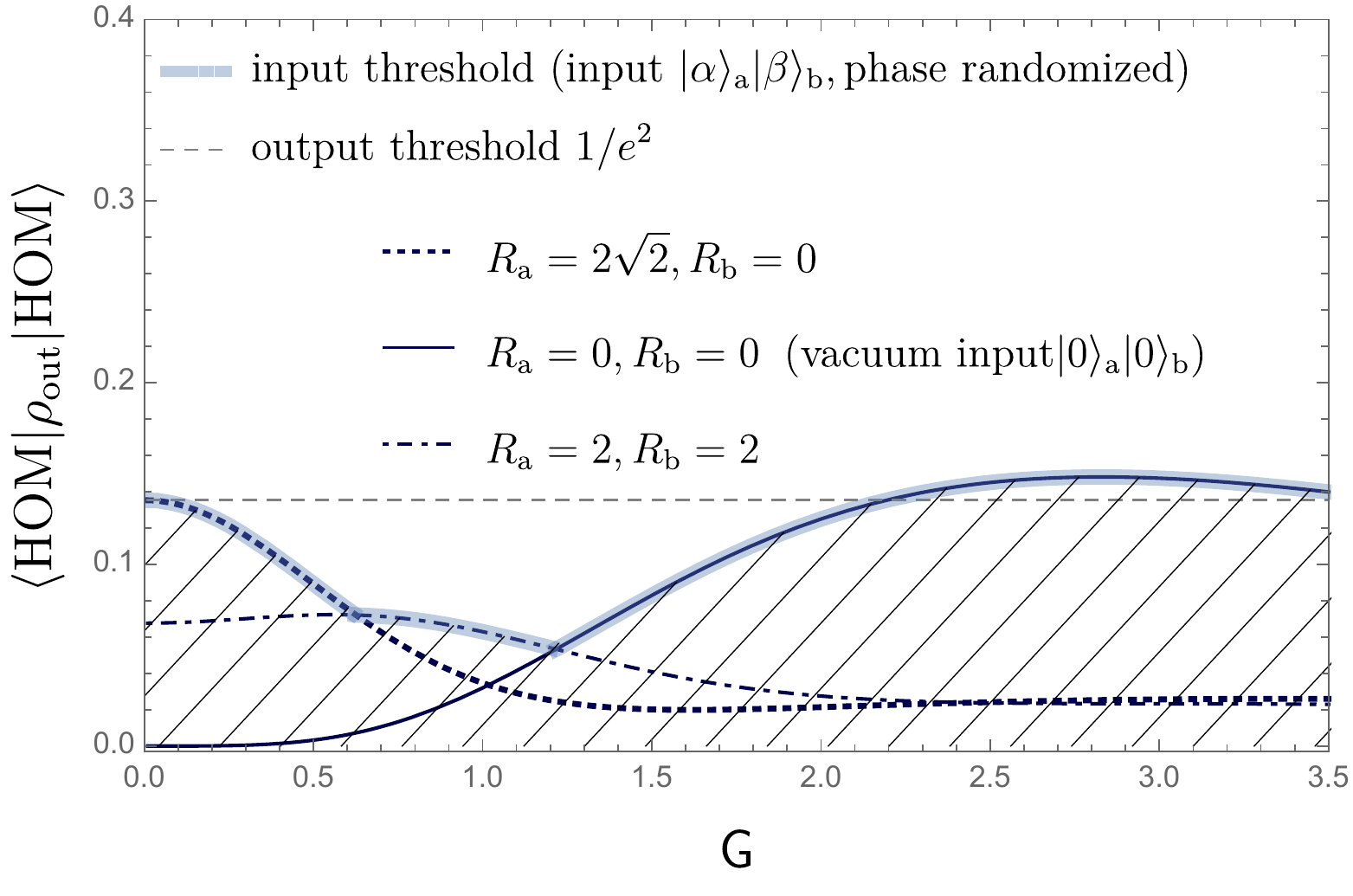}}

  \end{minipage}
\begin{minipage}[ht]{0.45\linewidth}
  \center{b\\ \includegraphics[width =\linewidth]{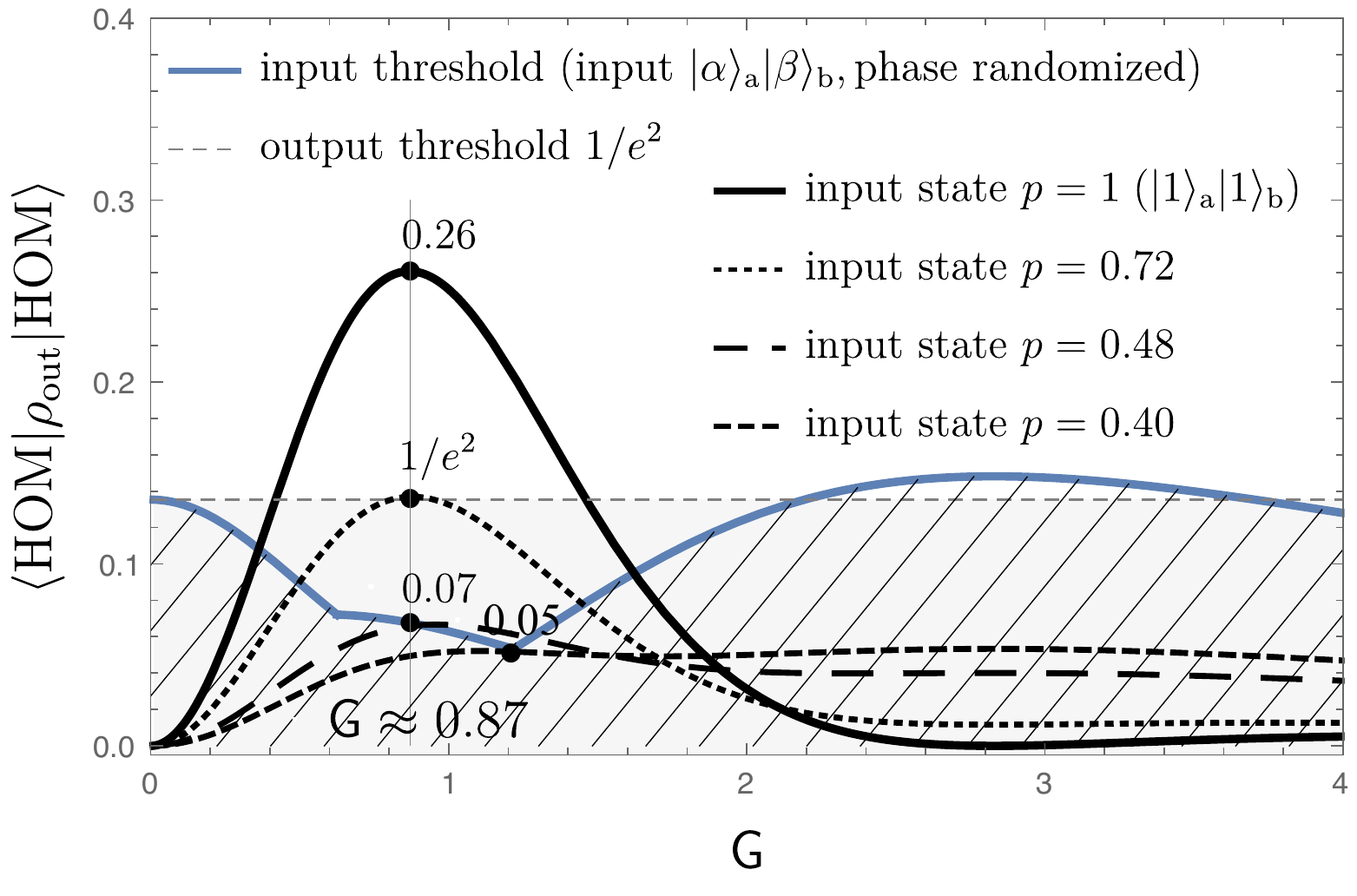}}
  \end{minipage}
\caption
{
a) Shape of the input coherent threshold.
b) $\langle \text{HOM}| \rho\s{out} |\text{HOM}\rangle$ matrix element of the output state for the ideal QND gate as a function of the gain $\mathsf{G}$ calculated for the different cases of the input:
quantum input $|1\rangle_\text{a} |1\rangle_\text{b}$ (solid black curves),
mixture input Eq.~(\ref{ME_superpositioninput_Eq1}) (dashed black curves, dashing scale indicates parameter $p$).
Dashed gray line is the {\em output threshold}.
Blue curve is the {\em input thresholds} (phase randomized) restricting area
that covers all the possible values of the matrix elements of the output state of the gate in the case of the random coherent input with averaged phases.
}
\label{FigSuperpositionSup}
\end{center}
\end{figure}

Figure~\ref{FigSuperpositionSup}(b) demonstrates $\langle \text{HOM}| \rho\s{out} |\text{HOM}\rangle$ depending on the gain, assuming $p\s{a}=p\s{b}=p$,
compared with the case of the pure input $|1\rangle_a|1\rangle_b$.
%%%%%%%%
Expectedly, as the parameter $p$ decreases, the contribution from $|1\rangle\s{a}|1\rangle\s{b}$ term decreases, while the contribution from $|0\rangle\s{a}|0\rangle\s{b}$ term increases.
Visually, it is reflected in the gradual change of the curves' shape --
for relatively high $p$ maximum first decreases, then smoothly shifts to the right.
Thus, the maximum of the HOM element decreases from $0.26$ at $p=1$ to $1/e^2$ at $p \approx 0.7$, which corresponds to the output threshold.
At $p\approx0.48$ it already crosses the input threshold, so for $p<0.48$ the HOM element lies below the input threshold at the gain $\mathsf{G}\approx0.87$.
For $p<0.40$ the HOM element lies below the input threshold for any gain.

%%%%%%%%
\begin{figure}[ht]
\begin{center}
 \includegraphics[width =0.9\linewidth]{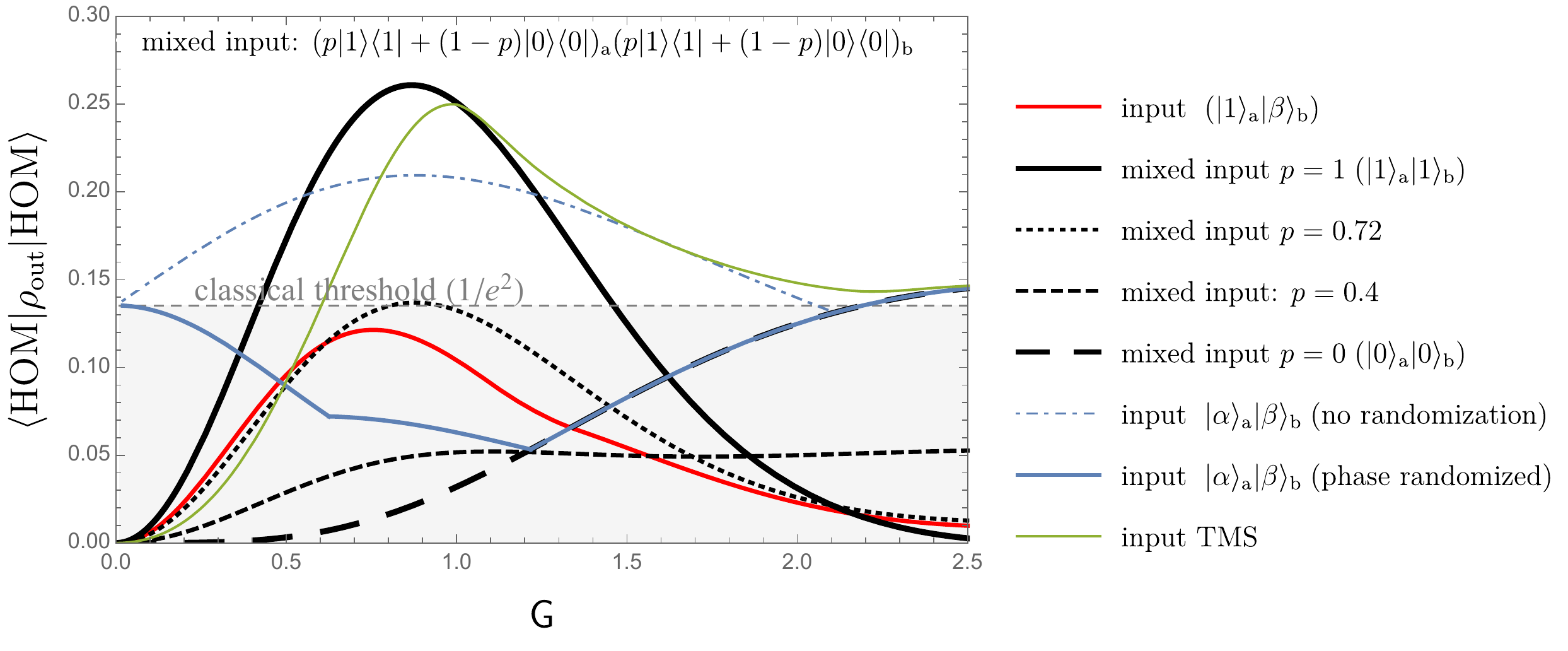}
\caption
{
 $\langle \text{HOM}| \rho\s{out} |\text{HOM}\rangle$ matrix element of the output state for the ideal QND gate as a function of the gain $\mathsf{G}$ calculated for the different cases of the input:
 the vacuum input $|0\rangle_\text{a}|0\rangle_\text{b}$(black dashed),
the state with one-boson at the first input and coherent at the second $|1\rangle_\text{a}|\beta\rangle_\text{b}$ (red),
two independent coherent states at the input $|\alpha\rangle_\text{a}|\beta\rangle_\text{b}$  (blue dot-dashed),
phase averaged coherent states at the input $|\alpha\rangle_\text{a}|\beta\rangle_\text{b}$  (blue),
two-mode squeezed state TMS (green).
The curve of the vacuum input depends only on the gain, all other inputs depend on many parameters (phases, displacements, squeezings etc), so their curves are the highest possible values (thresholds) of the HOM element calculated over all their parameters.
}
\label{App9}
\end{center}
\end{figure}

\end{document}